\documentclass[twocolumn,preprintnumbers,amsmath,amssymb,superscriptaddress,prb]{revtex4}
\usepackage{graphicx}
\usepackage{amsmath}
\usepackage{amssymb}
\usepackage{appendix}
\graphicspath{{figures/}}
\newcommand{\bra}[1]{\langle#1|}
\newcommand{\ket}[1]{|#1\rangle}

\renewcommand\d{\mathrm{d}}

\begin{document}

\title{Multi-hole models for deterministically placed acceptor arrays in silicon}
\author{Jianhua Zhu}\email{ucapjhz@ucl.ac.uk}
\affiliation{UCL Department of Physics and Astronomy and London Centre for Nanotechnology,\\
University College London, Gower Street, London WC1E 6BT, United Kingdom}
\author{Wei Wu}\email{wei.wu@ucl.ac.uk}
\affiliation{UCL Institute for Materials Discovery, University College London,\\
Gower Street, London WC1E 6BT, United Kingdom}
\author{Andrew J. Fisher}\email{andrew.fisher@ucl.ac.uk}
\affiliation{UCL Department of Physics and Astronomy and London Centre for Nanotechnology,\\
University College London, Gower Street, London WC1E 6BT, United Kingdom}

\begin{abstract}

In this paper, we compute the electronic structure of acceptor clusters in silicon by using three different methods to take into account electron correlations: the full configuration interaction (full CI calculation), the Heitler-London approximation (HL approximation), and the unrestricted Hartree-Fock method (UHF method). We show that both the HL approach and the UHF method are good approximations to the ground state of the full CI calculation for a pair of acceptors and for finite linear chains along [001], [110] and [111]. The total energies for finite linear chains show the formation of a 4-fold degenerate ground state (lying highest in energy), below which there are characteristic low-lying 8-fold and 4-fold degeneracies, when there is a long (weak) bond at the end of the chain. We present evidence that this is a manifold of topological edge states.  We identify a change in the angular momentum composition of the ground state at a critical pattern of bond lengths, and show that it is related to a crossing in the Fock matrix eigenvalues. We also test the symmetry of the self-consistent mean-field UHF solution and compare it to the full CI; the symmetry is broken under almost all the arrangements by the formation of a magnetic state in UHF, and we find further broken symmetries for some particular arrangements related to crossings (or potential crossings) between the Fock-matrix eigenvalues in the [001] direction. We also compute the charge distributions across the acceptors obtained from the eigenvectors of the Fock matrix; we find that, with weak bonds at the chain ends, two holes are localized at either end of the chain while the others have a nearly uniform distribution over the middle; this also implies the existence of the non-trivial edge states. We also apply the UHF method to treat an infinite linear chain with periodic boundary conditions, where the full CI calculation and the HL approximation cannot easily be used. We find the band structures in the UHF approximation, and compute the Zak phases for the occupied Fock-matrix eigenvalues;  however, we find they do not correctly predict the topological edge states formed in this interacting system. On the other hand, we find that direct study of the quantum numbers characterising the edge states, introduced by Turner \textit{et al.},  provides a better insight into their topological nature.

\end{abstract}

\maketitle
\section{Introduction}\label{introduction}
In the last few decades, studies of defects in semiconducting systems have broadened to include applications to quantum computation and quantum simulation as well as their more traditional role in doping for classical electronics. Donors are especially well studied\cite{Koiller2002EiSBQCA,Pantelides1974ToLSiSINRUaOM}, but in materials such as Si having degenerate conduction-band minima they suffer from the disadvantage of inter-valley interferences causing rapid oscillations in the wave-functions and hence also in hopping or exchange interactions, leading to extreme sensitivity to the precise dopant position.  For this reason, and because the spin-orbit coupling present in the valence band provides some additional opportunities to interact with the spin degrees of freedom, acceptors have recently attracted increased attention\cite{Salfi_2016,Hendrickx2021AFQGQP,Wang2021OOPfUHCGHSOQ}.

The electronic structure of a single acceptor can most simply be described by the `spherical model'\cite{Luttinger1956QToCRiSGT,Baldereschi1973SMoSASiS,Baldereschi1974CCttSMoSAS,Lipari1978IoASiS}, which includes spin-orbit coupling but neglects the cubic anisotropy of the host semiconductor and offers reasonable results for the electronic structure of an isolated acceptor. The interactions between a pair of acceptors in the spherical model have been studied by Durst, et. al., in the framework of a Heitler-London model\cite{C.Durst2017HLMfAAIiDS}; later, they found that the inter-acceptor interactions in the same model are dominated at large distance by electric quadrupole moments \cite{Durst2020QIbAPiDS}. For linear chains of acceptors, an independent-hole model was developed, including the contribution of cubic terms, and the existence of non-trivial single-particle topological edge states was demonstrated for finite chains, and related to band invariants of the corresponding infinite systems \cite{Zhu2020LCoAOMfDPAAiS}. These investigations of pairs and linear arrays of acceptors suggest that the emerging techniques of deterministic doping \cite{Schofield2003APPoSDiS} could lead to interesting results if applied to acceptors \cite{Dwyer2021}. Advances in the experimental characterisation of acceptors in silicon include measurements of the optical transitions and spectra of acceptors\cite{Clauws1989OSoSISiGaS}, measurement of the coherence time of the excited state of acceptors\cite{Vinh2013TRDoSATiS}, and a study of transport properties of boron-doped material \cite{Dai1992ECoMSntMIT}, etc. The readout and control of the spin-orbit state of two coupled acceptors has also been demonstrated experimentally in silicon, suggesting a possible alternative route to quantum computing \cite{Litvinenko2015CCaDoOWiSwEaORO, Heijden2018RaCotSOSoTCAAiaST,Corna2018EDESRMbSVOCiaSQD,Crippa2018ESDbMMiSOQ,Crippa2019GRDRaCCoaSQiS}. Finally, there is great potential for applications in the simulation of fermionic strongly-correlated many-body systems using acceptors \cite{Salfi2016QSotHMwDAiS}. All these investigations imply that  a system of acceptors in well-defined locations could offer some unique properties in its electronic structure.

In this paper, we construct and solve multi-hole models (including hole-hole Coulomb interactions) for lines of acceptors in silicon with one hole per acceptor, along three high-symmetry directions ([001], [110] and [111]), based on three different methods: full configuration interaction calculation (full CI, in \S\ref{ci}), the Heitler-London approximation (HL approximation, in \S\ref{ci}) based on the full CI calculation but with a restricted basis, and the unrestricted Hartree-Fock method (UHF method, in \S\ref{uhf}) which represents the multi-hole state by a Slater determinant of one-hole states. Some limitations of the full CI calculation and the HL approximation are discussed in \S\ref{ci}. We study dimerised chains with staggered bond lengths $d_1$ and $d_2$ and concentrate on a 'small-separation' case with $d_1+d_2=3a_0$ and a 'large-separation' case with $d_1+d_2=6a_0$ where $a_0$ is the effective Bohr radius; we show that both the HL approach and the UHF method are accurate approximations to the ground state of the fully exact CI calculation for these finite-length linear chains. We investigate the energy spectrum obtained from full CI for a 4-acceptor chain and explain the ground state in terms of the formation of edge states; we also relate an anti-crossing in the [001] direction for the small-separation case to the behavior of the Fock matrix eigenvalues obtained from the UHF method.  We analyse the symmetries of the states produced by symmetry breaking in the UHF solution, and present evidence for the existence of non-trivial many-body edge states in the finite chain system. We point out that the UHF method can be applied to a linear chain with periodic boundary conditions, and calculate the band structure formed by the Fock matrix eigenvalues.   We also analyse the topological phases of the system based on two methods: first, a method focusing on the edge states of finite one-dimensional interacting Fermionic systems, and second, the Zak phase \cite{Zak1989BPfEBiS} for an infinite non-interacting system. 

\section{Computational details}\label{theory}

\subsection{Multi-hole models}\label{model}
In our previous paper\cite{Zhu2020LCoAOMfDPAAiS}, we developed a one-hole model to describe a pair of acceptors and a linear acceptor chain. Here, we use the same approach to describe the one-hole part of the Hamiltonian, including cubic anisotropy, but only considering the nearest transitions for the chain (see \S\ref{cutoff}).  We then combine this one-hole Hamiltonian with two-hole terms representing the inter-hole Coulomb repulsion, using methods described in Reference~\onlinecite{Clementi1966ESoLMS}.
Our units of energy and length are the effective Rydberg $R_0=\frac{e^4m_0}{2\hbar^2\epsilon^2_0\gamma_1}$ and the effective Bohr radius $a_0=\frac{\hbar^2\epsilon^2_0\gamma_1}{e^4m_0}$, respectively\cite{Baldereschi1973SMoSASiS}.  We use parameters appropriate for silicon throughout; however, our methods are easily transferable to other cubic semiconductors.  With these silicon parameters, $R_0=24.8\,\mathrm{meV}$ and $a_0=2.55\,\mathrm{nm}$.  In all cases we report our results for lines oriented along the three highest-symmetry directions of the cubic host: [001], [110] and [111].

\subsubsection{Full configuration interaction calculation (full CI calculation) and Heitler-London approximation (HL approximation)}\label{ci}
The configuration interaction calculation (full CI) retains a basis of Slater determinants corresponding to all possible configurations of the holes distributed across basis states on all acceptors, and the Hamiltonian is
\begin{equation}
\hat{H}_{CI}=\sum_{i}\hat{H}_i-\sum_{i<j}\frac{2}{r_{ij}}\label{e-2-1-1-1},
\end{equation} 
where $\hat{H}_i$ is the one-hole Hamiltonian from our previous paper\cite{Zhu2020LCoAOMfDPAAiS}, $\frac{2}{r_{ij}}$ is the hole-hole interaction in effective Rydberg units, and $i,j$ label the holes. The interaction appears with a minus sign because the Hamiltonian is expressed for electron states.  Therefore, throughout this paper, the most favourable states for occupation by holes are those with the \textit{highest} energy---we refer to the highest-energy state as the `ground state'.  The overlap matrix is also needed and can be written as
\begin{equation}
\hat{S}_{CI}=\hat{S}_1\otimes\hat{S}_2\otimes\cdots\otimes\hat{S}_N\label{e-2-1-1-2},
\end{equation}
where $\hat{S}_i$ is the overlap matrices for the one-hole model, and $N$ is the number of holes.

The full CI calculation is exact for a given choice of single-particle basis, but scales very badly (super-exponentially) with the size of the system. Also, the total energy expression is not extensive so it cannot be implemented under periodic boundary conditions.  The first problem is ameliorated by restricting the set of configurations to those with exactly one hole per acceptor; we call this the Heitler-London (HL) approximation because it is in the same spirit as the Heitler-London treatment of the $\mathrm{H}_2$ molecule, and has been used for acceptors pairs in Reference~\onlinecite{C.Durst2017HLMfAAIiDS}. The many-particle basis set now grows more slowly (although still exponentially), but the difficulty in treating the infinite system still remains.

\subsubsection{Unrestricted Hartree-Fock method}\label{uhf}
To handle the infinite system we employ an unrestricted Hartree-Fock (UHF) method, where the many-hole wave-function is optimised over single Slater determinants constructed from a set of one-hole functions, without any restriction on the spin components of each function.   The optimisation of the one-hole functions results in a self-consistent-field (SCF) approach, where each hole can be understood to experience the average interaction of the others. The one-hole functions are eigenfunctions of the Fock matrix $\hat{F}$, which is given by
\begin{equation}
\hat{F}=\hat{H}^{\mathrm{core}}+\hat{G}\label{e-2-1-2-1},
\end{equation}
where $\hat{H}^{\mathrm{core}}$ is the Hamiltonian for the one-hole model (including spin-orbit coupling), and $\hat{G}$ is a matrix reflecting the self-consistent influence from other holes. If we expand all quantities in terms of a set of single-hole basis functions $|\phi_{\mu}\rangle$, $G$ is given by
\begin{equation}
G_{\mu\nu}=\sum_{\lambda\sigma}P^{\lambda\sigma}\left(\left(\mu\nu\|\sigma\lambda\right)-\left(\mu\lambda\|\sigma\nu\right)\right)\label{e-2-1-2-2},
\end{equation}
where $\mu,\nu,\sigma,\lambda$ are labels running over all basis functions on all acceptors, 
\begin{equation}
\left(\mu\nu\|\sigma\lambda\right)=\int\d\mathbf{x}_1\d\mathbf{x}_2\,\phi_\mu^*(\mathbf{x}_1)\phi_\sigma^*(\mathbf{x}_2)\frac{-2}{|\mathbf{r}_1-\mathbf{r}_2|}\phi_\nu(\mathbf{x}_1)\phi_\lambda(\mathbf{x}_2)
\end{equation}
(where $\mathbf{x}=(\mathbf{r},\tau)$ is a composite coordinate for position $\mathbf{r}$ intrinsic angular momentum $\tau$, and $\left(\mu\nu\|\sigma\lambda\right)$ is the notation used in Reference~\onlinecite{Szabo1982MQCItAEST}) are matrix elements of the Coulomb interaction, and $P$ is the one-hole density matrix which can be constructed as
\begin{equation}
P^{\mu\nu}=\sum^N_{i}C_{i}^{\mu}{C^*}_{i}^{\nu}\label{e-2-1-2-3},
\end{equation}
where $C_i$ is an eigenvector of the generalised eigenproblem
\begin{equation}
F\cdot C_i=\epsilon_i S\cdot C_i,
\end{equation} 
$N$ is the number of holes (hence the number of occupied eigenvectors), and $i$ goes through all eigenvector labels. Once again, because our calculation is describing holes, the single particle states are occupied according to the \textit{aufbau} principle from the highest eigenvalue downwards.  The total energy can then be written as
\begin{equation}
E_{\textrm{tot}}=\frac{1}{2}\sum_{\mu\nu}P^{\nu\mu}\left(H^{\mathrm{core}}_{\mu\nu}+F_{\mu\nu}\right)\label{e-2-1-2-4},
\end{equation}
The self-consistent calculation continues until the output density matrix (\ref{e-2-1-2-3}) is similar to the input one used in (\ref{e-2-1-2-2}). Further details can be found in Reference~\onlinecite{Szabo1982MQCItAEST}; however, in contrast to the conventional case, our system contains spin-orbit coupling and therefore we cannot separate the single-particle functions into separate sets corresponding to each spin component. So it is necessary to include exchange interactions between \textit{all} pairs of single-hole states, not just those of the same spin.

\subsubsection{Periodic boundary conditions}
Although less accurate than the CI method, the UHF method does not have the limitations mentioned in \S\ref{ci}.  It scales polynomially, rather than exponentially, as the system size increases, and the total energy expression (\ref{e-2-1-2-4}) is extensive. So it is possible to apply it to a linear chain with periodic boundary conditions. In this case, the Fock matrix $\hat{F}_k$ at a particular Bloch wavevector $k$ will be
\begin{equation}
\hat{F}_k=\sum_{X}e^{ikX}\hat{F}_X=\sum_{X}e^{ikX}(\hat{H}_X^{\mathrm{core}}+\hat{G}_X)=\hat{H}^{\mathrm{core}}_k+\hat{G}_k\label{e-2-1-2-5},
\end{equation}
where $X$ labels lattice displacements of a single unit cell, $\hat{F}_X$, $\hat{H}_X^{\mathrm{core}}$ and $\hat{G}_X$ are the elements of $F$, $H$ and $G$ connecting different cells separated by $X$, and $\hat{H}^{\mathrm{core}}_k$ and $\hat{G}_k$ are the matrices of $\hat{H}^{\mathrm{core}}$ and $\hat{G}$ in momentum space.  The Fock matrix $\hat{F}_k$ can be diagonalised to find a set of eigenvectors $C_{ki}$, and the corresponding contribution $P_k$ to the the one-particle density matrix is
\begin{equation}
P^{\mu\nu}_{k}=\sum^N_{i}C_{ki}^{\mu}{C^*}_{ki}^{\nu}\label{e-2-1-2-6}.
\end{equation}
The real-space form of $P^{\mu\nu}$ can then be recovered by inverse Fourier transformation, and re-inserted into the SCF procedure as previously.

\subsubsection{Spatial cut-offs}\label{cutoff}
In practice, the sums in equations (\ref{e-2-1-2-2}) and (\ref{e-2-1-2-5}), as well as the corresponding sums over acceptor cores in $\hat{H}^{\mathrm{core}}$ have to be truncated.  For the results in this paper we have performed this truncation after nearest neighbours; for exchange and hopping terms which involve transferring a single hole from site to site, this is justified by the relatively well localised acceptor wave-functions (this means that the relevant matrix elements will decay exponentially with hopping distance).  The Coulomb terms (both the hole-hole interaction and the hole-core interaction) decay much more slowly, like $\frac{1}{R}$ (where $R$ is the separation between the charges), but will cancel one another out provided the system is approximately charge neutral at all points.  We have checked that the key findings of this paper are reproduced in an extended model which includes all the next-nearest-neighbor transitions but only the largest next-nearest-neighbor hole-hole interactions, for both the finite length chain and periodic boundary case. There, all the key features that we are going to discuss in this paper are kept. And considering introducing the next nearest hole-hole interactions will more than double the time of the calculations, it is a wise choice to only consider the nearest neighbours.

\subsubsection{Computing the Zak phase from the unrestricted Hartree-Fock calculation}\label{zak}
The Zak phase \cite{Zak1989BPfEBiS} is a bulk quantity that indicates whether a non-interacting insulator is topologically trivial or non-trivial (supporting edge states): when it is $0$ modulo $2\pi$ the system should be trivial, when it is $\pm\pi$ modulo $2\pi$ the system becomes non-trivial.  We can obtain this quantity from the total density matrix, which is available during the SCF procedure of the UHF calculation.  We follow a recent paper\cite{Le2020TPoaDFHMfSNL}, in calculating the Zak phase for in a general situation is using the formula
\begin{equation}
Z=arg\left[tr\left(\prod_{k}S_k P_k\right)\right]\label{e-2-3-1},
\end{equation}
where $S_k$ is the overlap matrix transformed into Fourier space, $P_k$ is the single-particle density matrix as defined above, and $k$ is the wavevector going through the first Brillouin zone.  However, we have to remember that Coulomb interactions can change the topological classification \cite{Fidkowski2010EoIotTCoFFS,Turner2011TPoODFaEPoV} so we cannot necessarily expect the Zak phase to predict correctly the presence or absence of topological edge states; indeed, we show evidence in \S\ref{edge-states} that the Zak phases do not correspond to the topological property of the edge states.


\begin{table}
\centering
\caption{The eigenenergy of the 4-fold-degenerated ground state ($\Gamma^{+}_{8}$) obtained from the Gaussian expansion with 21 Gaussian parameters and 5 Gaussian parameters for Si and the difference between them; the energy unit is the effective Rydberg $R_0$, and the difference is shown in the percentage of the original 21-parameter result.}\label{t-1}
\begin{tabular}{|c|c|c|}
\hline
21-parameter result&5-parameter result&difference\\
\hline
1.868314$R_0$&1.854034$R_0$&0.7644$\%$\\
\hline
\end{tabular}
\end{table}
\begin{figure}
\centering
\includegraphics[scale=0.26]{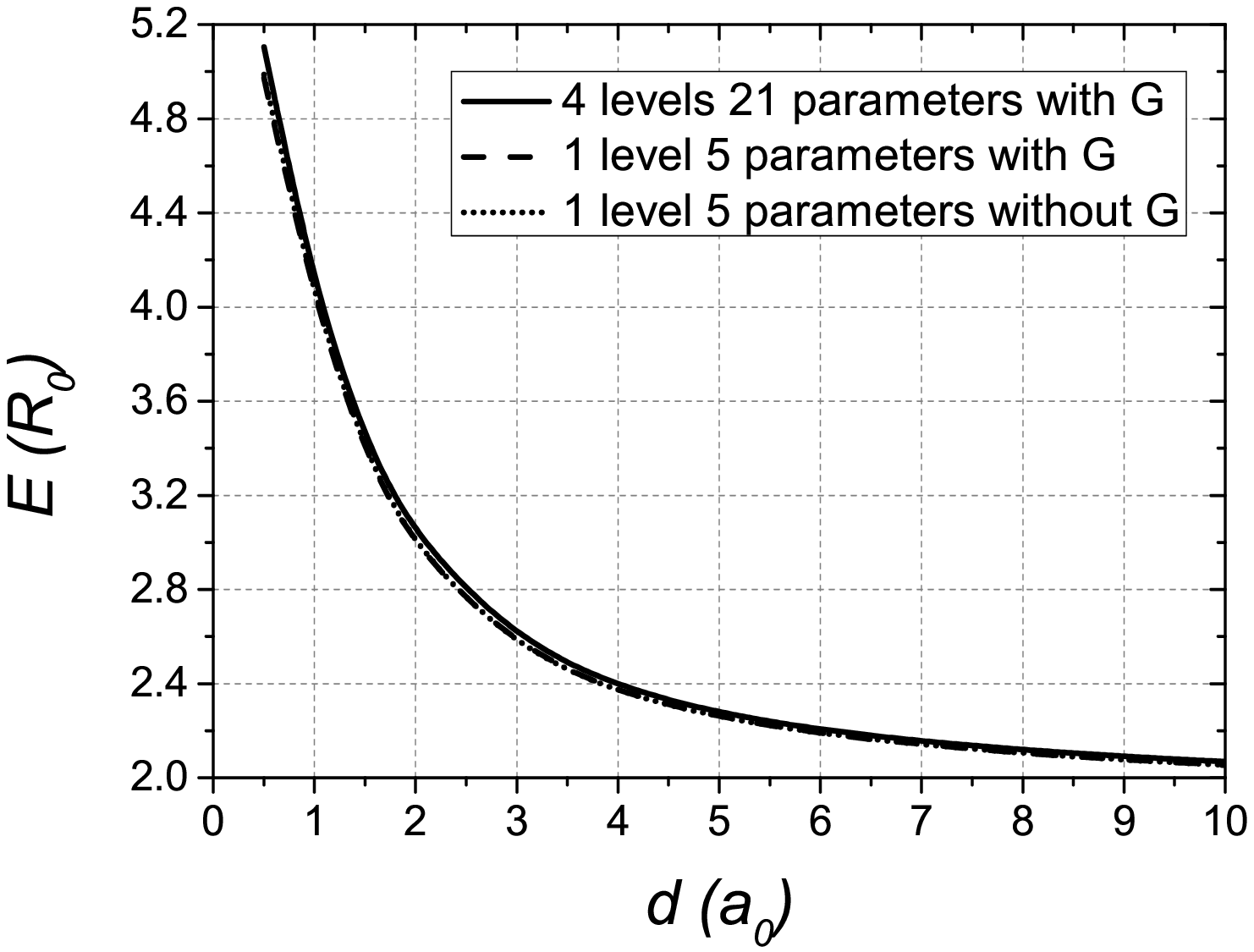}\\(a)\\
\includegraphics[scale=0.26]{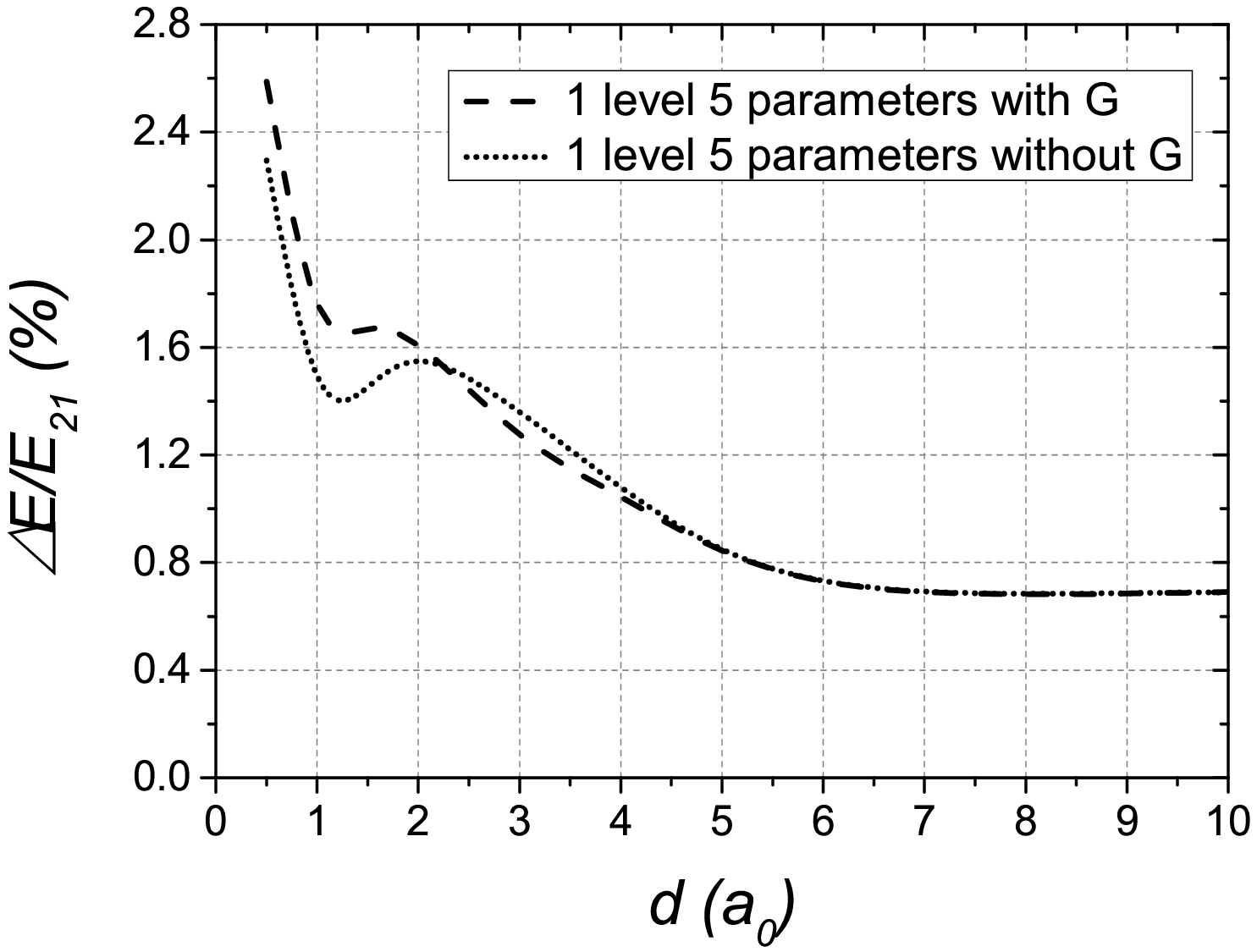}\\(b)
\caption{The behavior of the doubly-degenerate ground state energies with different approximations in the [001] direction for a pair of acceptors in Si under the one-hole model: (a) the ground state eigen energies, (b) the differences between the ground state eigen energies with different approximations. The solid line is the result of our earlier paper\cite{Zhu2020LCoAOMfDPAAiS} with 21 Gaussians, the dash lines are for the ground state ($1\Gamma^{+}_{8}$) with 5 Gaussian parameters but including $G$-orbitals, the dotted lines are for the ground state ($1\Gamma^{+}_{8}$) with 5 Gaussian parameters excluding $G$-orbitals. In (b), energy differences with respect to energy $E_{21}$ (the solid line in (a)) are shown as percentages of the energy $E_{21}$ (the solid line in (a)).}\label{f-0}
\end{figure}

\subsection{Single-particle basis}
It remains to specify the basis for the single-particle states on each acceptor.  As in our previous paper\cite{Zhu2020LCoAOMfDPAAiS}, we decompose the spatial parts of the acceptor states into linear combinations of Gaussian orbitals.  However, as we are interested in the behavior of the low-lying states of the linear chain, we make several changes.  First, we consider only the 4-fold-degenerate ground state manifold ($1\Gamma^{+}_{8}$) of an isolated acceptor. We expand the radial parts as
\begin{equation}
f_0(r)=r^l\sum_{i}A_ie^{-\alpha_ir^2}\label{e-3-1},
\end{equation}
where $l$ is the orbital angular momentum of the envelope function and $\alpha_i$ is a Gaussian exponent. Second, because we only need to describe the ground state, we use only five Gaussian functions, with exponents $\alpha_i=\{100.0, 25.0, 6.25, 1.5625, 0.390625\}$, rather than 21 as in our previous paper \cite{Zhu2020LCoAOMfDPAAiS}; the single-acceptor ground-state energies in silicon computed with 5 and 21 Gaussians are compared in Table \ref{t-1} and found to differ by less than 1\%. The reduction in the number of Gaussians saves time in the evaluation of matrix elements for the subsequent calculations. 

Finally, we remove the admixture of $G$-orbital Gaussian components ($l=4$) in the ground-state manifold, to limit the size of the matrices involved in the calculation, and re-normalize the remaining parts of the wavefunction. As an example, we compare the energy of the doubly-degenerate ground state for a single hole bound to a pair of acceptors in the [001] direction with and without the $G$-orbitals in Figure \ref{f-0}.  It can be seen that omitting the $G$-orbitals leads to errors in the energy of 1--2\%.

For convenience in the discussion of results in \S\ref{4aresult}, we assign labels to the states of the 4-fold-degenerate ground $\Gamma^{+}_{8}$ manifold so that we can distinguish them. The main contribution is from the $S_{\frac{3}{2}}$ state with total angular momentum $F=\frac{3}{2}$; we therefore use the values of the angular momentum  projections $m_F=\{\frac{3}{2}, \frac{1}{2},-\frac{1}{2},-\frac{3}{2}\}$ to label the different rows of the irreducible representation.  (The total angular momentum $\vec{F}=\vec{L}+\vec{I}+\vec{S}$, where $\vec{I}$ is the intrinsic orbital angular momentum of the $p$ states in the valence band, is as defined in Reference~\onlinecite{Zhu2020LCoAOMfDPAAiS}.)

\section{Results and discussion}\label{result}

\subsection{A pair of acceptors}\label{pairresult}

\begin{figure*}
\centering
\includegraphics[scale=0.25]{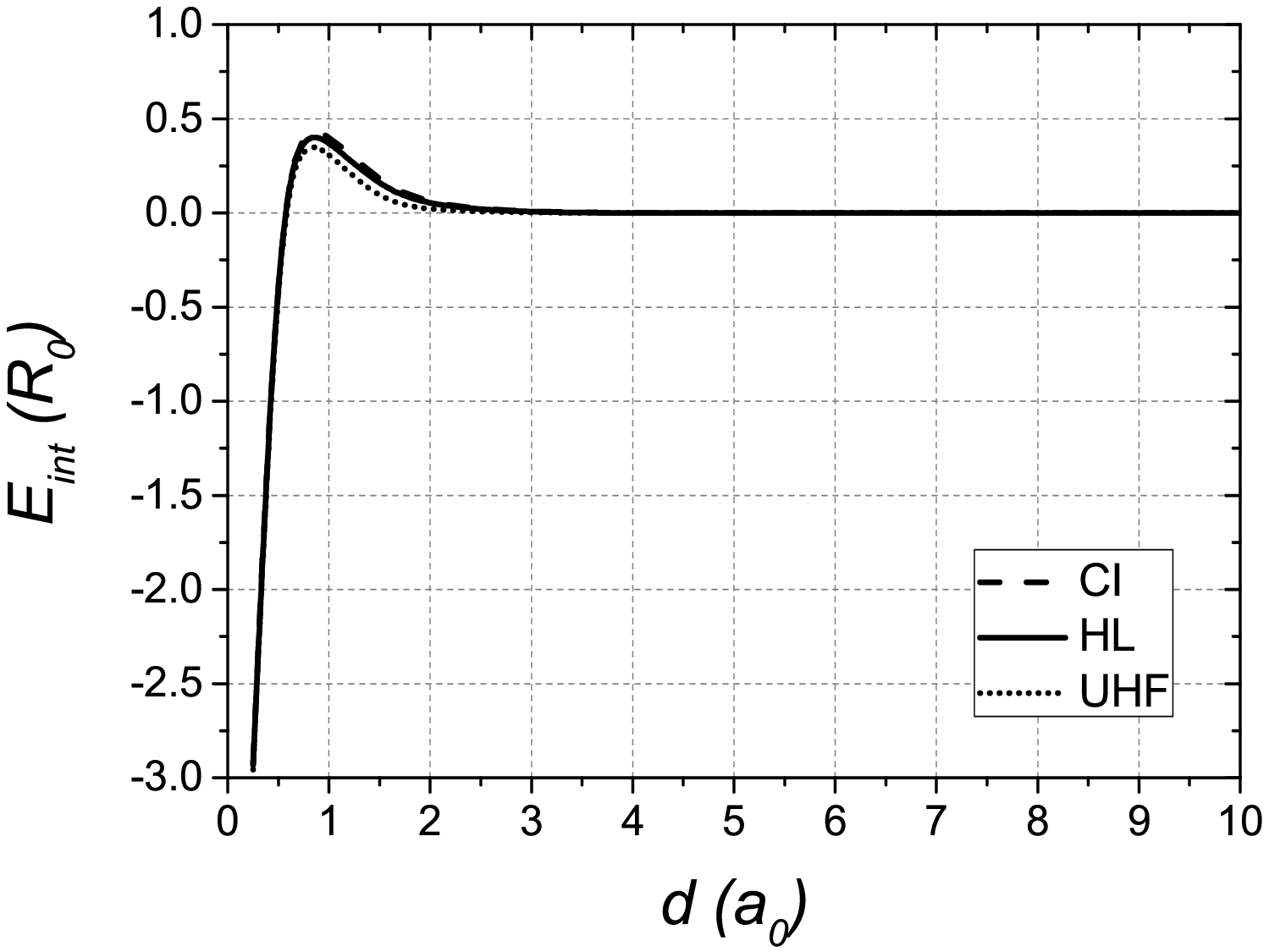}
\includegraphics[scale=0.25]{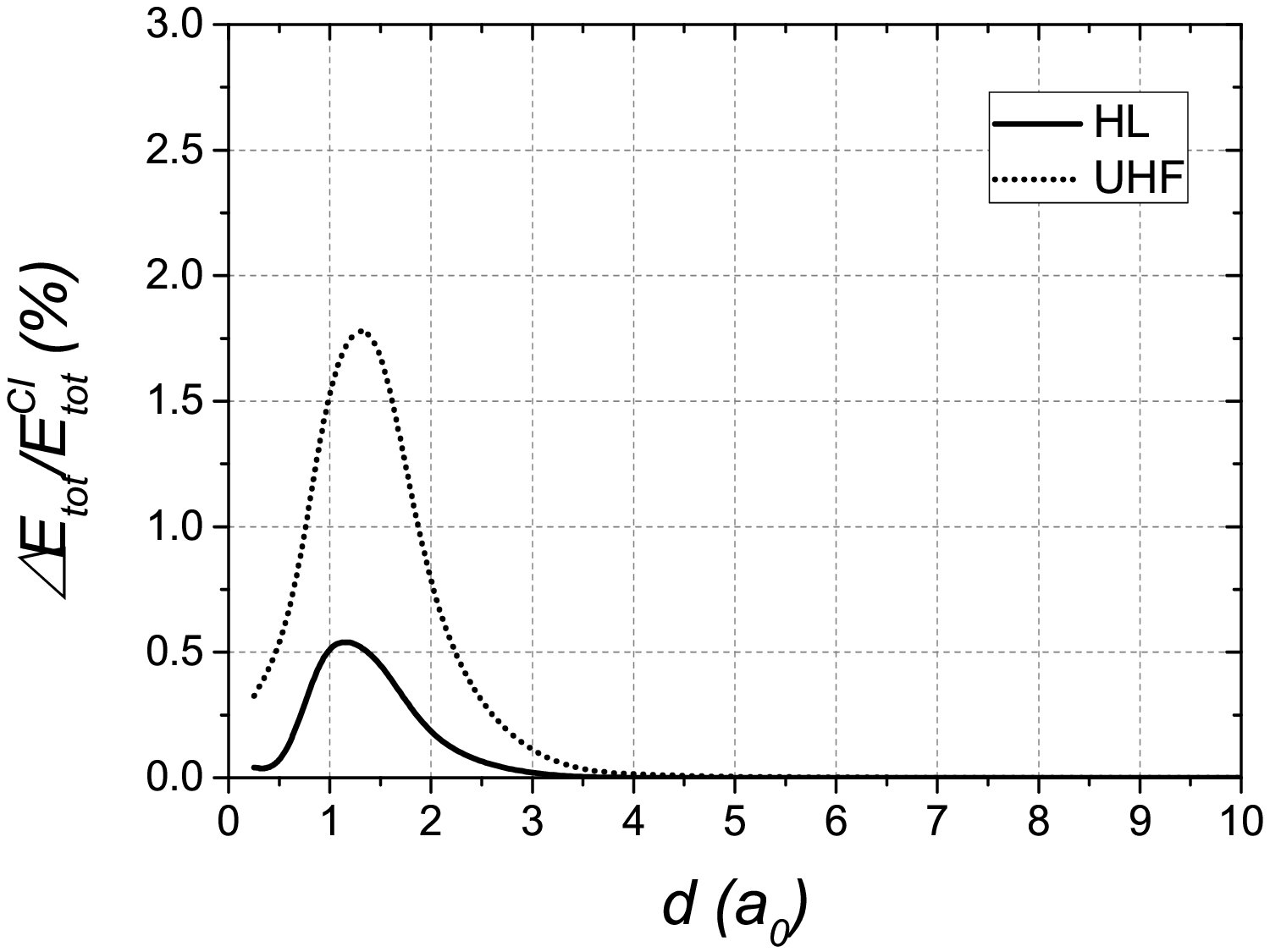}\\(a) [001] \qquad\qquad\qquad\qquad\qquad\qquad\qquad\qquad\qquad\quad (b) [001]\\
\includegraphics[scale=0.25]{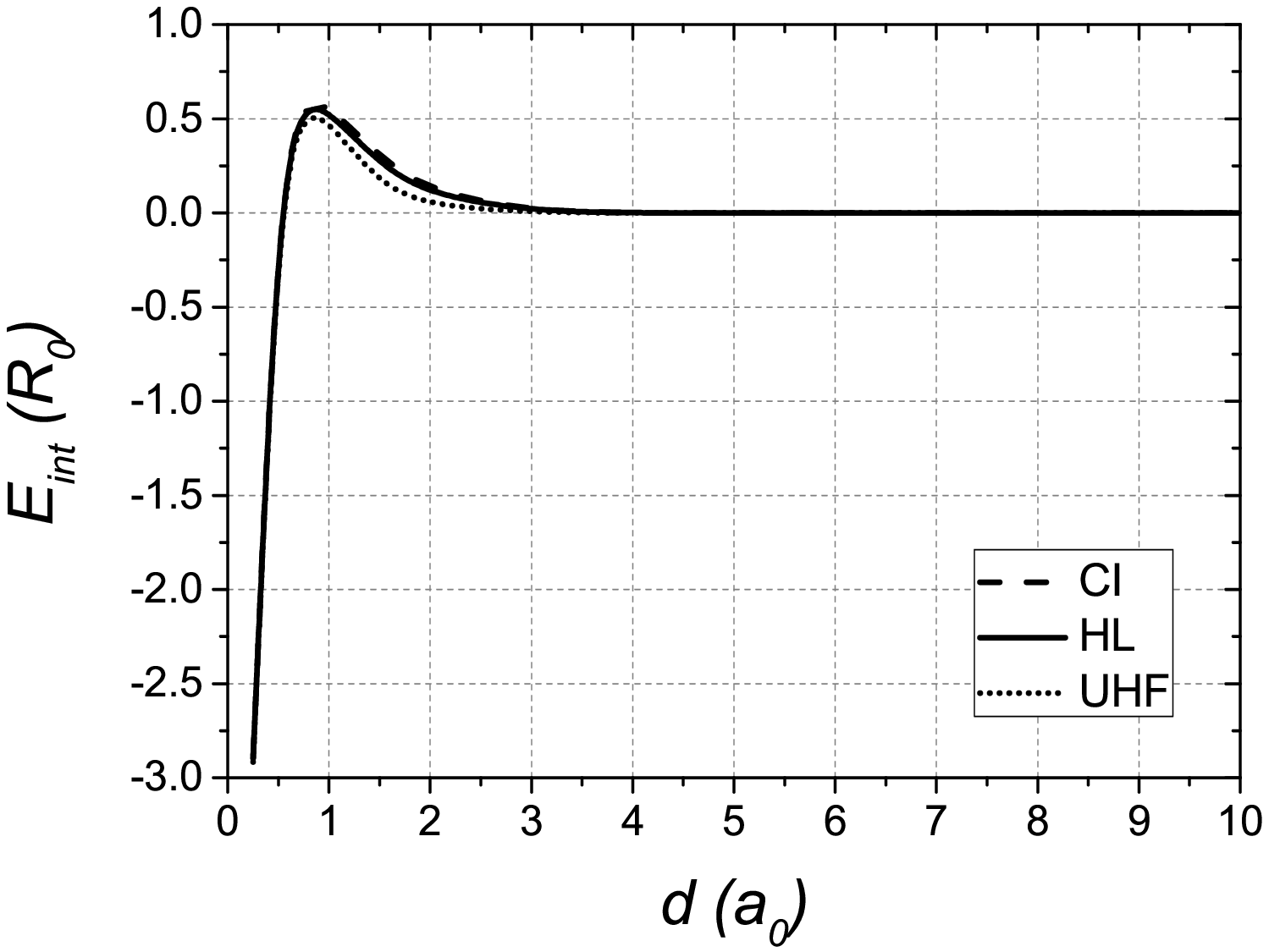}
\includegraphics[scale=0.25]{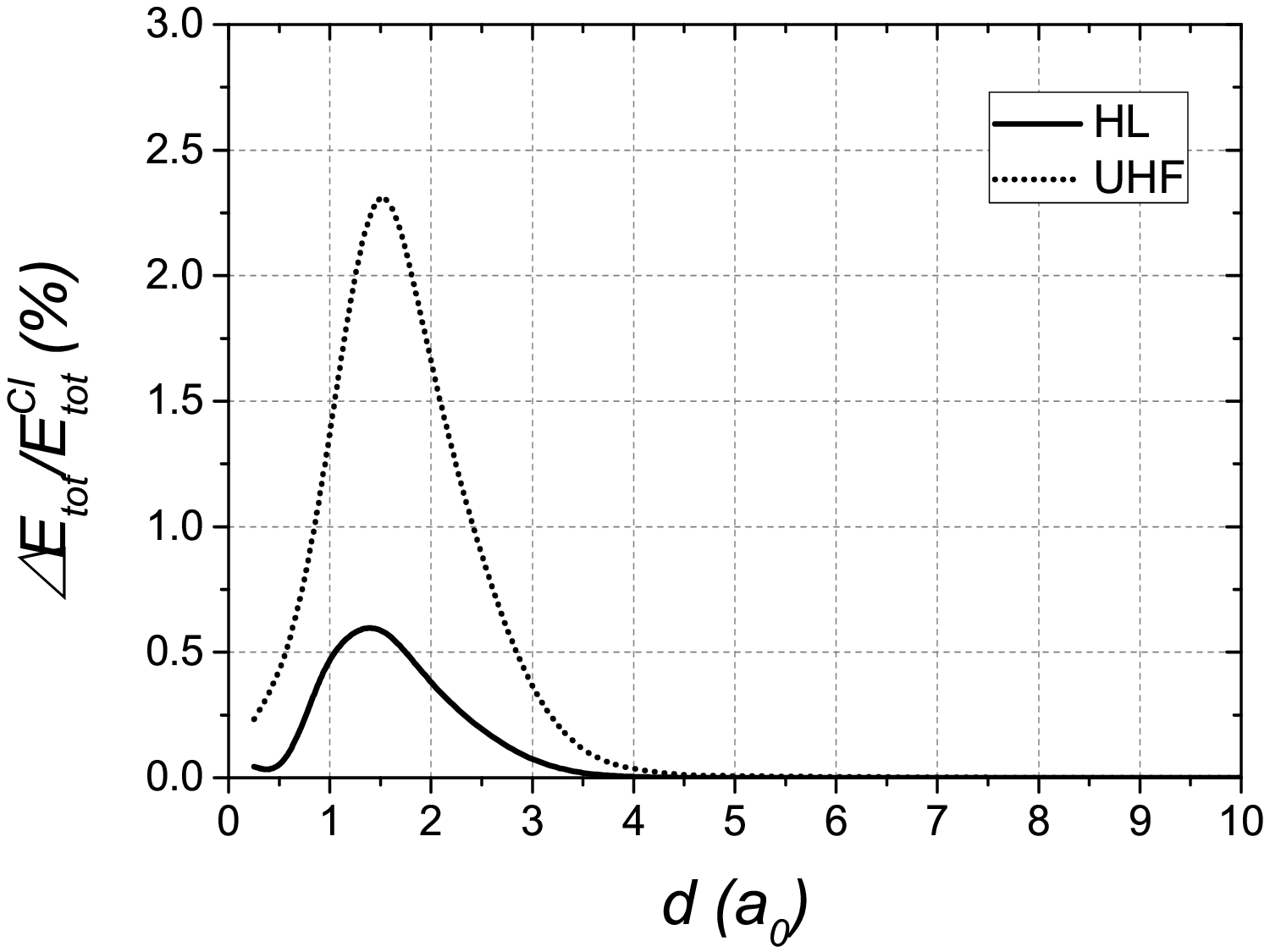}\\(c) [110] \qquad\qquad\qquad\qquad\qquad\qquad\qquad\qquad\qquad\quad (d) [110]\\
\includegraphics[scale=0.25]{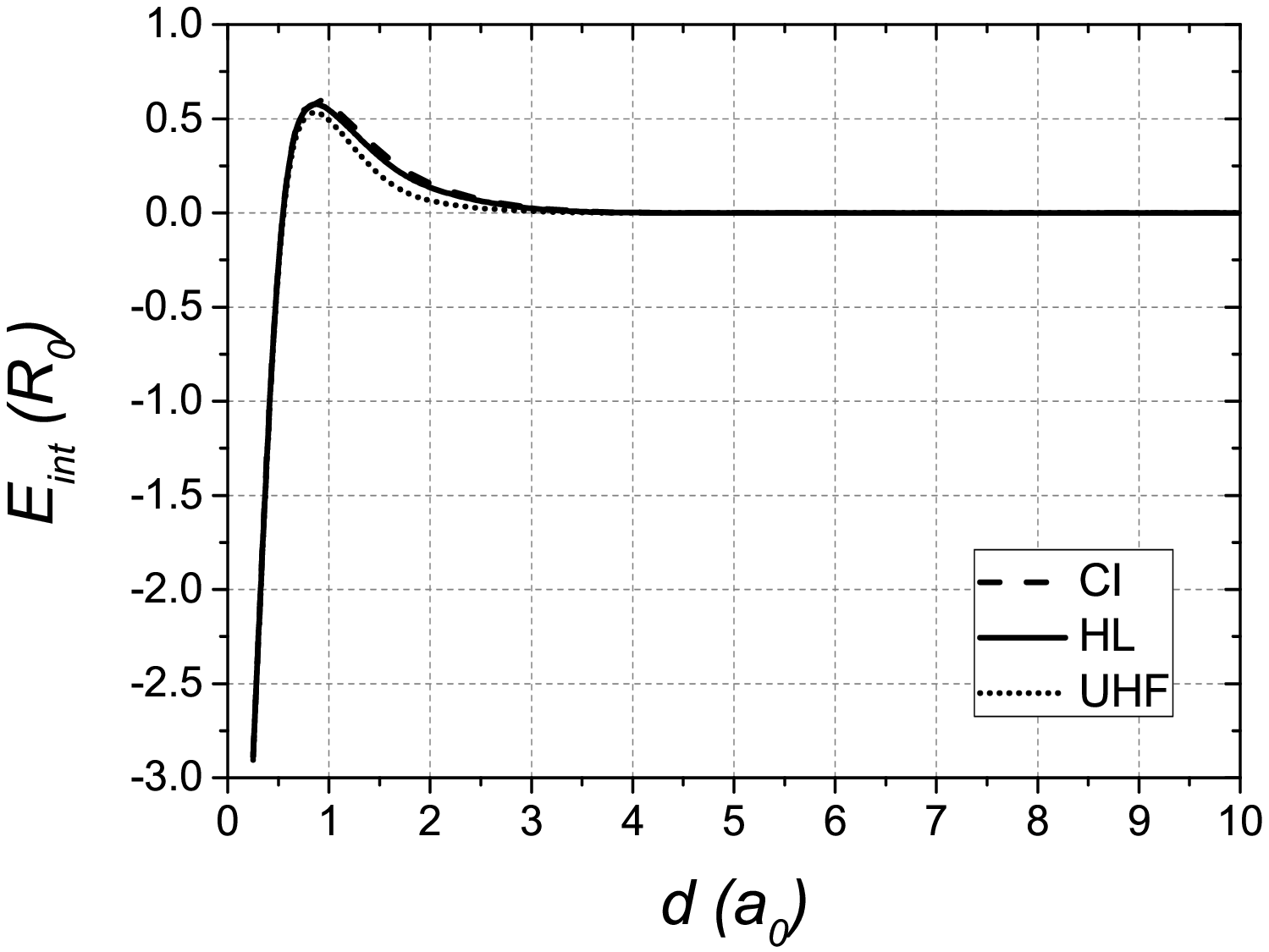}
\includegraphics[scale=0.25]{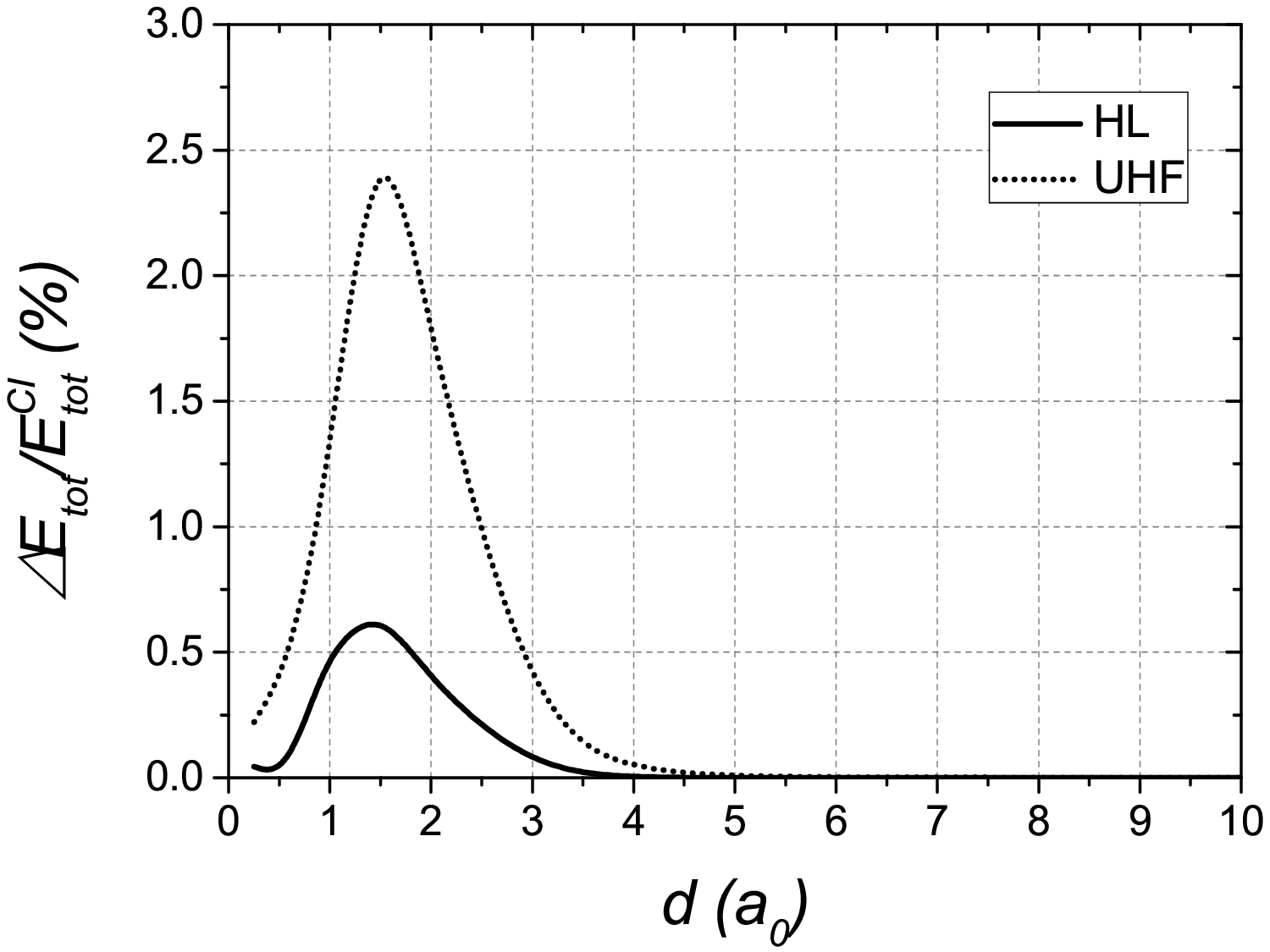}\\(e) [111] \qquad\qquad\qquad\qquad\qquad\qquad\qquad\qquad\qquad\quad (f) [111]
\caption{The interaction energy $E_{\textrm{int}}$ of the ground state and the difference of the total energy $E_{\textrm{tot}}$ towards the full CI calculation in three typical directions for a pair of acceptors: (a) the interaction energy $E_{\textrm{int}}$ in the [001] direction, (b) the difference of the total energy $E_{\textrm{tot}}$ in the [001] direction, (c) the interaction energy $E_{\textrm{int}}$ in the [110] direction, (d) the difference of the total energy $E_{\textrm{tot}}$ in the [110] direction, (e) the interaction energy $E_{\textrm{int}}$ in the [111] direction, (f) the difference of the total energy $E_{\textrm{tot}}$ in the [111] direction. For (a), (c), (e), the dashed line is for the full CI calculation, the solid line is for the HL approximation, the dotted line is for the UHF method. For (b), (d), (f), the solid line is for the HL approximation, the dotted line is for the UHF method, all the differences are in the percentage of the full CI result.}\label{f-1}
\end{figure*}
\begin{figure}
\centering
\includegraphics[scale=0.25]{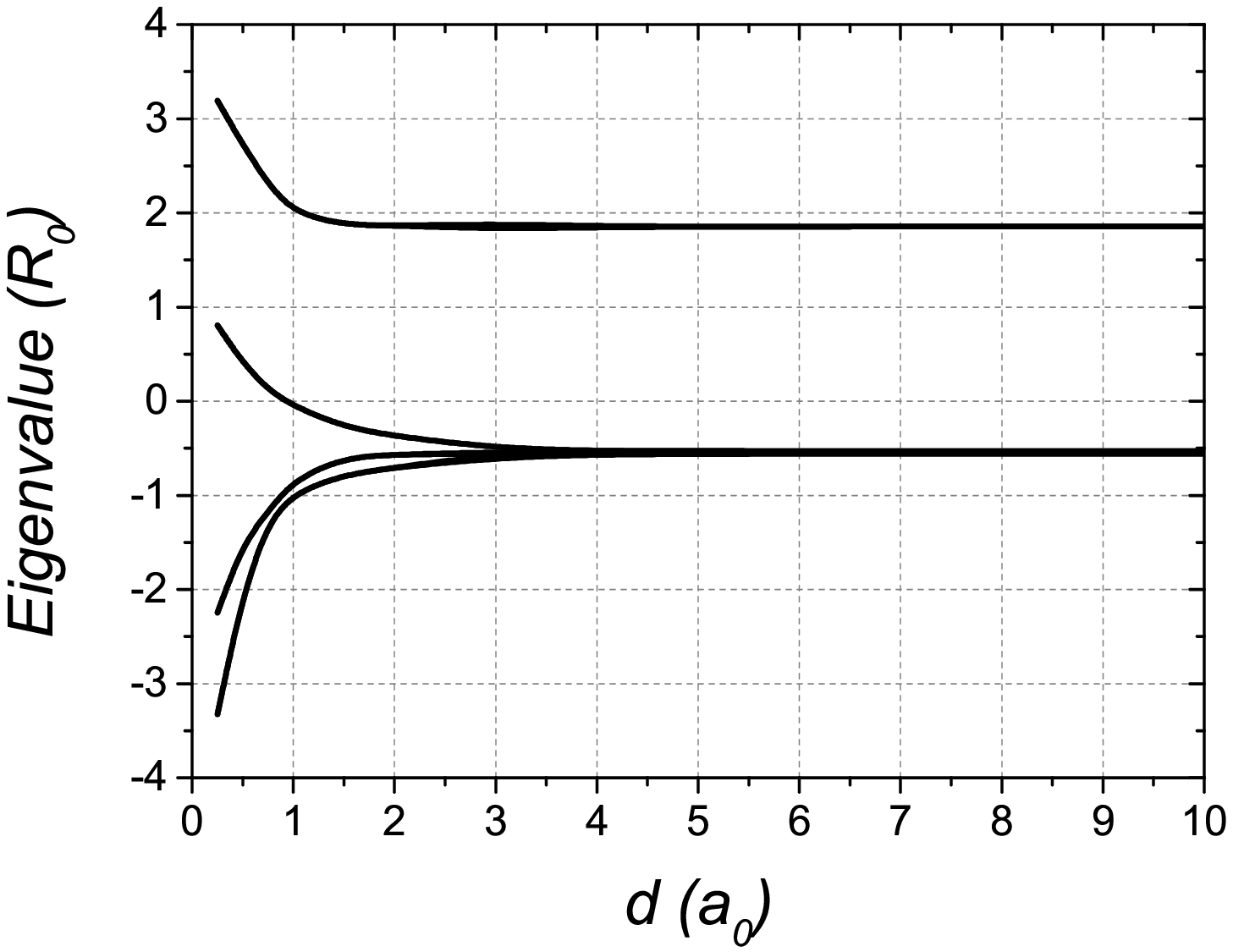}\\(a) [001]\\
\includegraphics[scale=0.25]{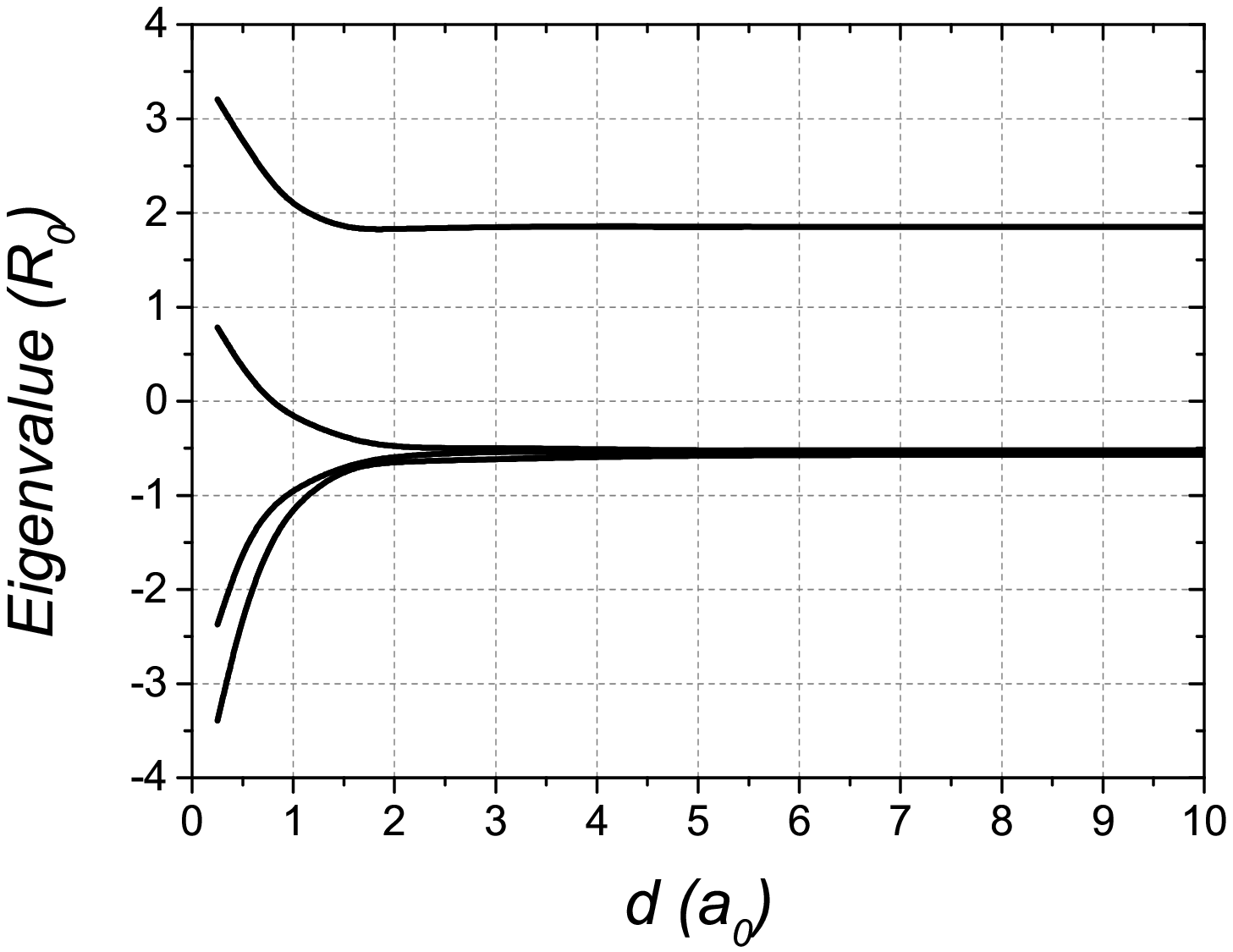}\\(b) [110]\\
\includegraphics[scale=0.25]{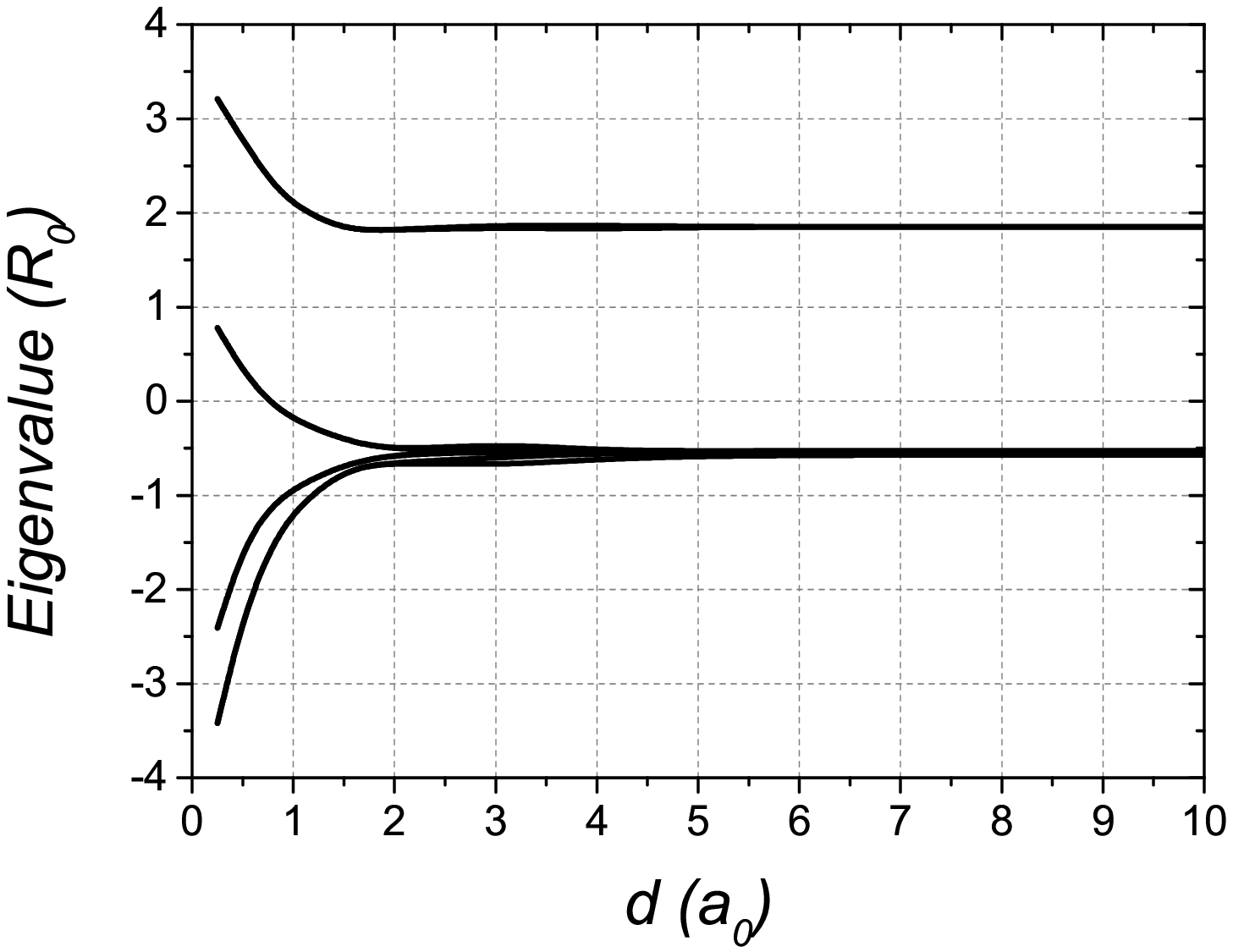}\\(c) [111]
\caption{The behavior of the Fock matrix eigenvalues in different directions for a pair of acceptors: (a) the [001] direction, (b) the [110] direction, (c) the [111] direction.}\label{f-2}
\end{figure}

For a pair of acceptors, all the methods and approximation mentioned in \S\ref{model} can be applied. To show the long-range behavior clearly, we calculate the interaction energy
\begin{equation}
E_{\textrm{int}}=E_T-2E_{\textrm{single}}=E_{\textrm{tot}}-\frac{2}{R}-2E_{\textrm{single}}\label{e-3-1-1},
\end{equation}
where $E_{\textrm{single}}$ is the single-acceptor energy, $E_T$ is the total energy including the core-core interaction, $E_{\textrm{tot}}$ is the total energy for the holes only (directly obtained from the Hamiltonian (\ref{e-2-1-1-1})), and $\frac{2}{R}$ is the core-core interaction term (appearing with a minus sign to be consistent with our convention for the hole energy). We did not consider the core-core interaction term in our previous paper \cite{Zhu2020LCoAOMfDPAAiS}; we refer to $E_{\textrm{tot}}$ as the `total energy' for the rest of this paper. The interaction energies $E_{\textrm{int}}$ of the ground state from three different models in three high-symmetry directions are shown in the left column of Figure \ref{f-1}; they appear as the negatives of standard molecular binding-energy curves. We also show the difference in the total energy $E_{\textrm{tot}}$ between the full CI calculation and the other approaches (as a percentage of the full CI result) in the right column. Both the HL approximation and the UHF method are good approximations to full CI for all directions, but the differences are greatest at small separations; the HL approach generally provides a better energy than UHF (since they involve variational approximations to the true wave-function, both methods give a lower bound on the true ground-state energy in the hole system). For the [001] direction, the differences reach a maximum around $1.5a_0$ and can be ignored when the separation $d\ge4a_0$; for the [110] and [111] directions, they peak around $1.5a_0$ and could be ignored for $d\ge5a_0$.

For the convenience of further discussion in \S\ref{4aresult}, Figure \ref{f-2} shows the Fock matrix eigenvalues for pairs oriented along different directions. The ground state appears at the top of the pictures as this is a calculation for acceptors. Each line represents a pair of almost doubly-degenerate states; since there are two holes, only the doubly-degenerate ground state at the top of the diagram will be filled. There is a large gap between the filled and empty states at all separations; this is generated by the strong hole-hole repulsion within the self-consistent field. We will see that this feature persists in the calculations on larger systems.

In the absence of cubic anisotropy, Durst \textit{et al.} \cite{Durst2020QIbAPiDS} argue that the long-range interaction between two acceptors is dominated by quadrupolar effects, which they find favour a doubly degenerate state with total angular momentum $M_F=\pm 2$ about the core axis.  This corresponds to partially aligned pairs of holes, with $m_F=\pm(\frac{3}{2},\frac{1}{2})$ on the two acceptors.  However, with the inclusion of significant cubic anisotropy appropriate for Si ($\delta>0$ and indeed $\delta\sim\mu$) we find that the pair ground state in the quantized direction (the [001] direction) only crosses over to this form for very large separations $d> 5a_0$; for smaller separations, the ground state is non-degenerate and dominated by anti-ferromagnetically coupled configurations such as $m_F=\pm(\frac{3}{2},-\frac{3}{2})$ (for $d<3a_0$) and $m_F=\pm(\frac{1}{2},-\frac{1}{2})$ for $3a_0\le d\le5a_0$.  

\subsection{Finite dimerised linear chains}\label{4aresult}

\begin{figure}
\centering
\includegraphics[scale=0.3]{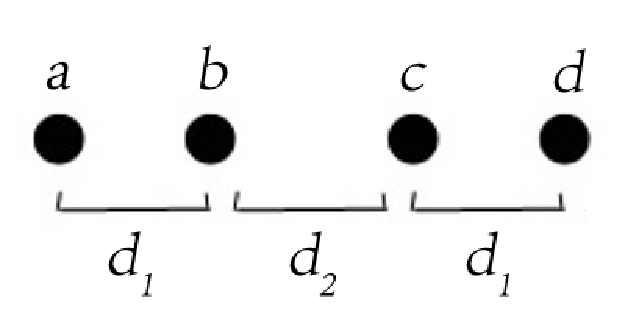}\\(a) 4-acceptor linear chain\\
\includegraphics[scale=0.3]{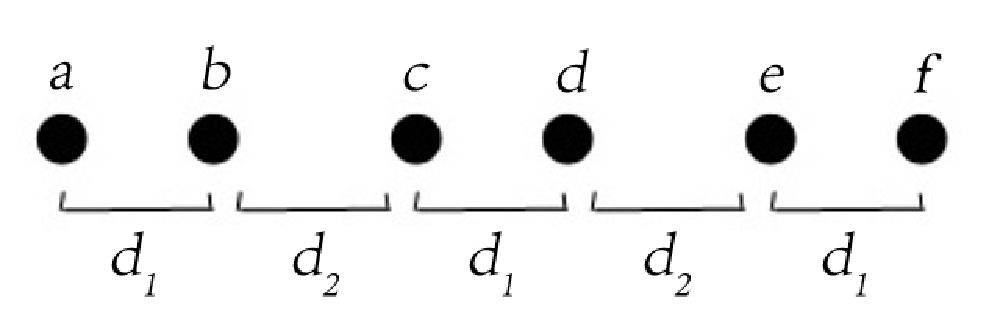}\\(b) 6-acceptor linear chain
\caption{Schematics of the linear chains studied in this paper. a, b, c, d, e, f are the labels of acceptors; $d_1<d_2$ is known as the `short-long arrangement', $d_1>d_2$ is known as the `long-short arrangement'.}\label{f-3}
\end{figure}

We next consider chains of 4 and 6 acceptors, with one hole per acceptor and with the separations $(d_1,d_2)$ alternating to form a dimer chain as shown in Figure \ref{f-3}. When  $d_1<d_2$, we will refer to a `short-long arrangement' throughout the rest of the paper, while when $d_1>d_2$ we will call it a `long-short arrangement'. We investigate two different regimes, each defined by a fixed value of $d_1+d_2$: a `small-separation' case with $d_1+d_2=3a_0$, and a `large-separation' case with $d_1+d_2=6a_0$.

\subsubsection{Small-separation case ($d_1+d_2=3a_0$)}\label{4asmallresult}

\begin{figure}
\centering
\includegraphics[scale=0.25]{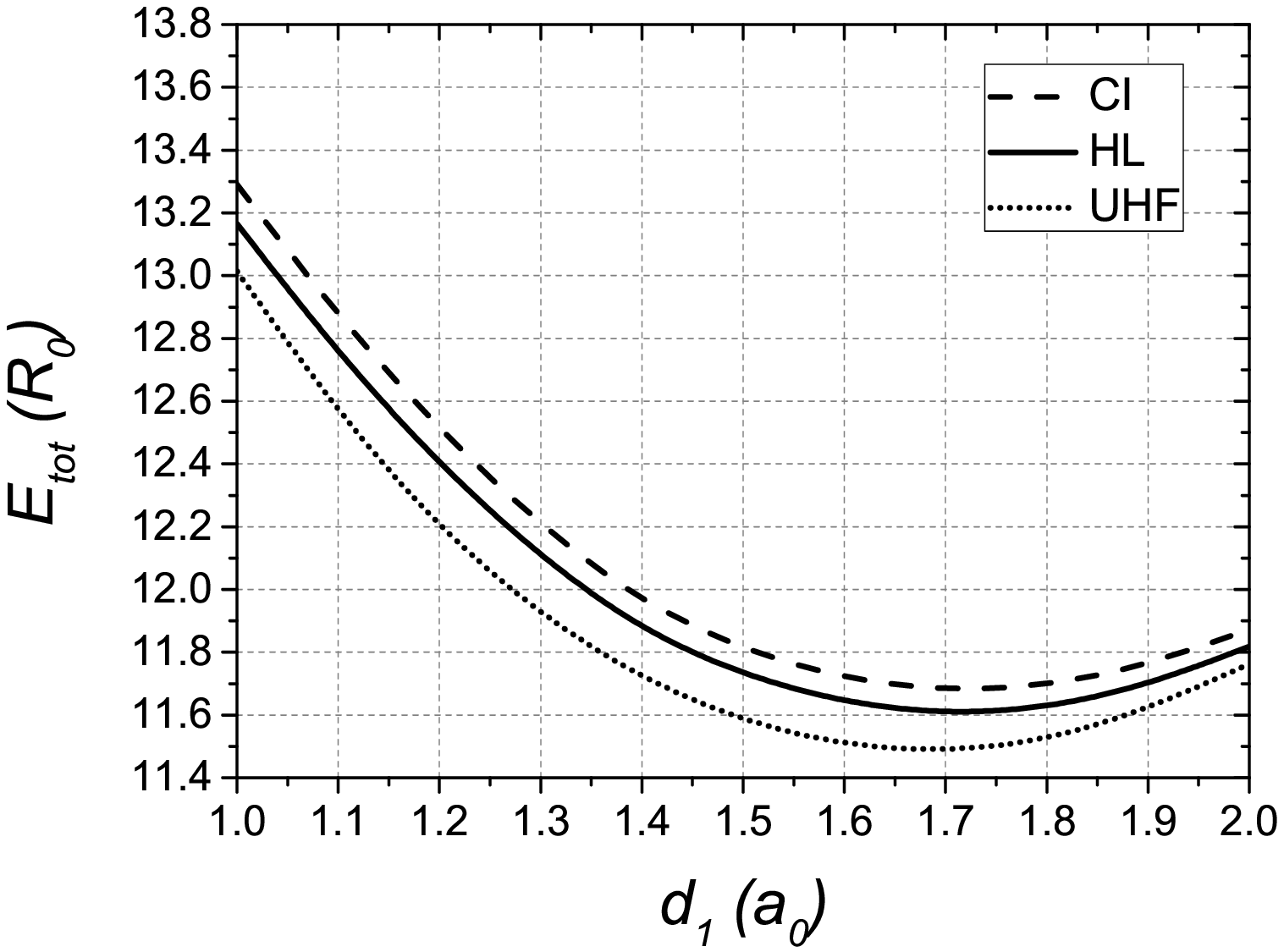}\\(a) [001]\\
\includegraphics[scale=0.25]{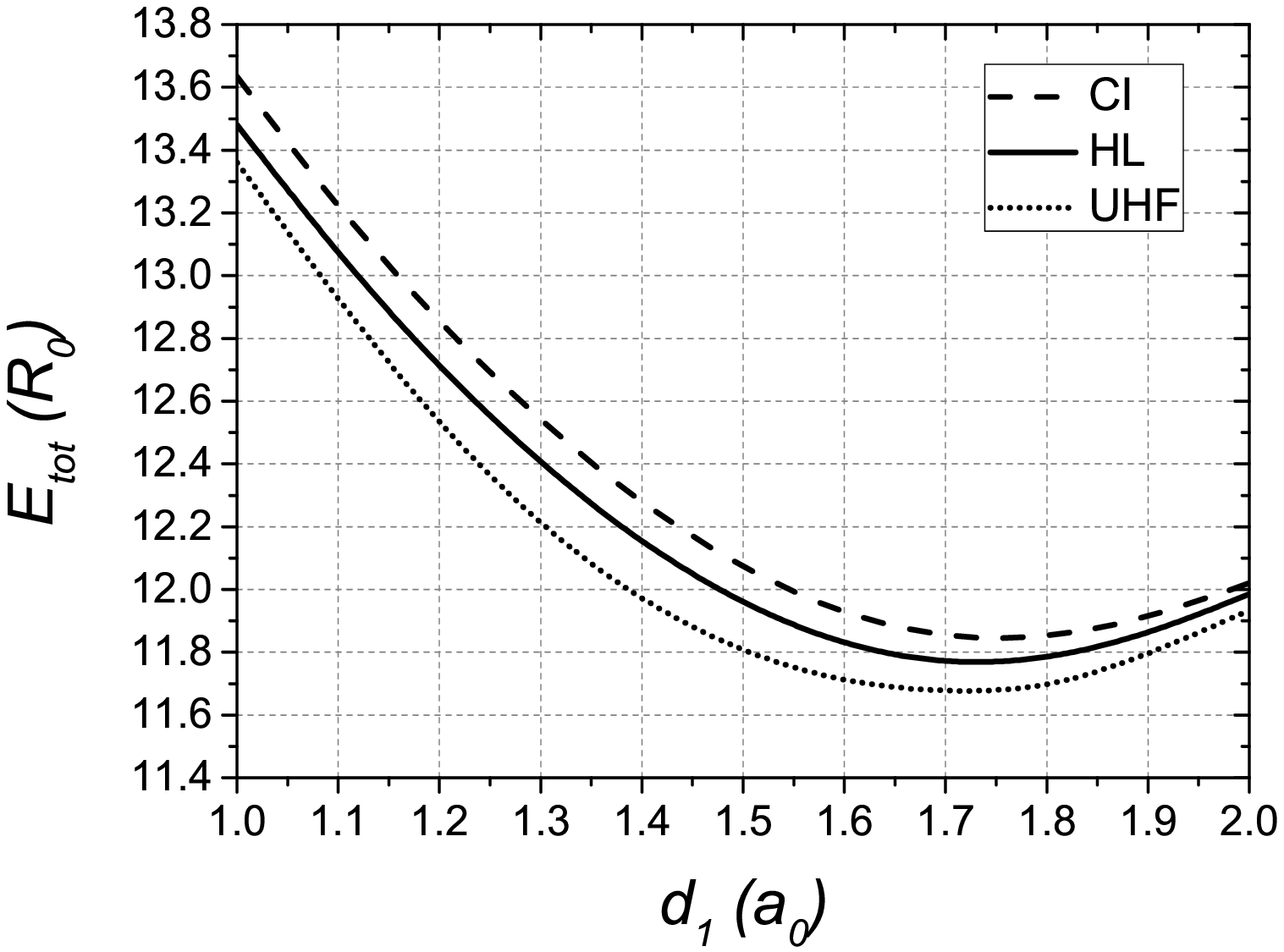}\\(b) [110]\\
\includegraphics[scale=0.25]{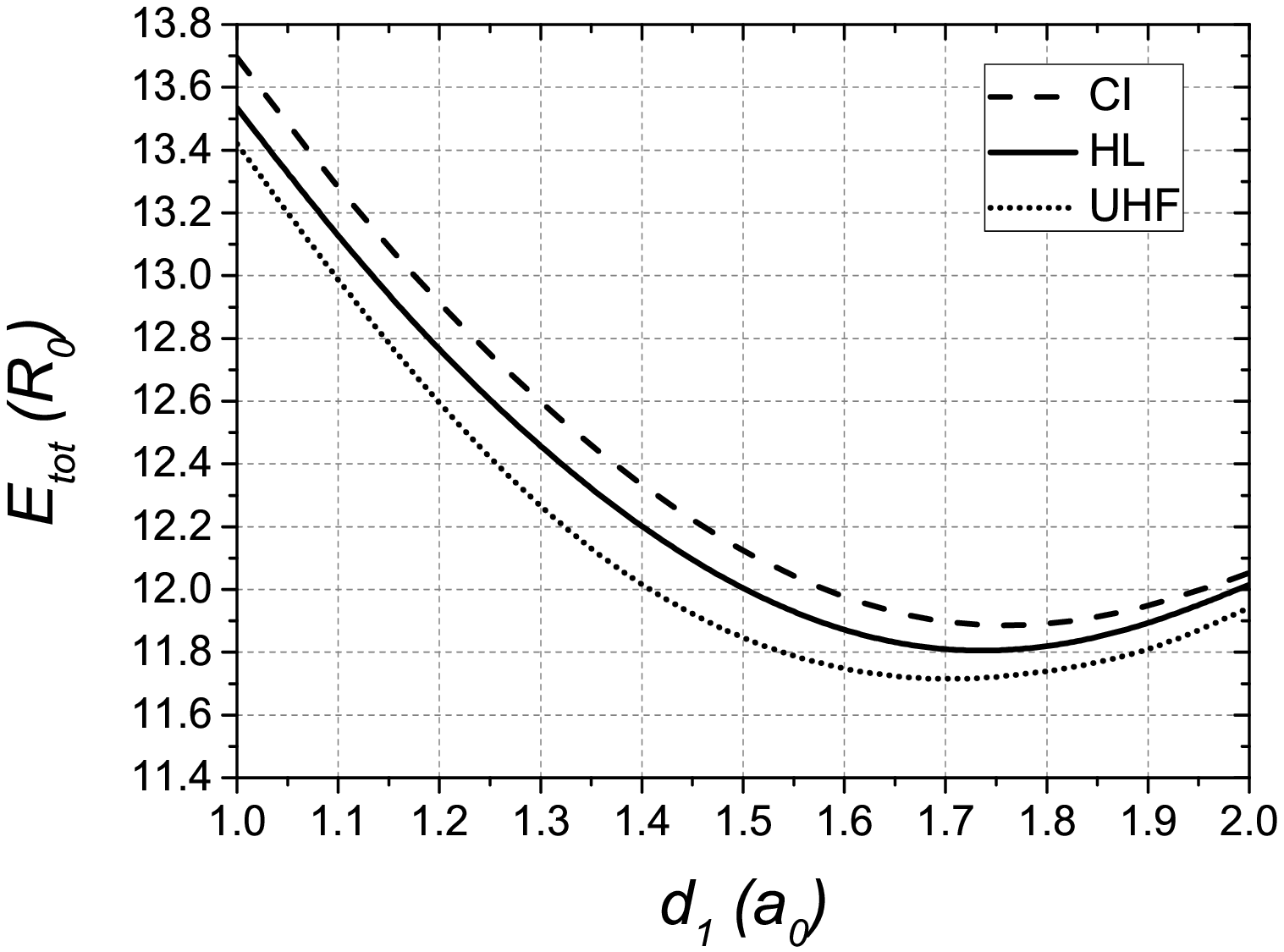}\\(c) [111]
\caption{The behavior of the total energy of the ground state under different arrangements in three typical directions for the small-separation case ($d_1+d_2=3a_0$) of the 4-acceptor linear chain: (a) the [001] direction, (b) the [110] direction, (c) the [111] direction. The dashed line is for the full CI calculation, the solid line is for the HL approximation, the dotted line is for the UHF method.}\label{f-4}
\end{figure}
\begin{figure*}
\centering
\includegraphics[scale=0.25]{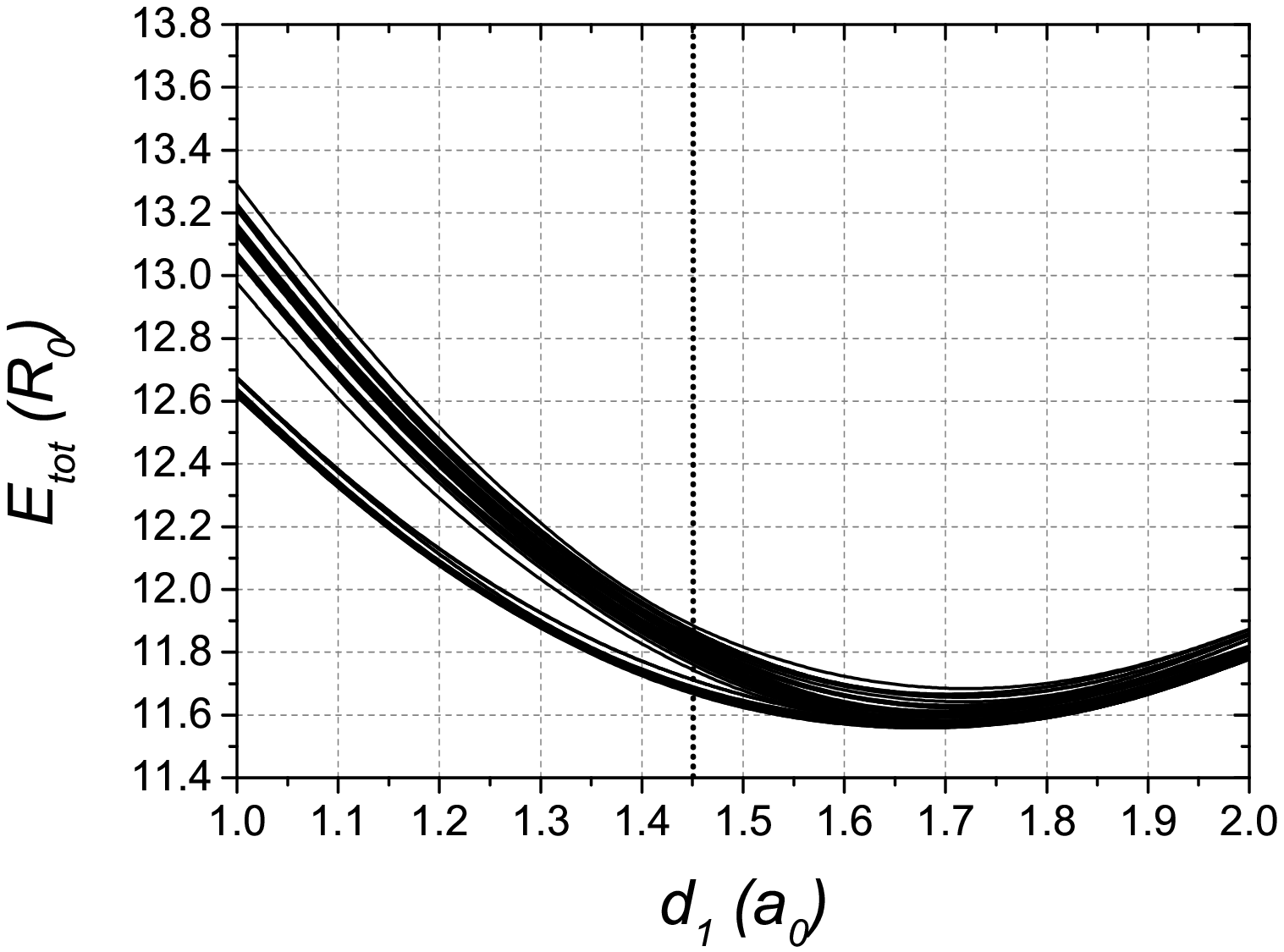}
\includegraphics[scale=0.25]{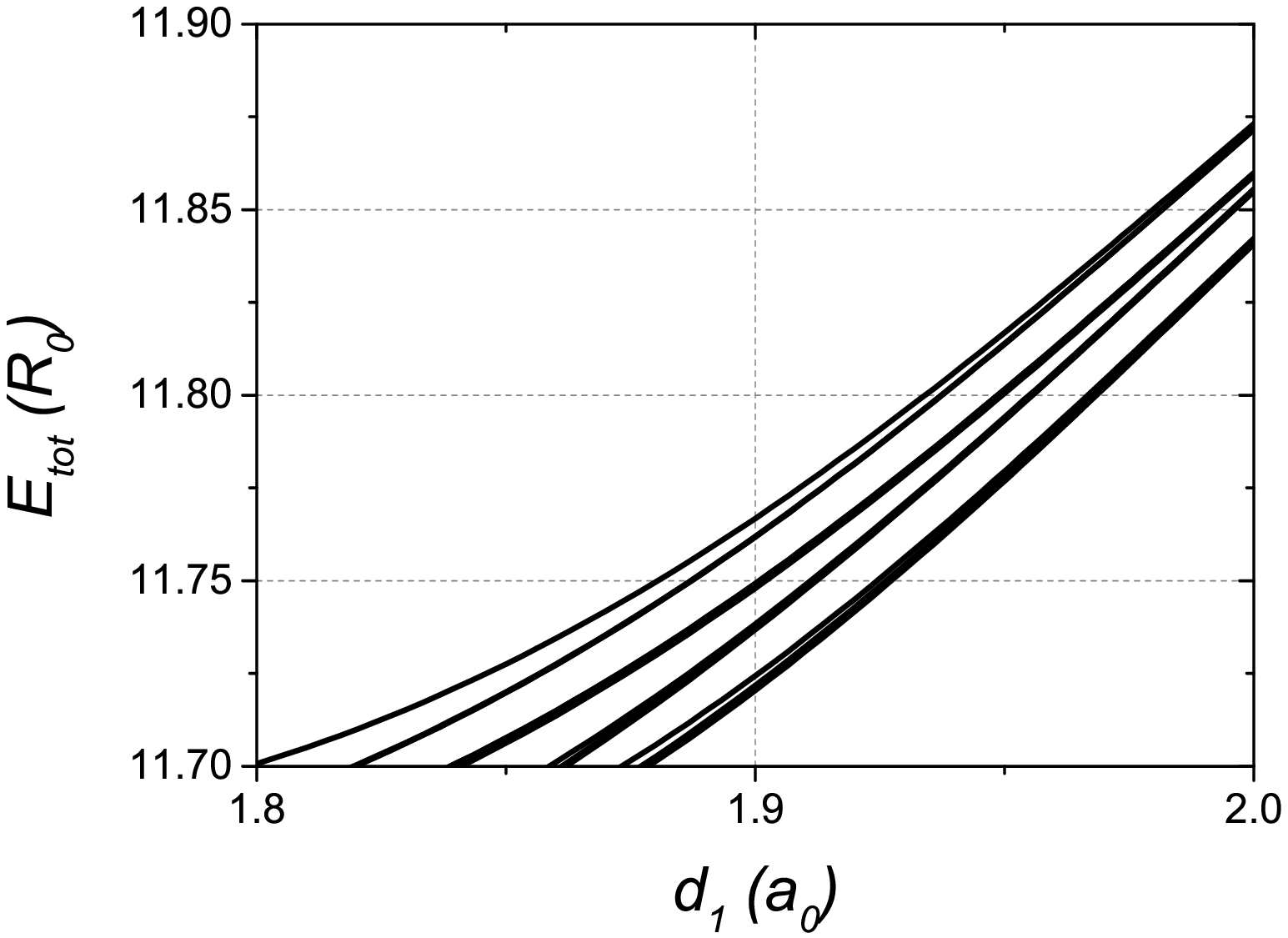}\\(a) [001] \qquad\qquad\qquad\qquad\qquad\qquad\qquad\qquad\qquad\quad (b) [001]\\
\includegraphics[scale=0.25]{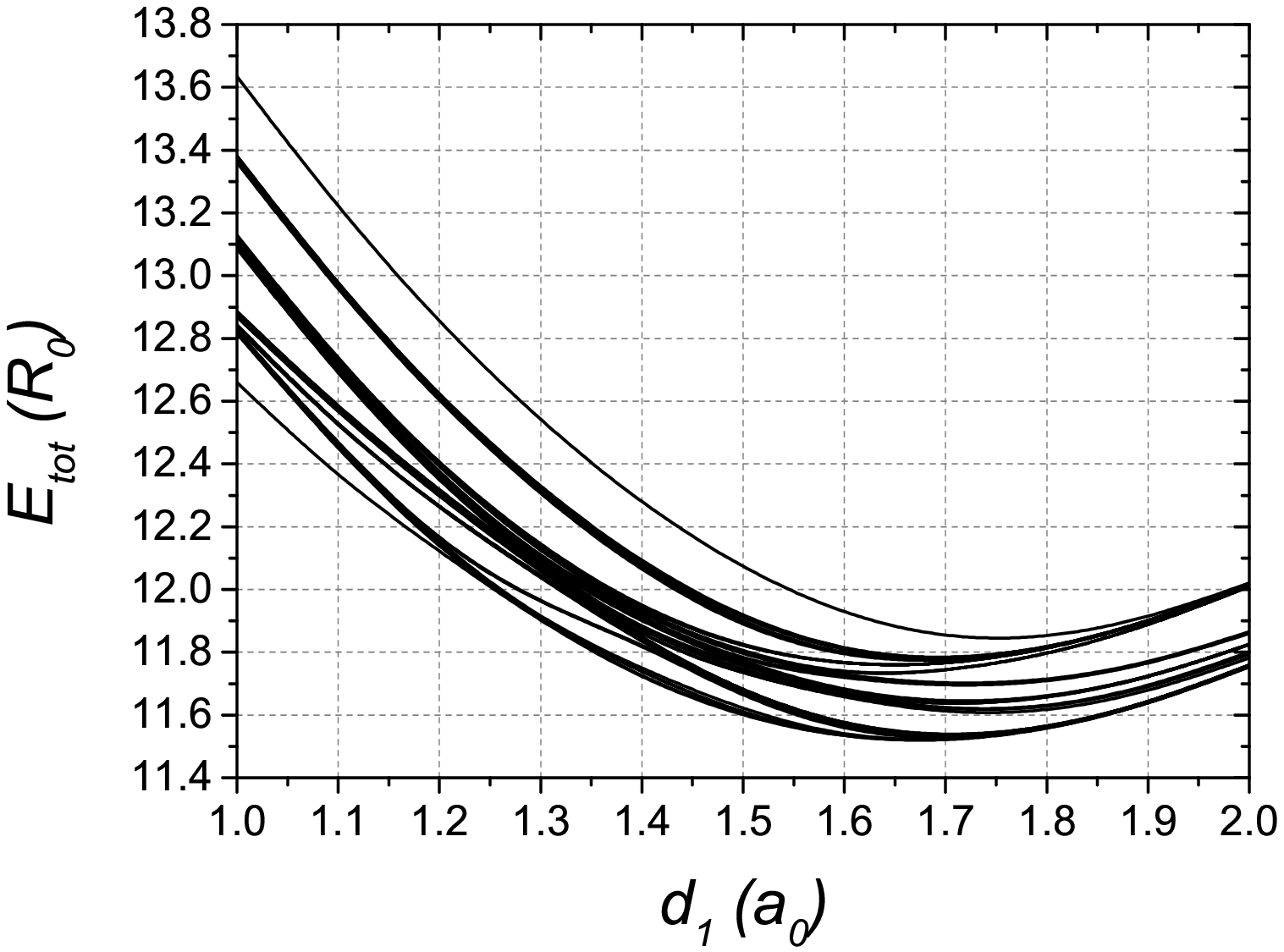}
\includegraphics[scale=0.25]{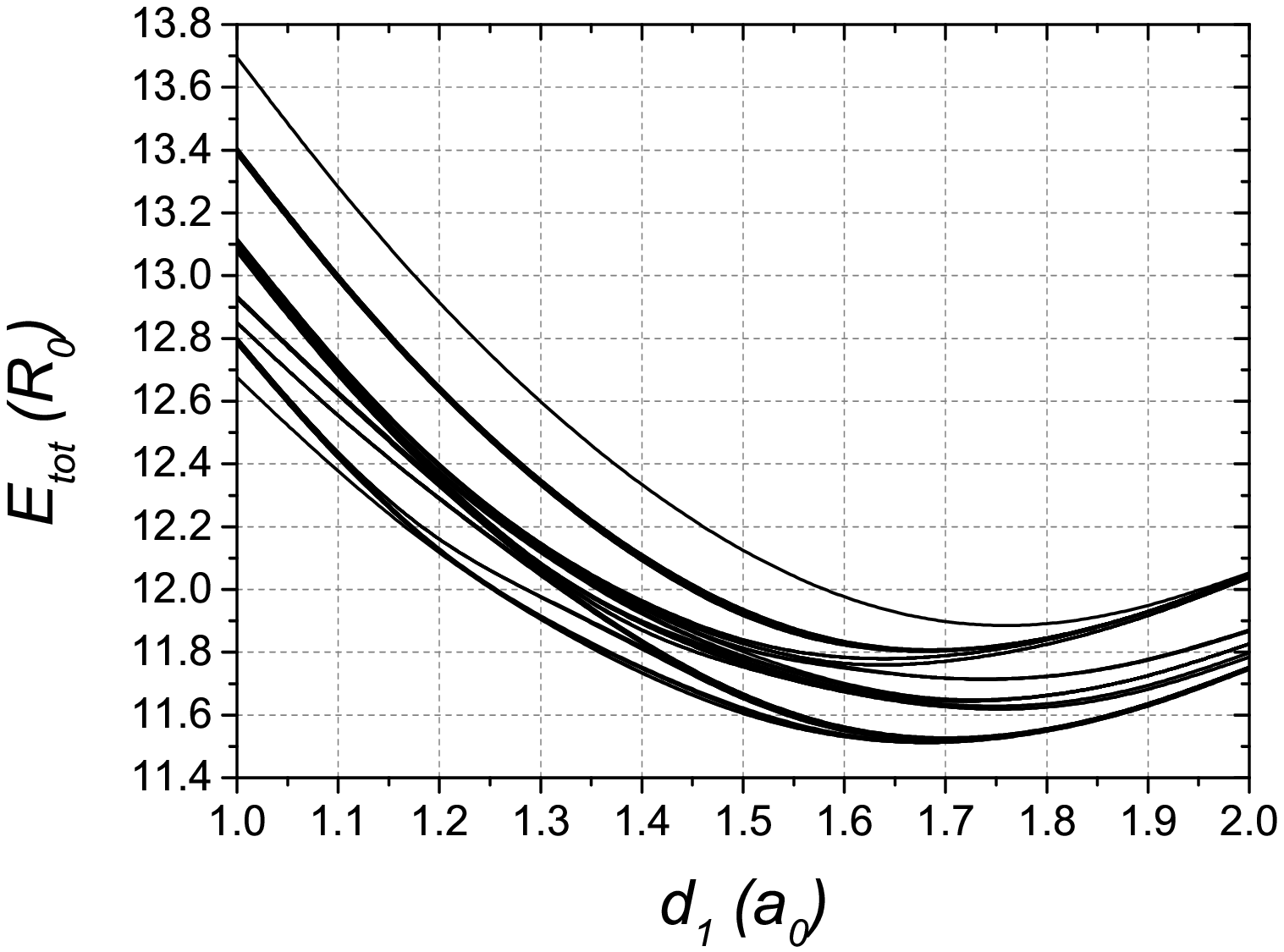}\\(c) [110] \qquad\qquad\qquad\qquad\qquad\qquad\qquad\qquad\qquad\quad (d) [111]
\caption{The behavior of the total energy of the highest 50 energy states of the full CI result under different arrangements in three typical directions for the small-separation case ($d_1+d_2=3a_0$) of the 4-acceptor linear chain with the changing point: (a) the [001] direction, (b) details of the long-short arrangement side in the [001] direction, (c) the [110] direction, (d) the [111] direction. In Picture (a), the dotted line is for the changing point.}\label{f-5}
\end{figure*}
\begin{figure*}
\centering
\includegraphics[scale=0.25]{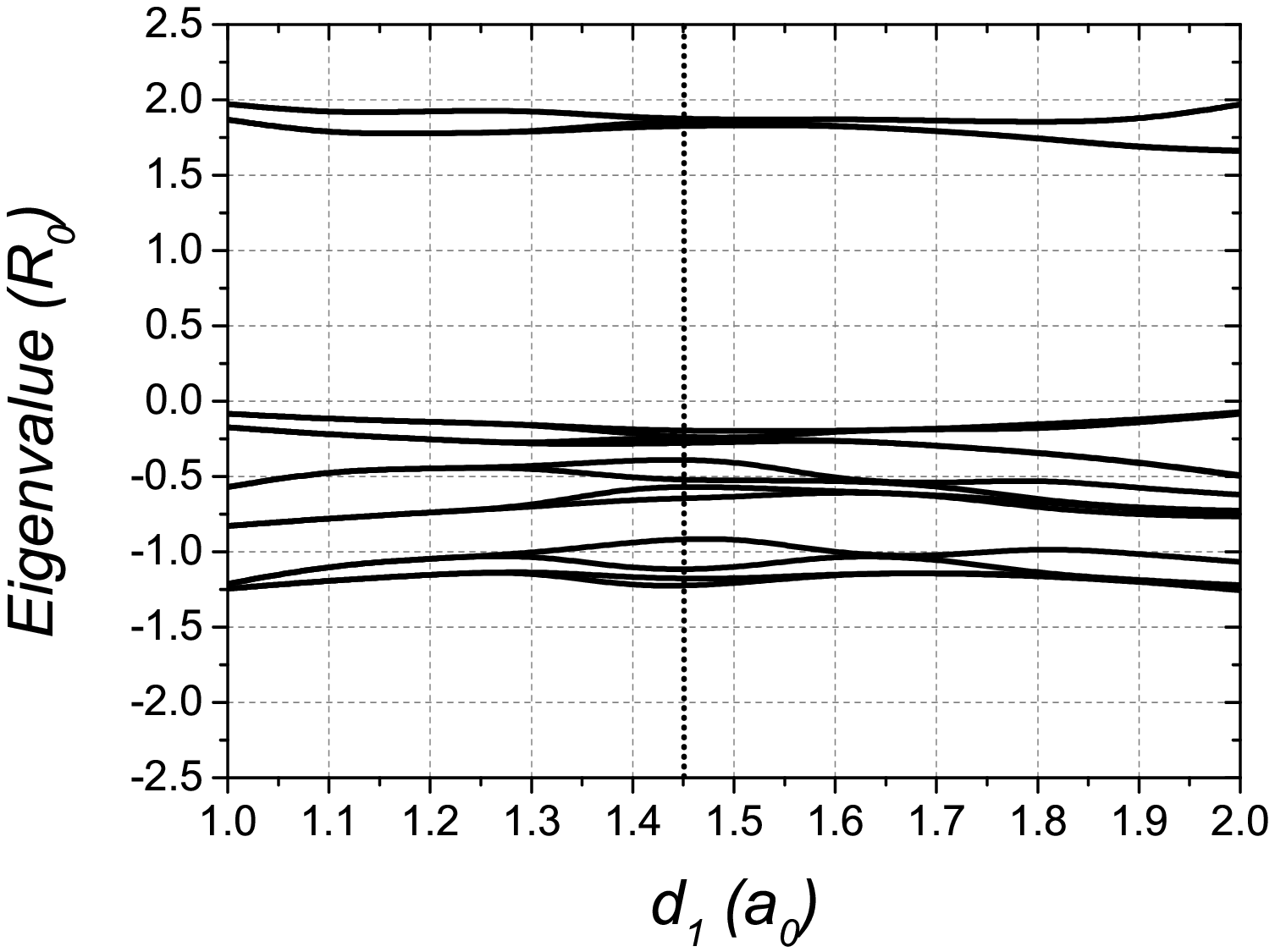}
\includegraphics[scale=0.25]{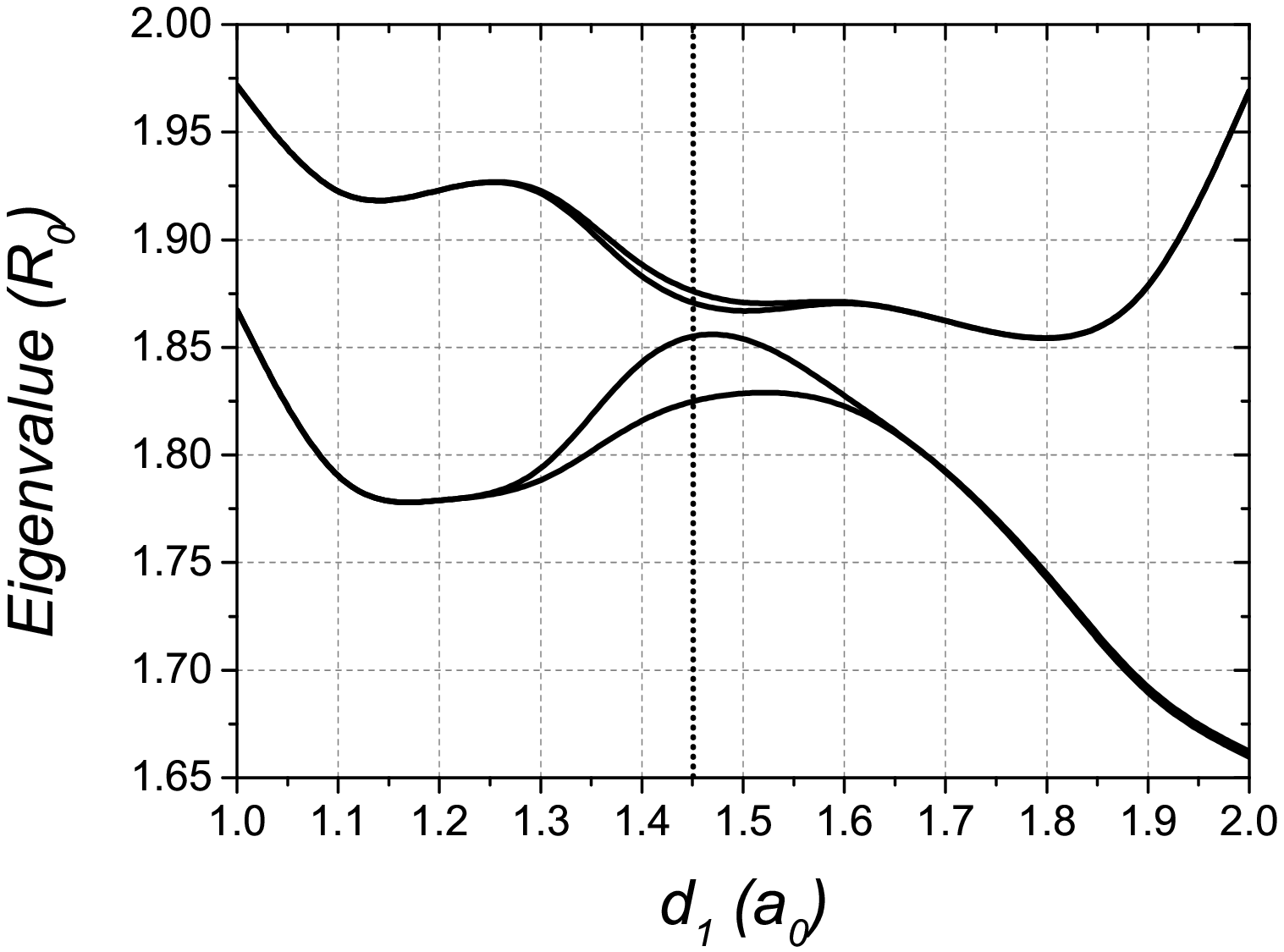}\\(a) [001] \qquad\qquad\qquad\qquad\qquad\qquad\qquad\qquad\qquad\quad (b) [001]\\
\includegraphics[scale=0.25]{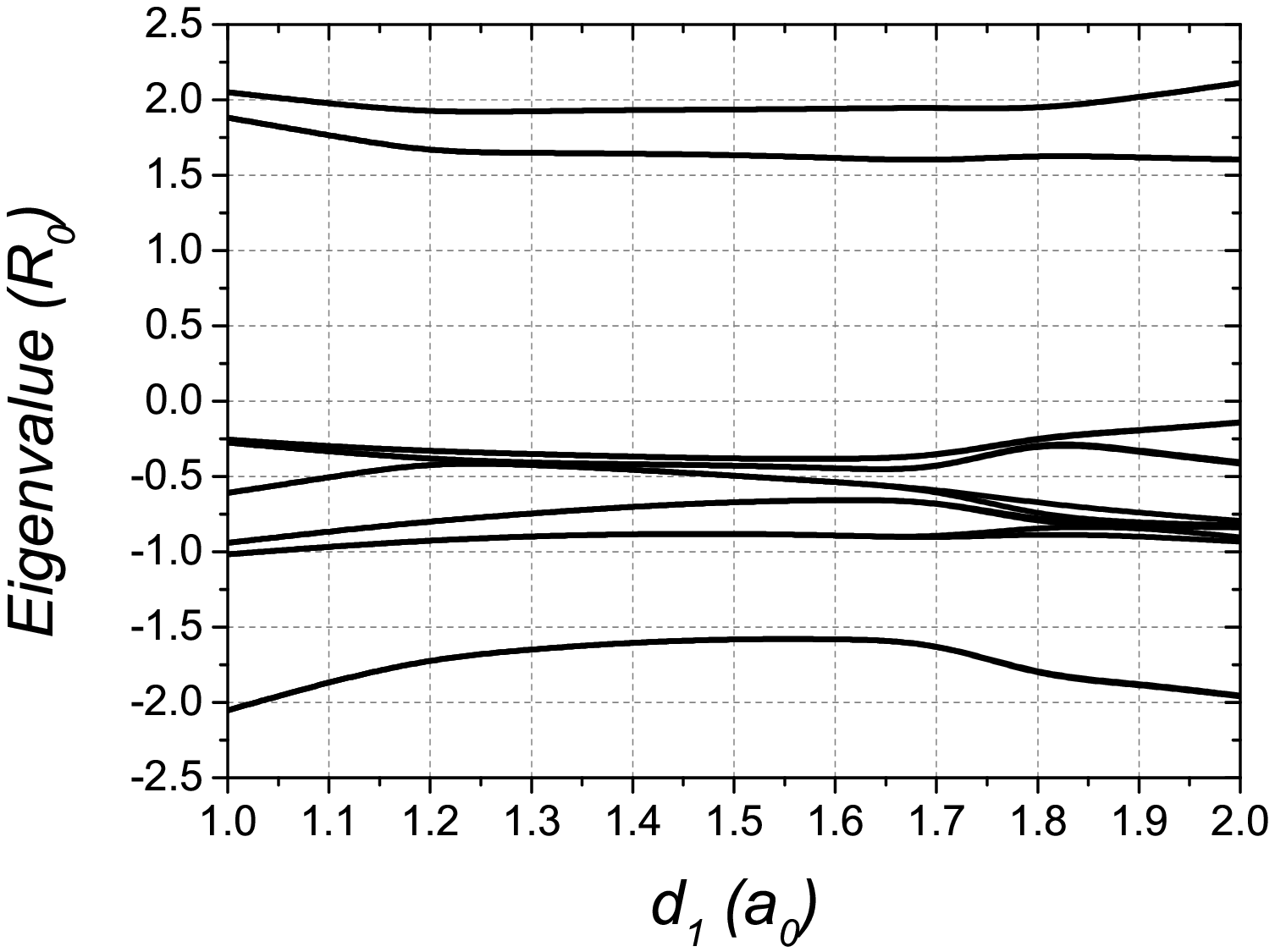}
\includegraphics[scale=0.25]{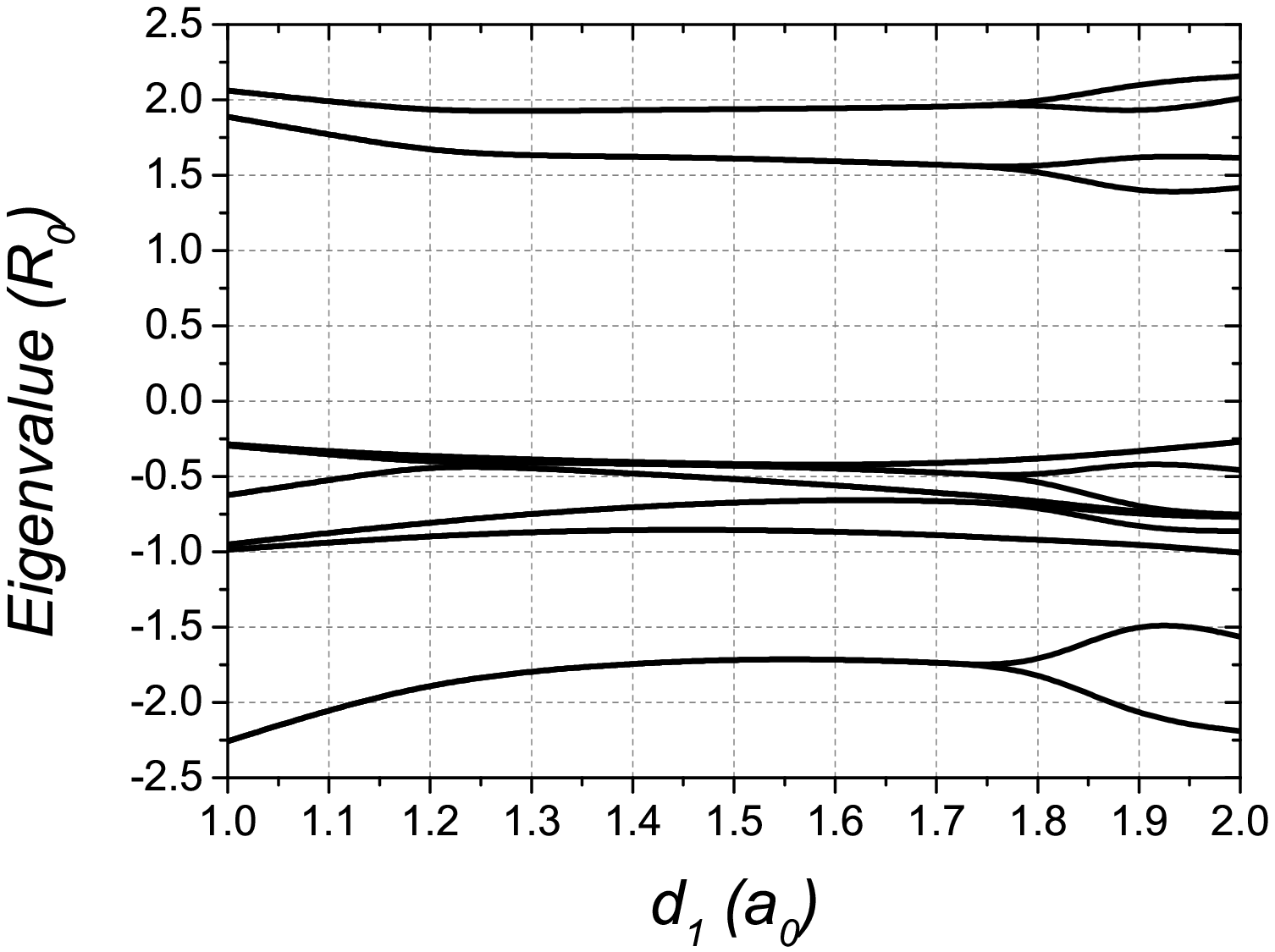}\\(c) [110] \qquad\qquad\qquad\qquad\qquad\qquad\qquad\qquad\qquad\quad (d) [111]
\caption{The behavior of the Fock matrix eigenvalues obtained from the UHF method under different arrangements in three typical directions for the small-separation case ($d_1+d_2=3a_0$) of the 4-acceptor linear chain with the changing point: (a) the [001] direction, (b) details of the highest 4 eigenvalues in the [001] direction,(c) the [110] direction, (d) the [111] direction. For (a) and (b), the dotted lines are for the changing points.}\label{f-6}
\end{figure*}
\begin{figure*}
\centering
\includegraphics[scale=0.25]{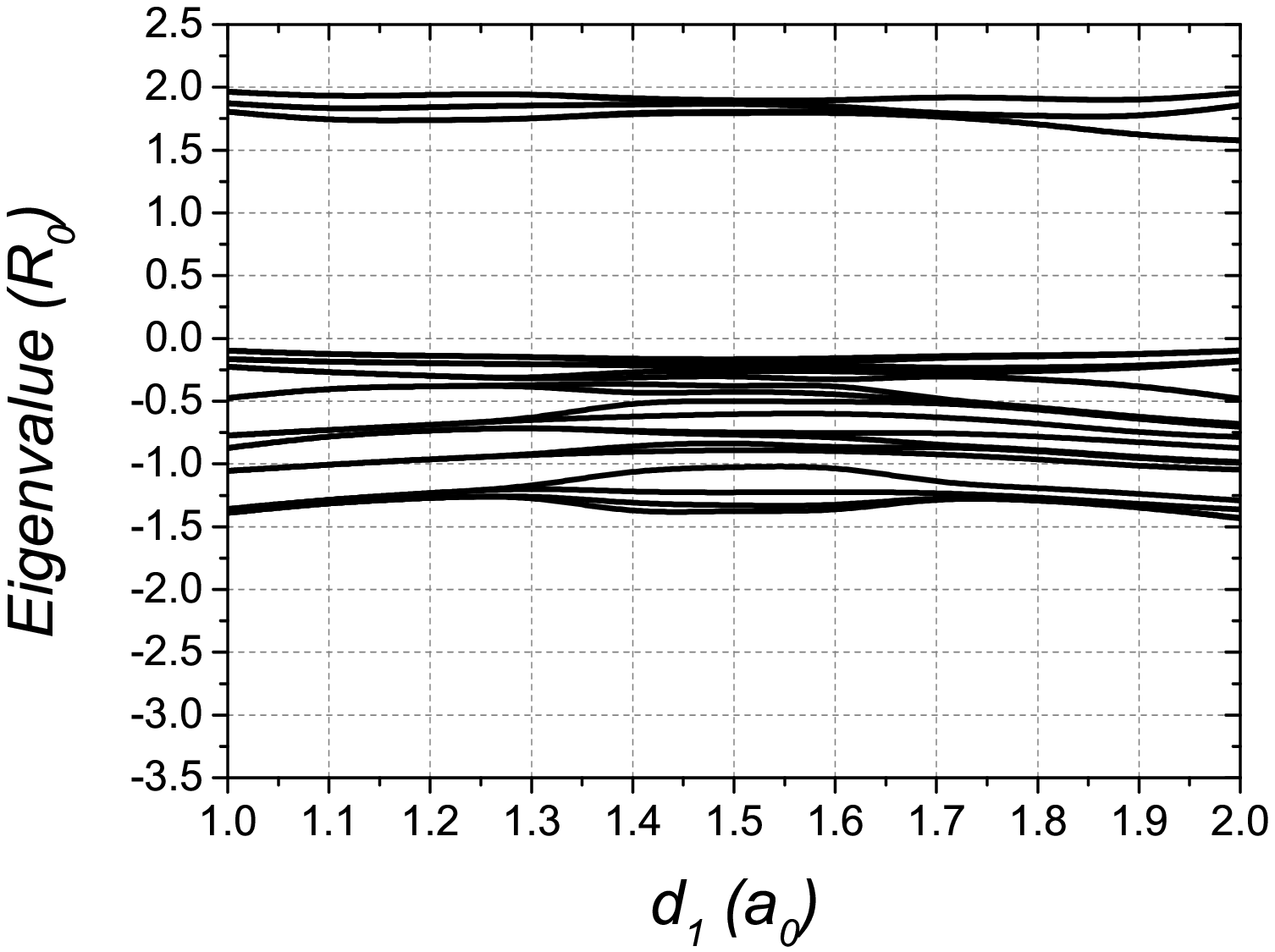}
\includegraphics[scale=0.25]{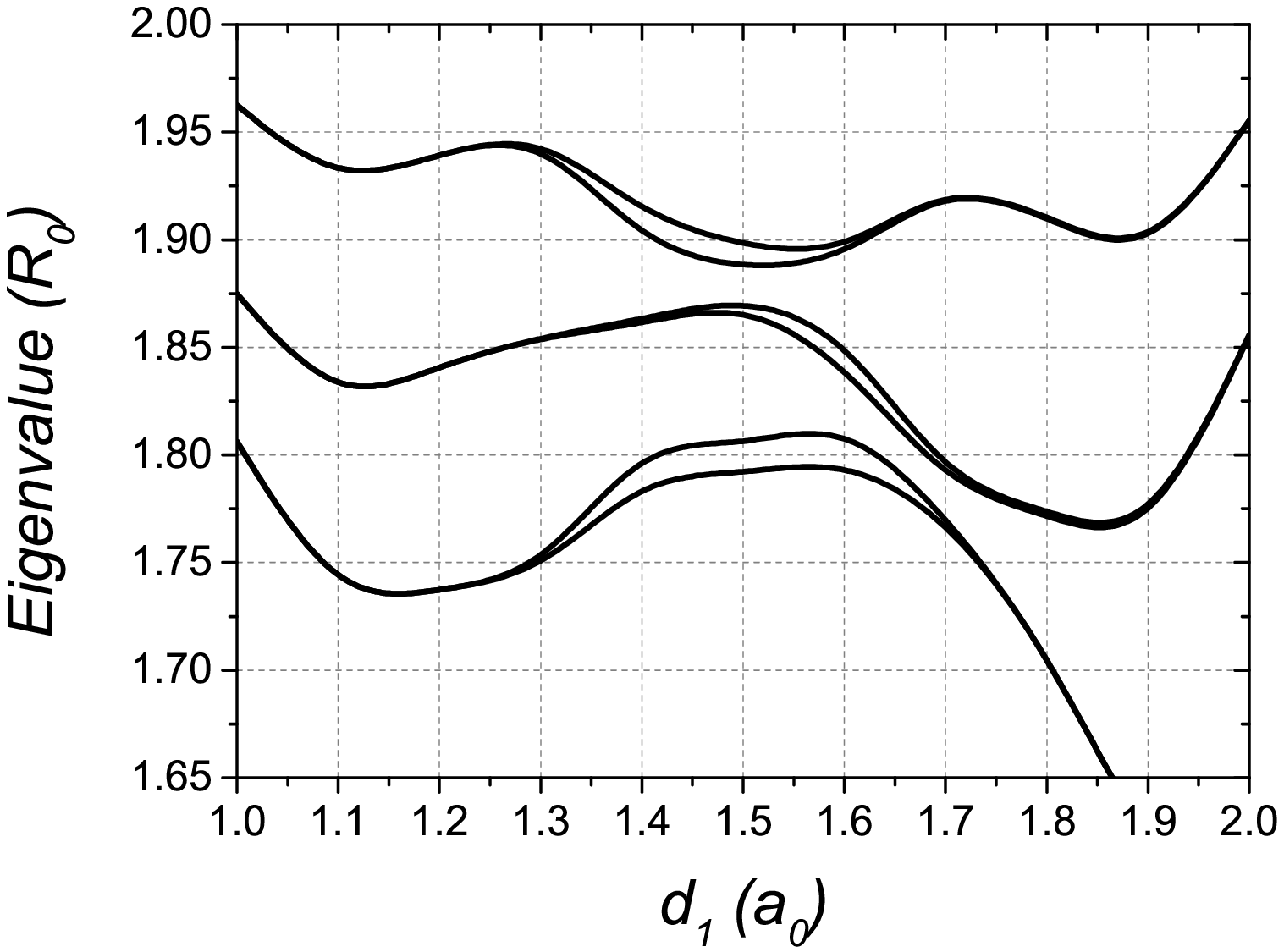}\\(a) [001] \qquad\qquad\qquad\qquad\qquad\qquad\qquad\qquad\qquad\quad (b) [001]\\
\includegraphics[scale=0.25]{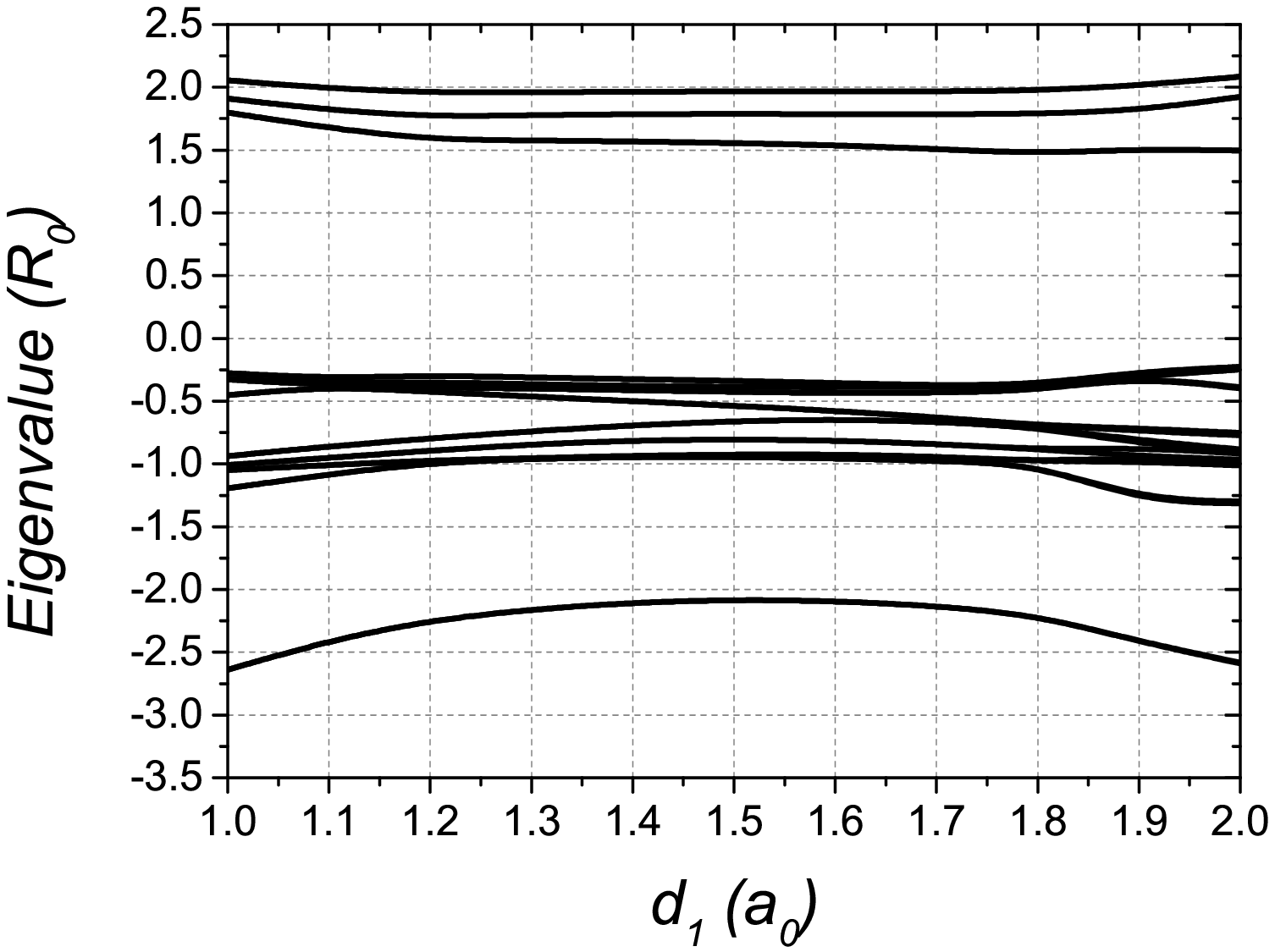}
\includegraphics[scale=0.25]{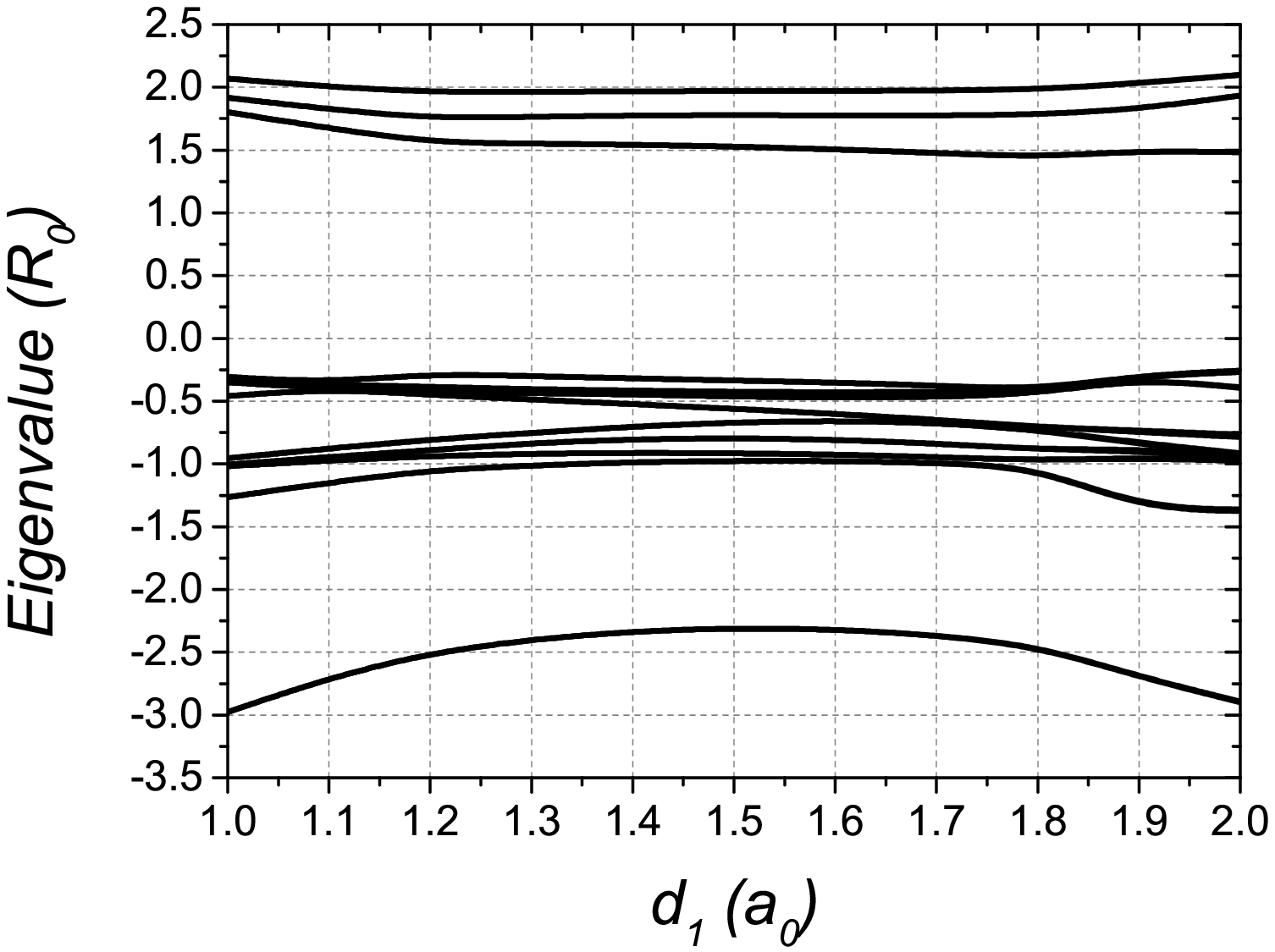}\\(c) [110] \qquad\qquad\qquad\qquad\qquad\qquad\qquad\qquad\qquad\quad (d) [111]
\caption{The behavior of the Fock matrix eigenvalues obtained from the UHF method under different arrangements in three typical directions for the 6-acceptor linear chain when $d_1+d_2=3a_0$: (a) the [001] direction, (b) details of the highest 6 eigenvalues in the [001] direction, (c) the [110] direction, (d) the [111] direction.}\label{f-11}
\end{figure*}
\begin{figure}
\centering
\includegraphics[scale=0.25]{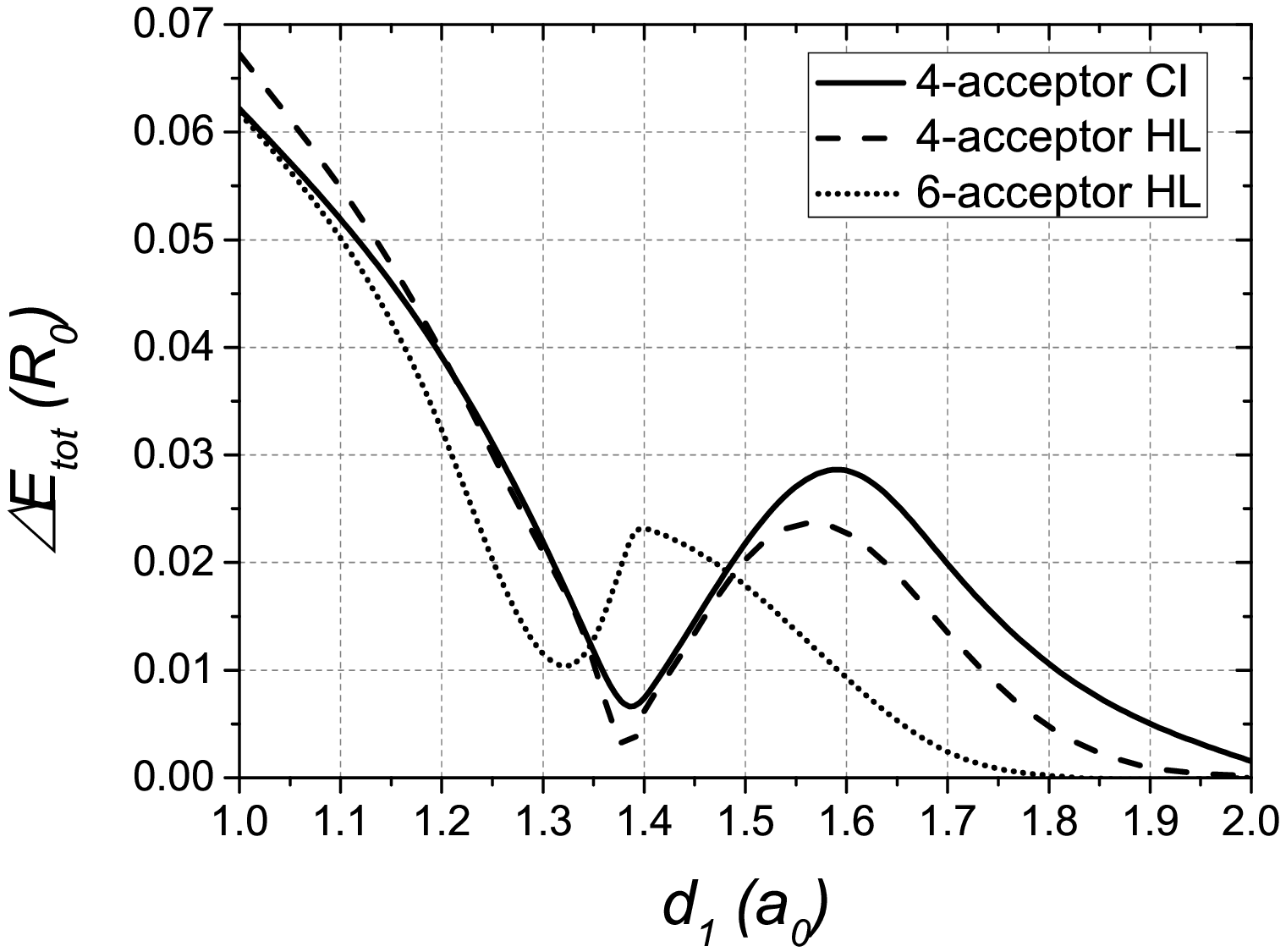}\\(a)\\
\includegraphics[scale=0.25]{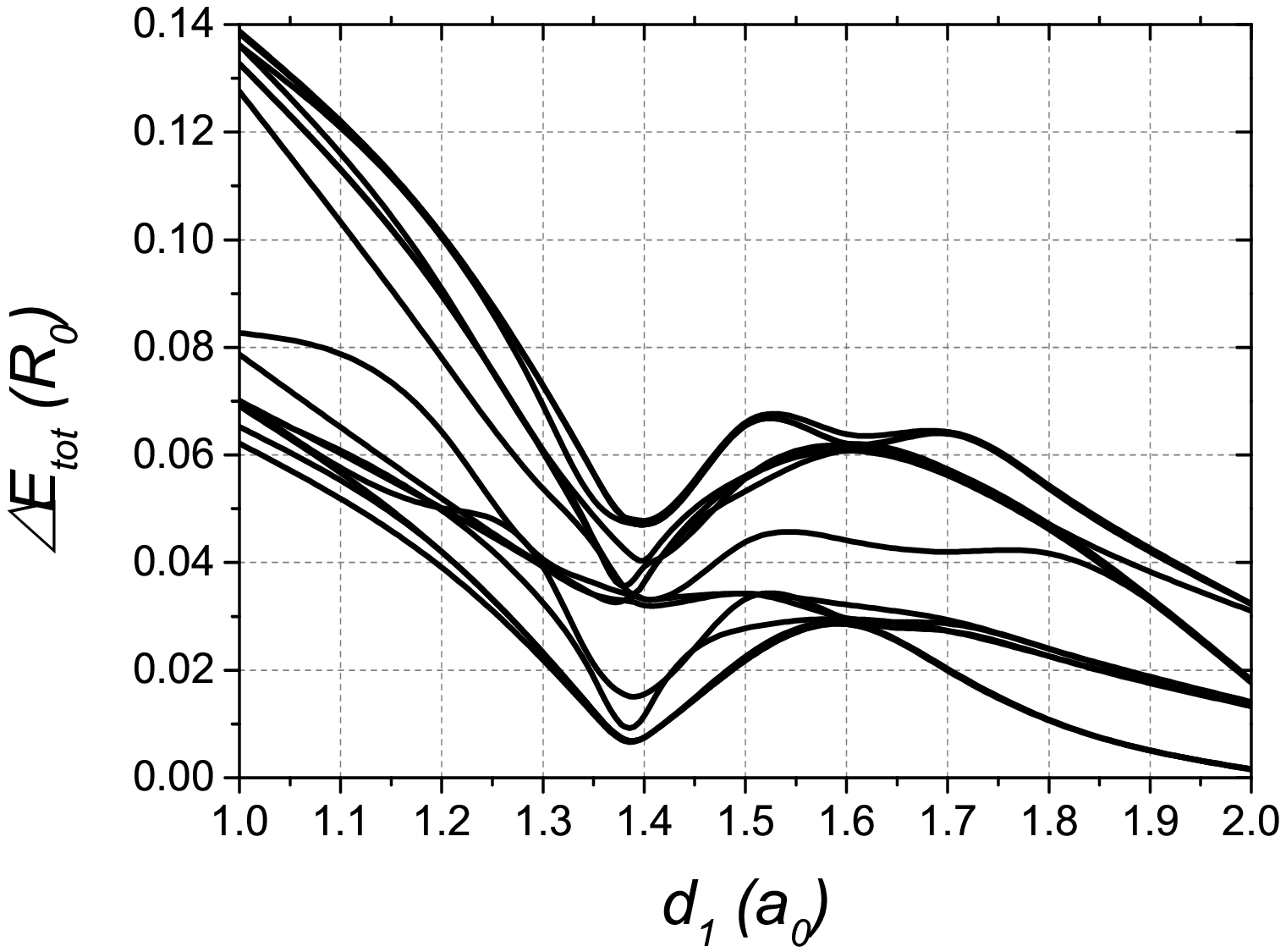}\\(b)\\
\includegraphics[scale=0.25]{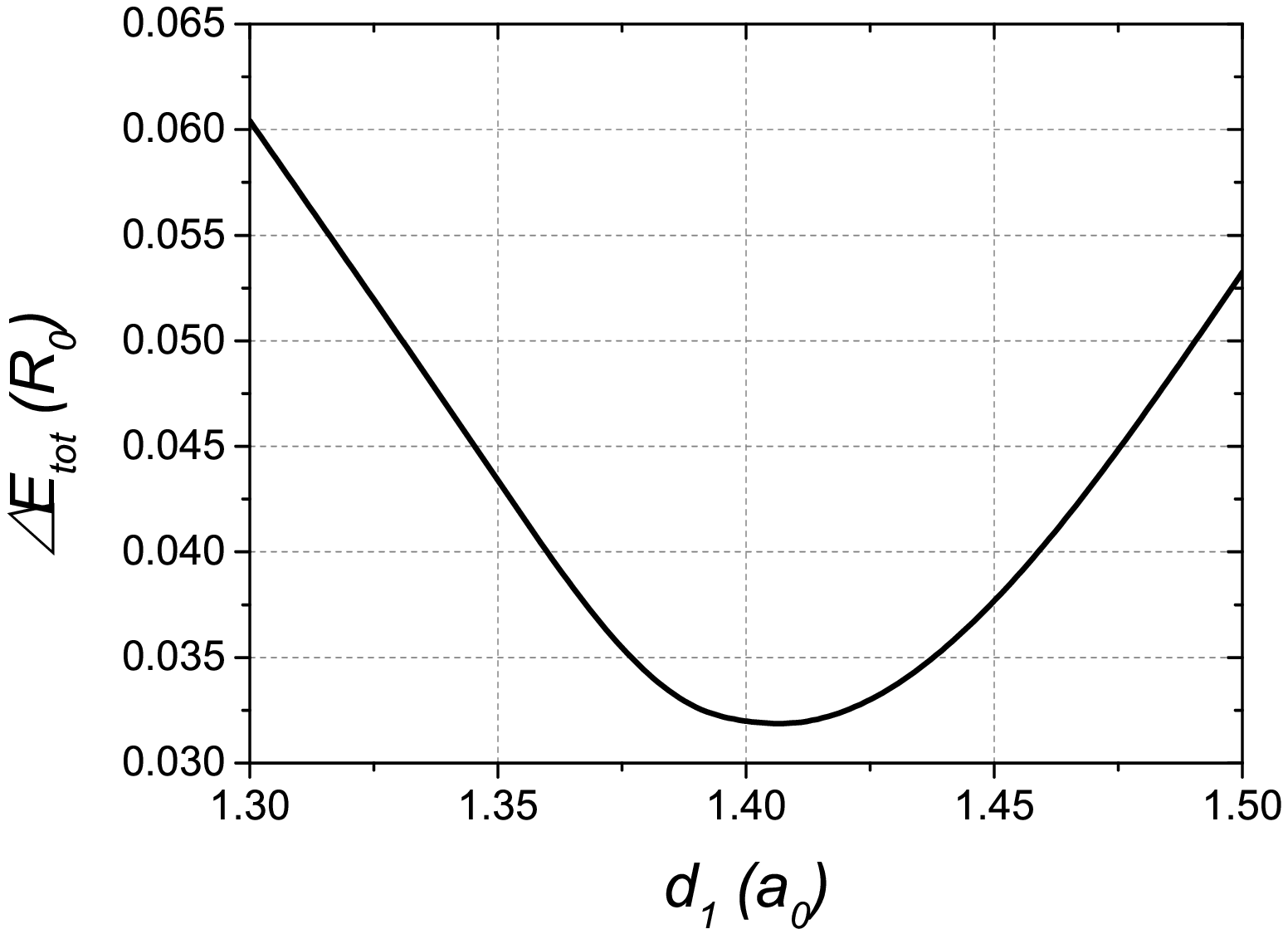}\\(c)
\caption{(a) The differences between the total energy ground state and the first excited state in different systems under different models in the [001] direction when $d_1+d_2=3a_0$: the solid line is for the 4-acceptor full CI calculation, the dashed line is for the 4-acceptor HL calculation, the dotted line is for the 6-acceptor HL calculation, (b) The differences between the total energy ground state and the first 15 excited states for the full CI calculation in the [001] direction when $d_1+d_2=3a_0$, (c) The differences between the old and new total energy ground state during the anti-crossing for the full CI calculation in the [001] direction when $d_1+d_2=3a_0$.}\label{f-12}
\end{figure}
\begin{figure}
\centering
\includegraphics[scale=0.25]{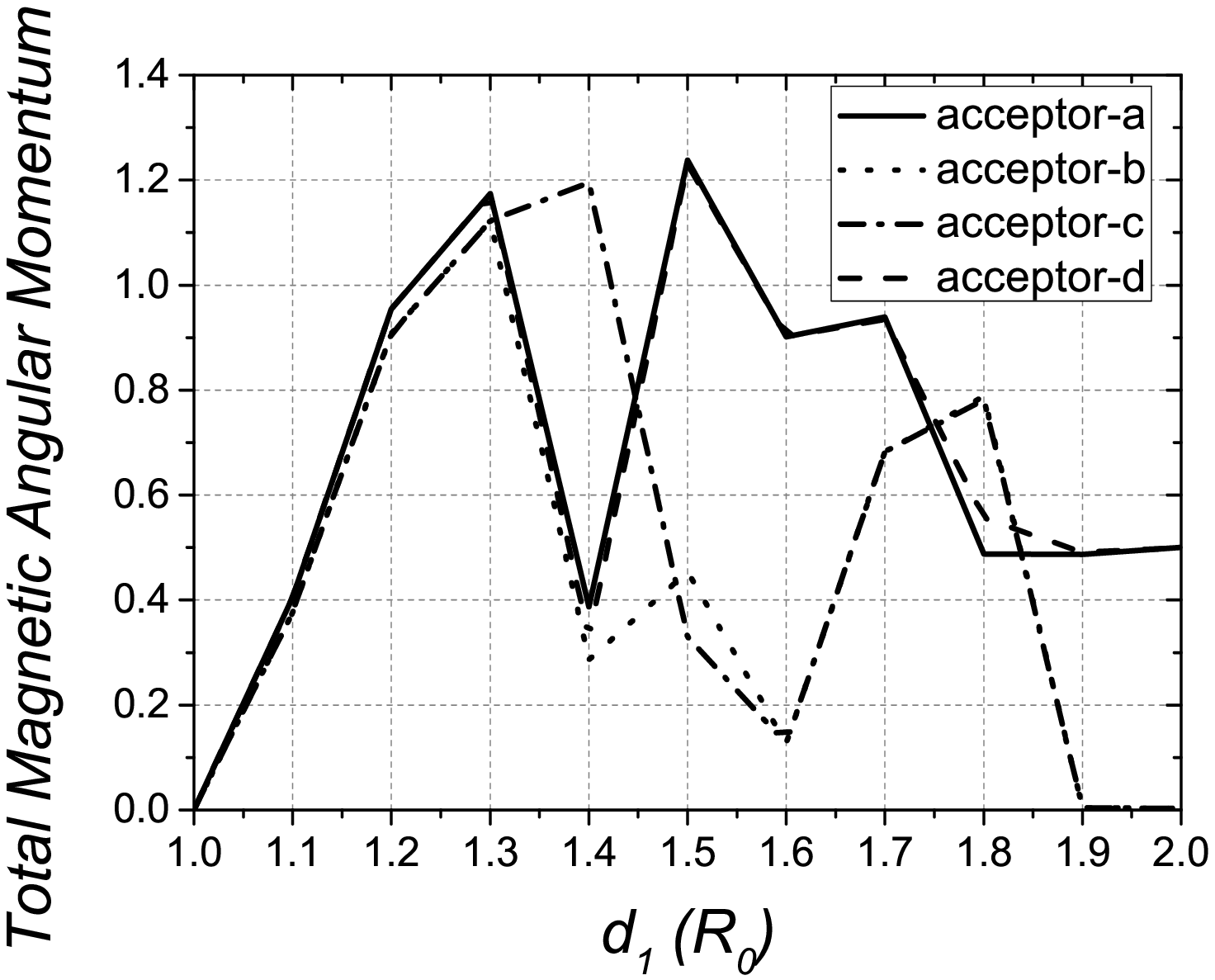}\\(a) [001]\\
\includegraphics[scale=0.25]{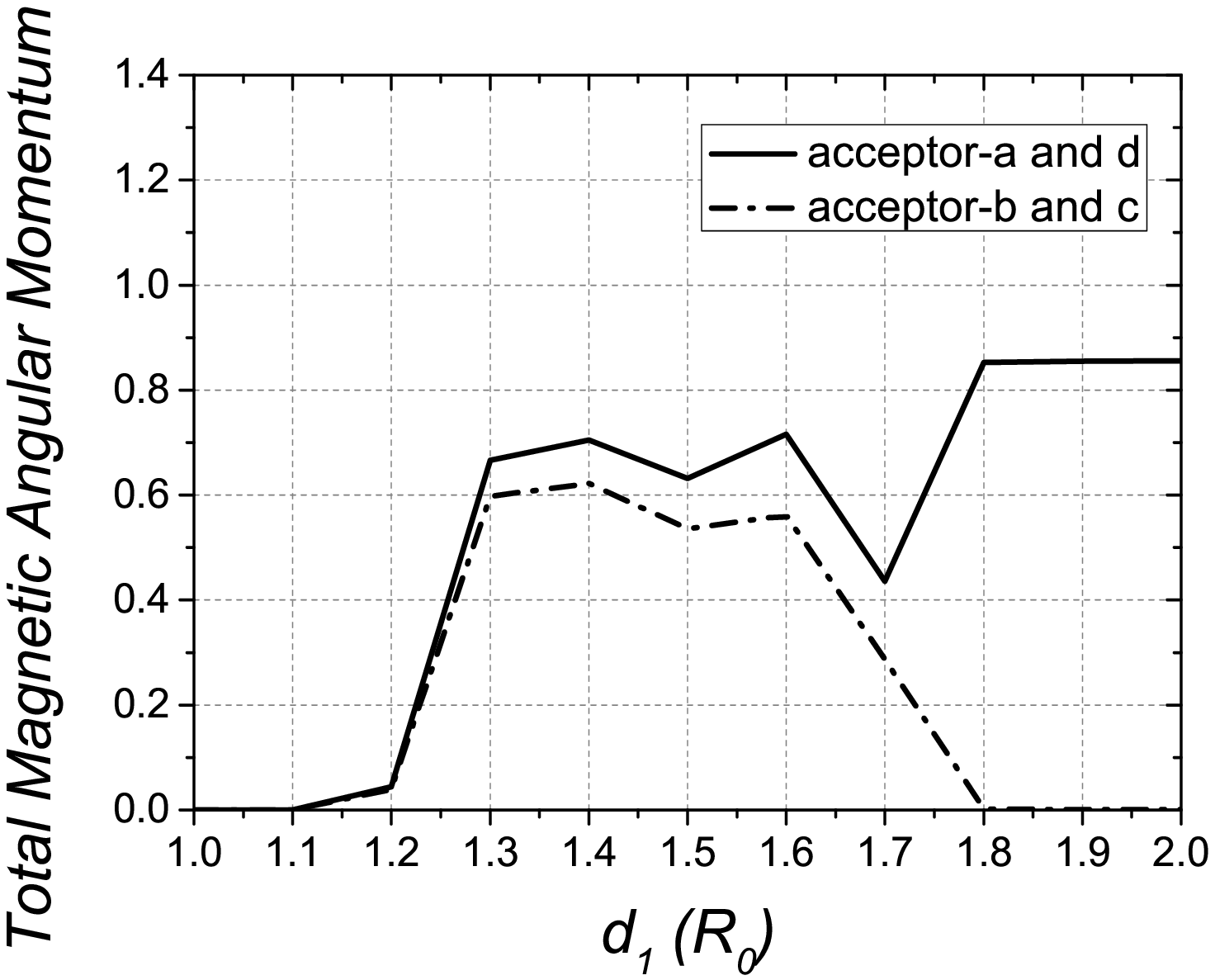}\\(b) [110]\\
\includegraphics[scale=0.25]{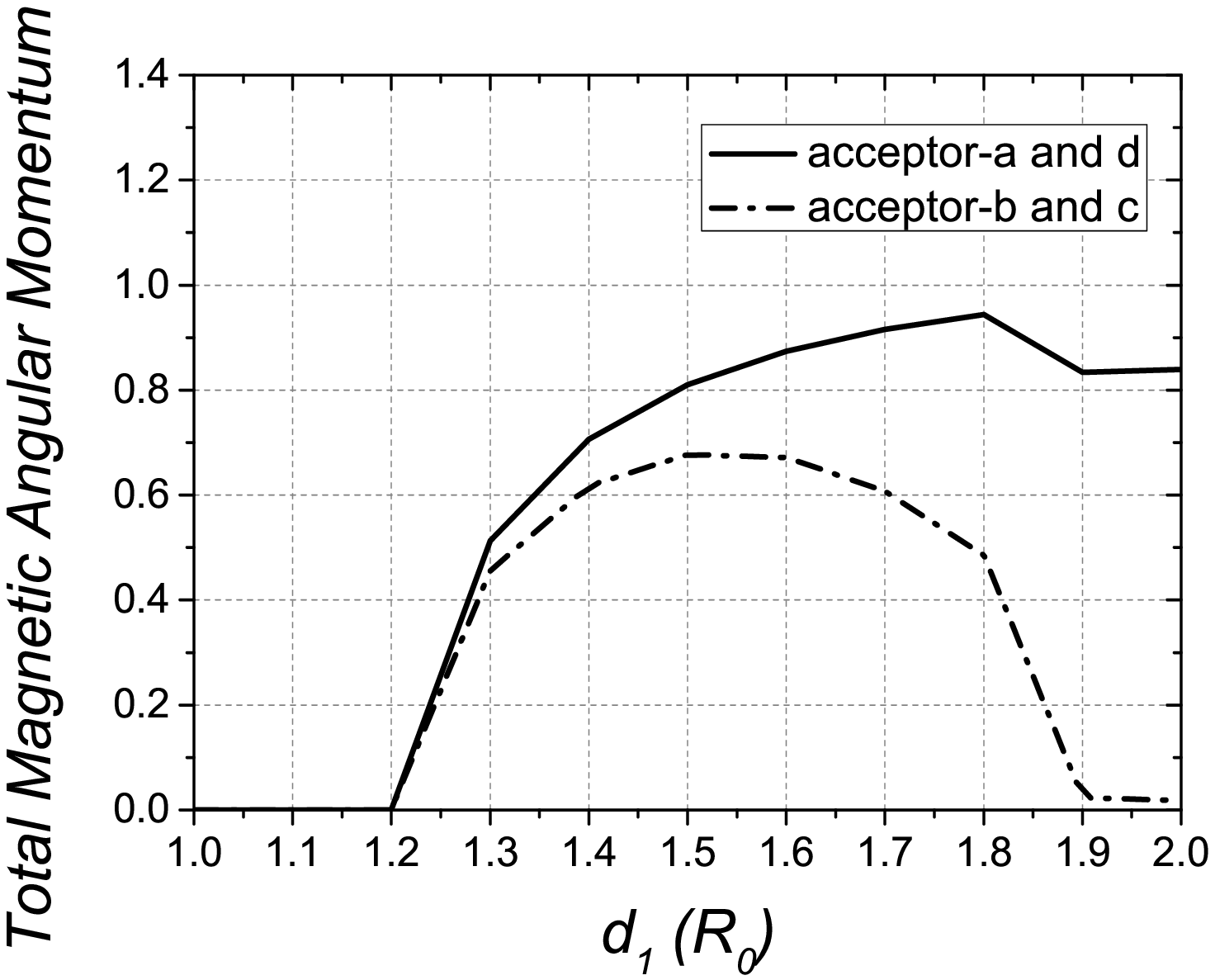}\\(c) [111]
\caption{The behavior of the total magnetic angular momentum obtained from the UHF method under different arrangements in three typical directions for the small-separation case ($d_1+d_2=3a_0$) of the 4-acceptor linear chain: (a) the [001] direction, (b) the [110] direction, (c) the [111] direction.}\label{f-6.1}
\end{figure}
\begin{figure*}
\centering
\includegraphics[scale=0.18]{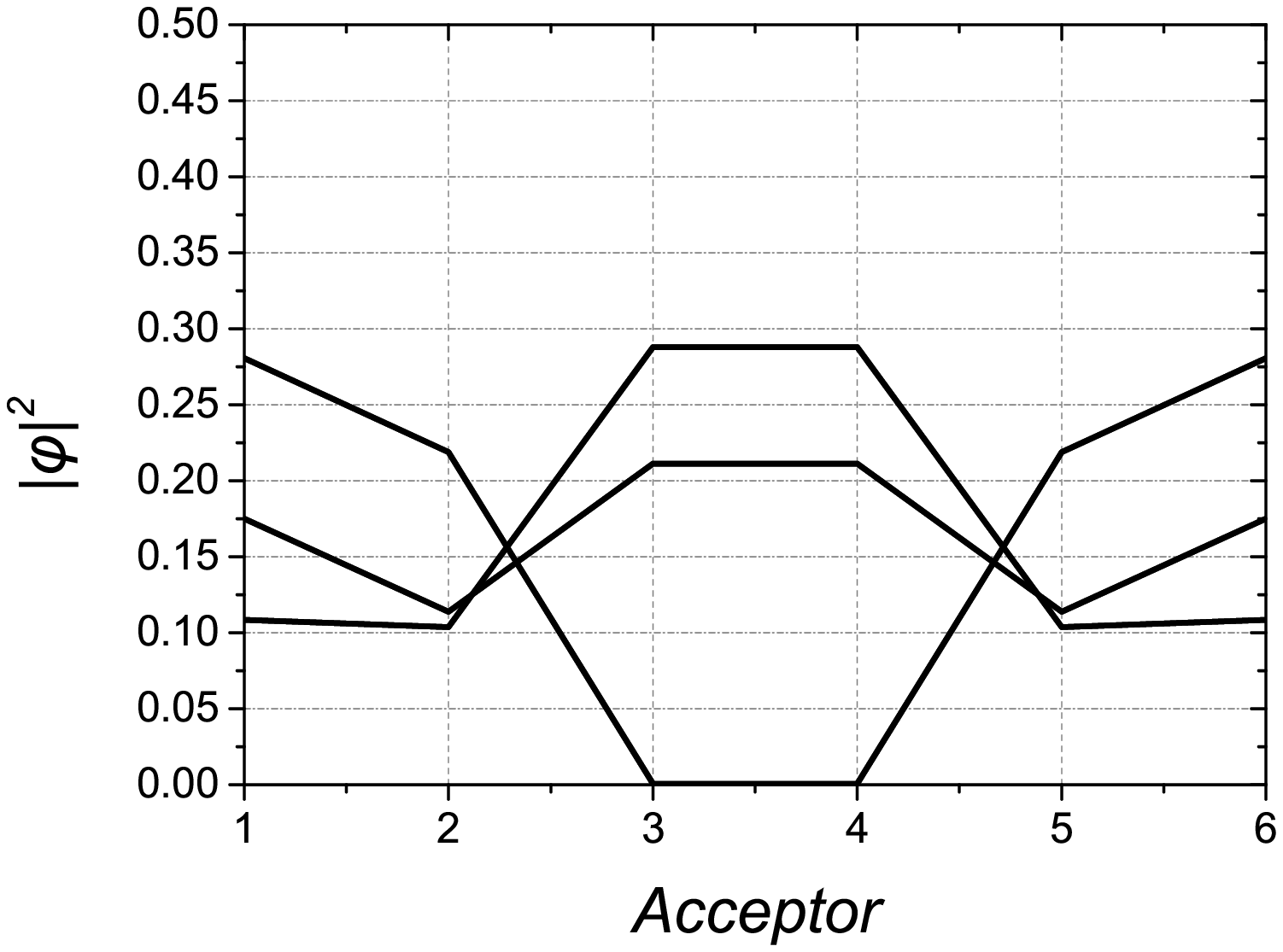}
\includegraphics[scale=0.18]{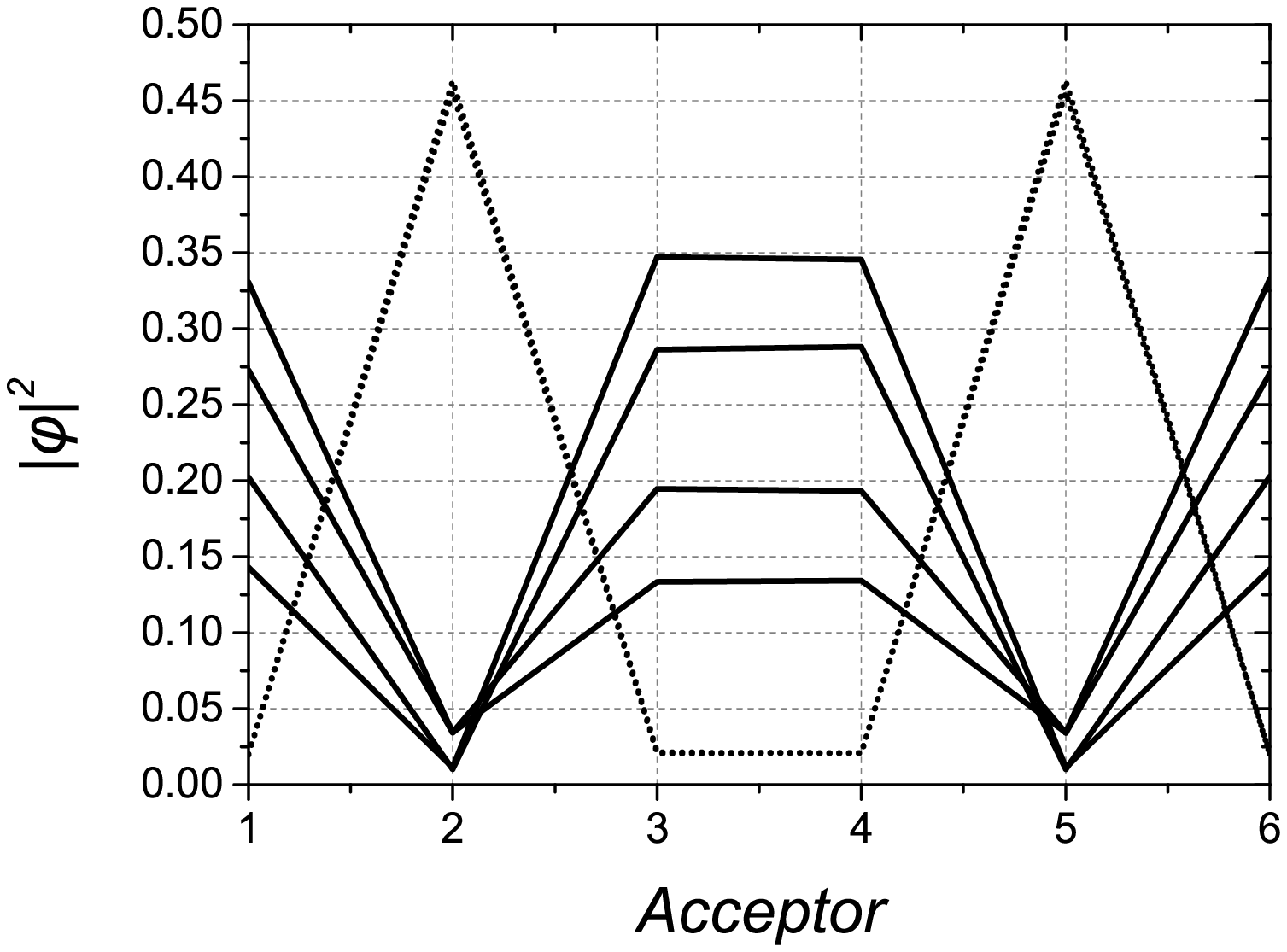}
\includegraphics[scale=0.18]{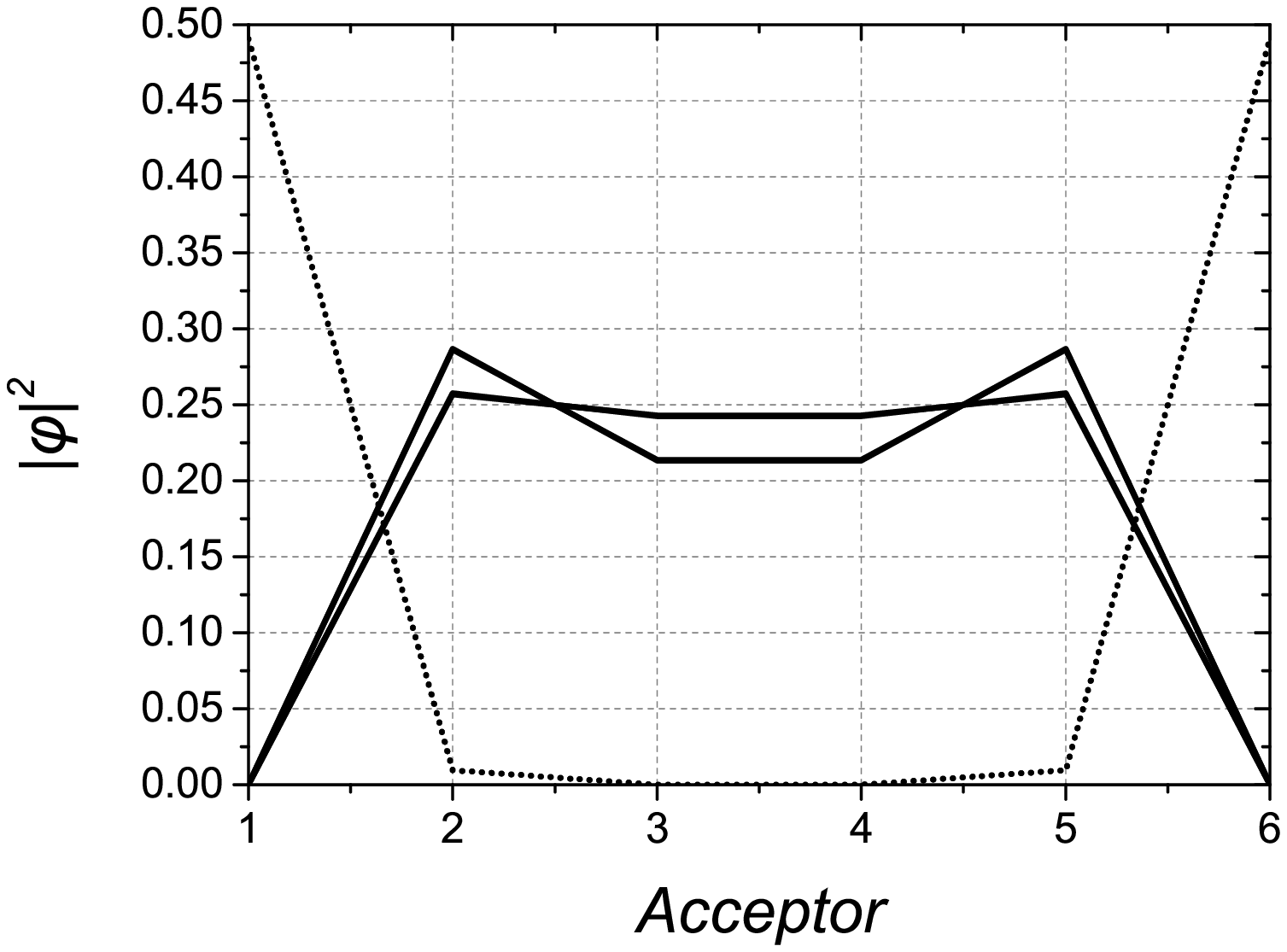}\\(a) [001] \qquad\qquad\qquad\qquad\qquad\qquad\quad (b) [001] \qquad\qquad\qquad\qquad\qquad\qquad\quad (c) [001]\\
\includegraphics[scale=0.18]{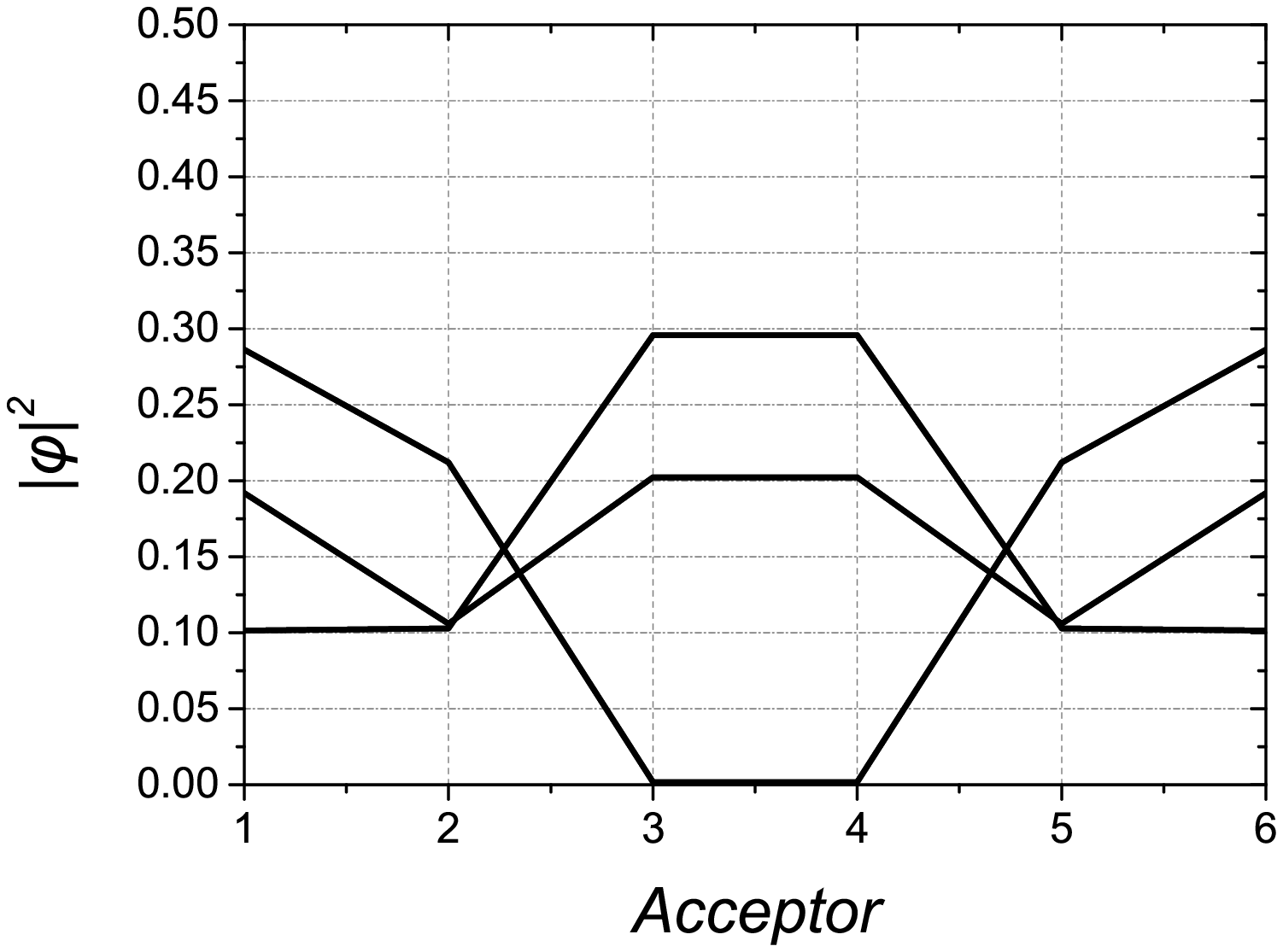}
\includegraphics[scale=0.18]{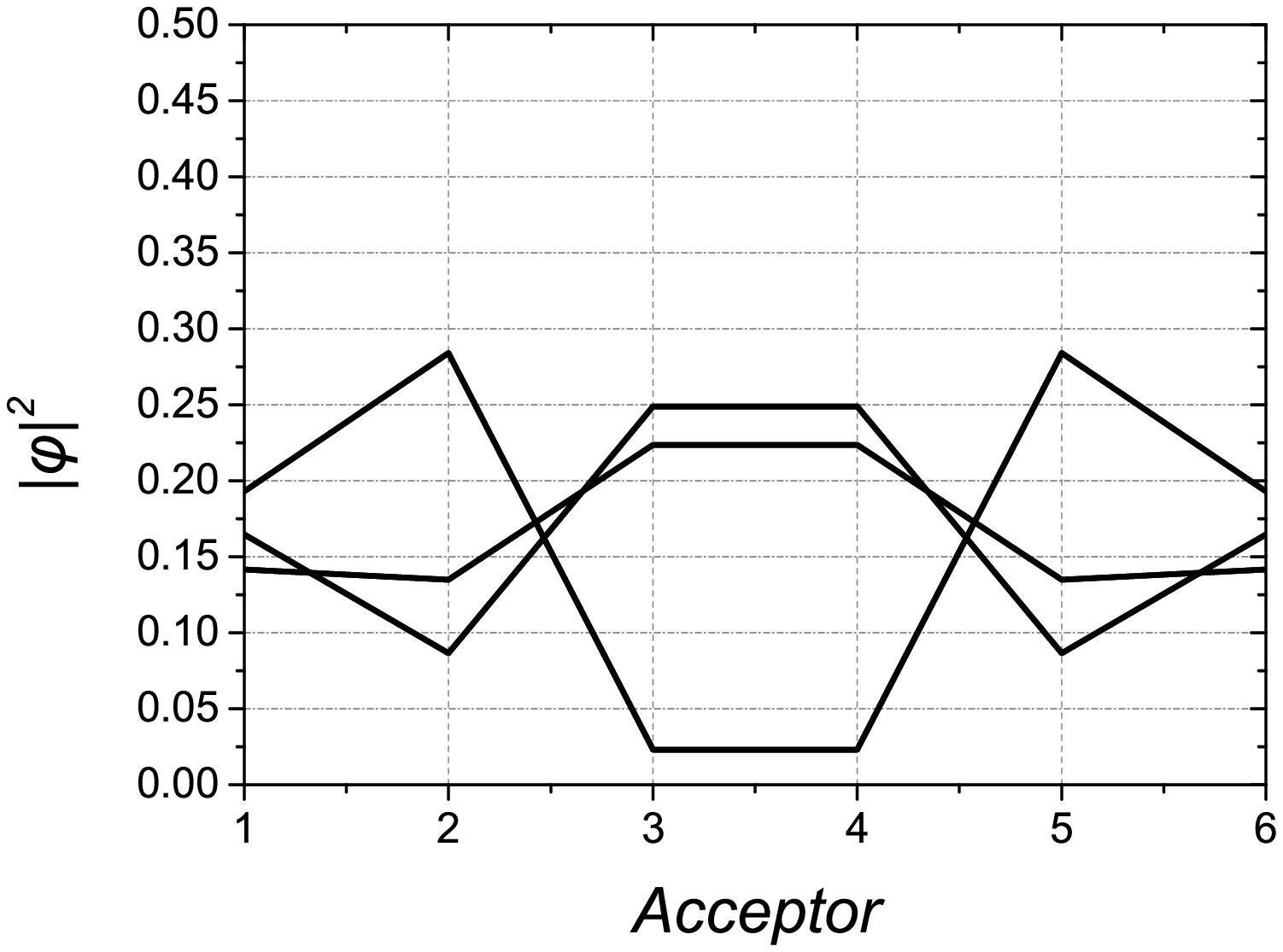}
\includegraphics[scale=0.18]{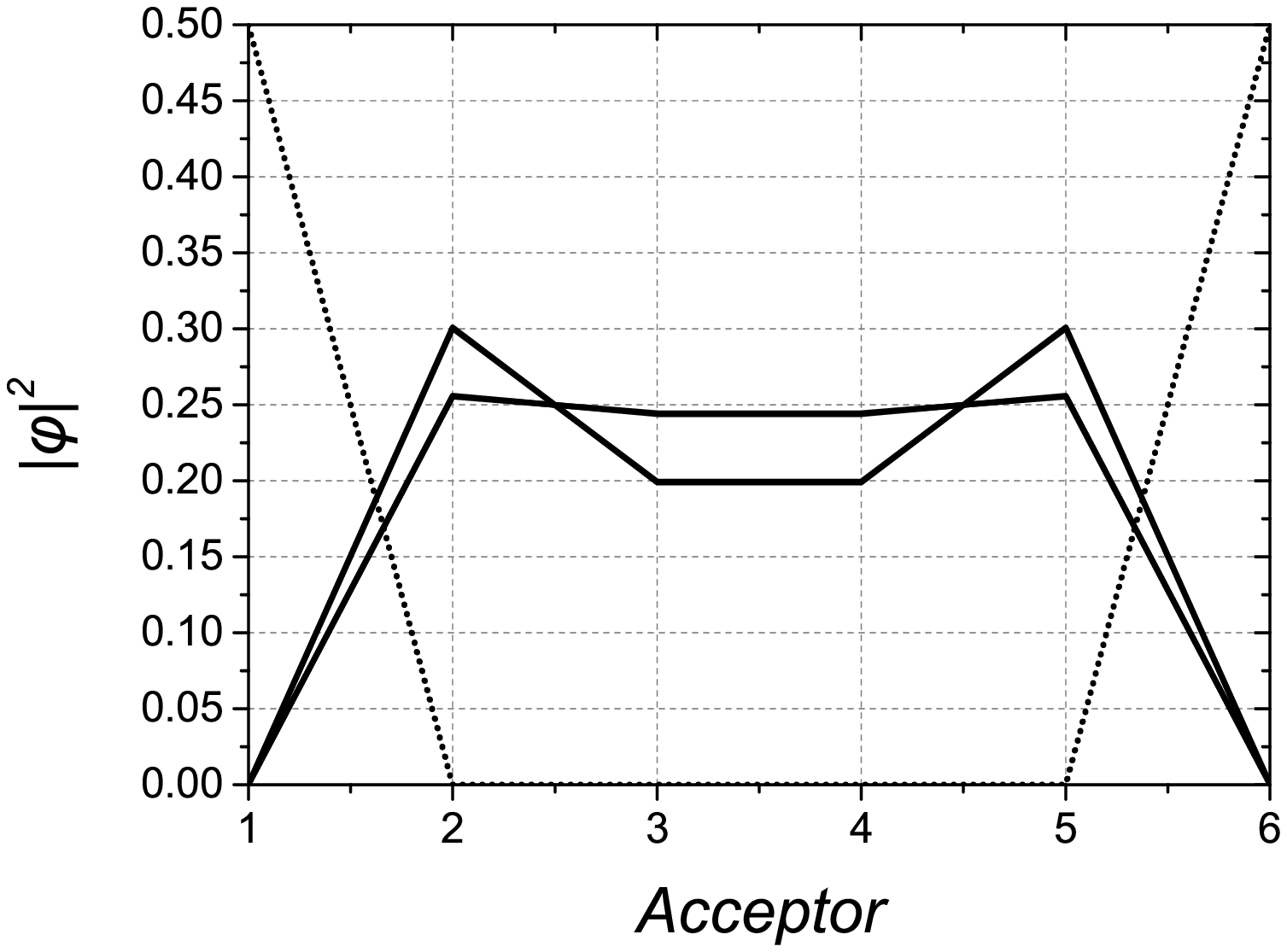}\\(d) [110] \qquad\qquad\qquad\qquad\qquad\qquad\quad (e) [110] \qquad\qquad\qquad\qquad\qquad\qquad\quad (f) [110]\\
\includegraphics[scale=0.18]{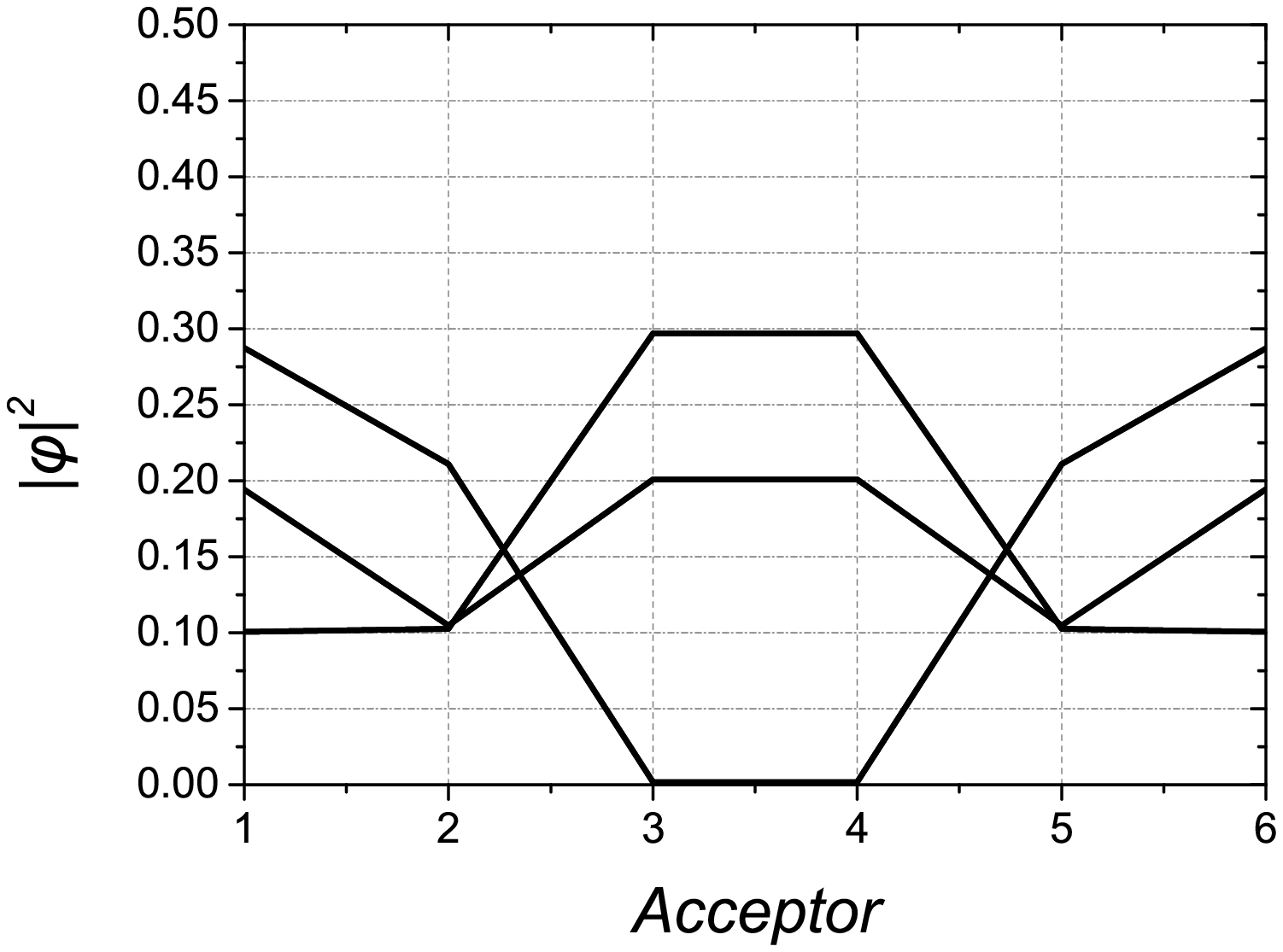}
\includegraphics[scale=0.18]{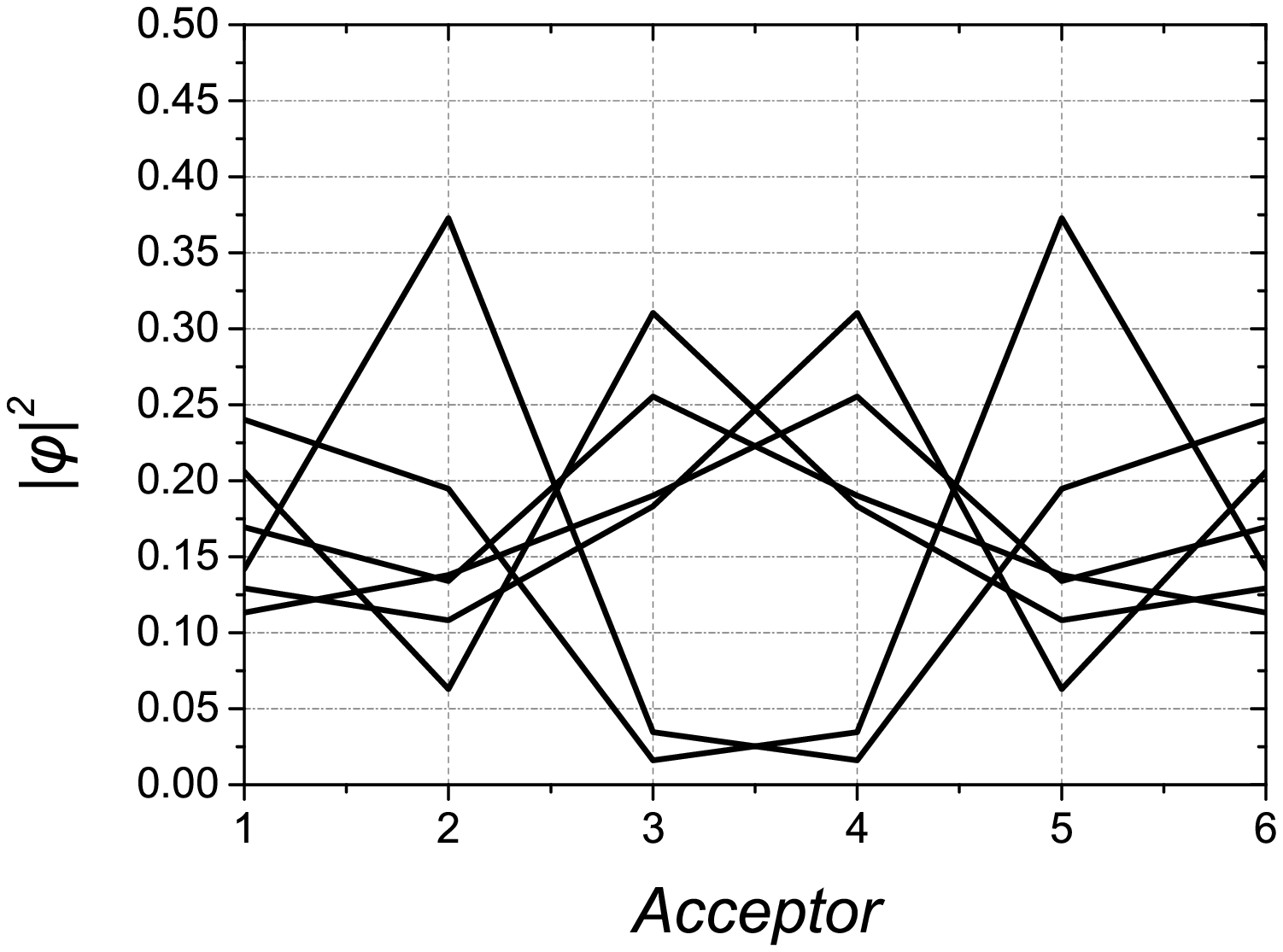}
\includegraphics[scale=0.18]{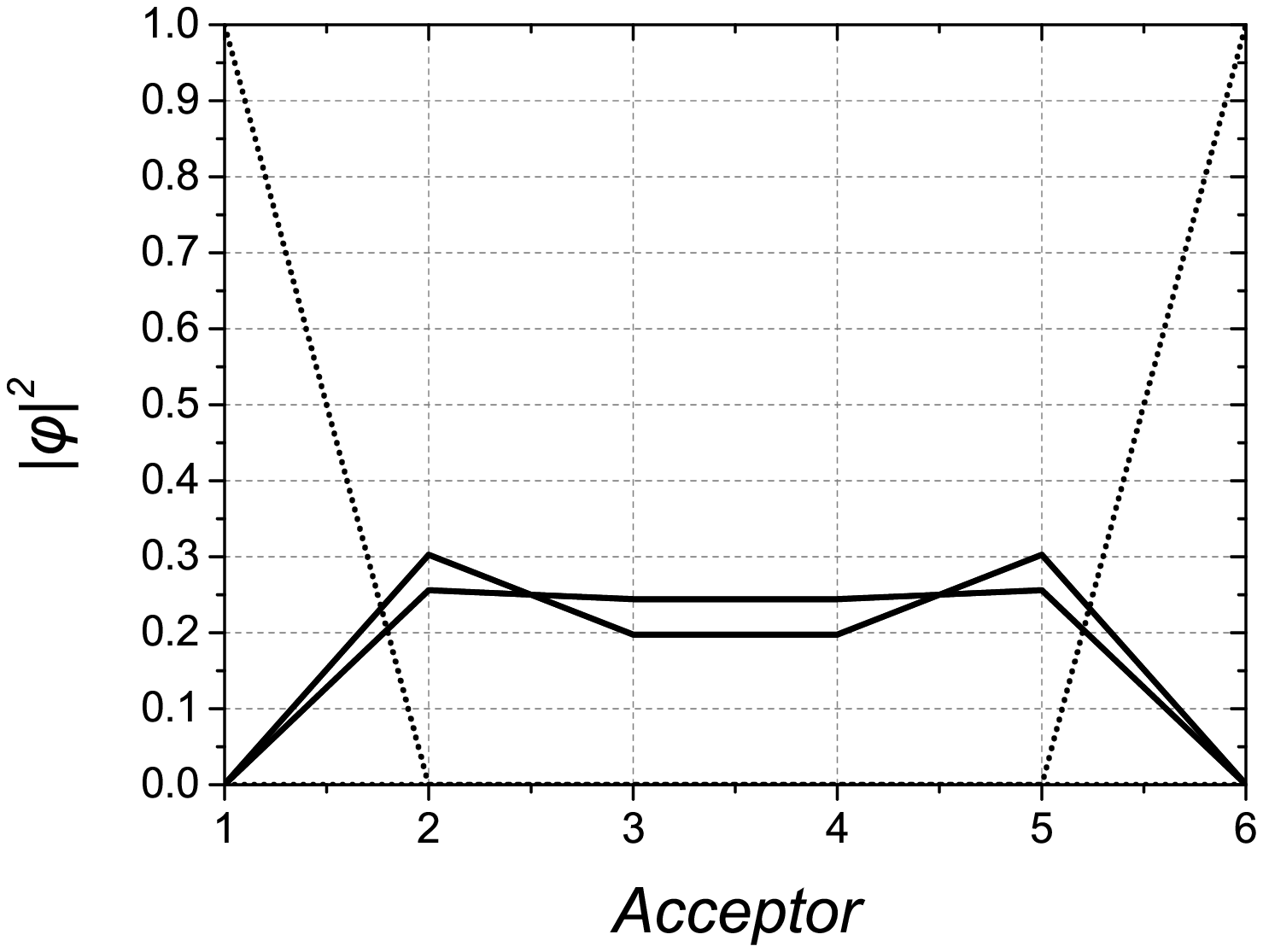}\\(g) [111] \qquad\qquad\qquad\qquad\qquad\qquad\quad (h) [111] \qquad\qquad\qquad\qquad\qquad\qquad\quad (i) [111]
\caption{The total charge distribution among acceptors under different arrangements in three typical directions for the 6-acceptor linear chain when $d_1+d_2=3a_0$: (a) the short-long arrangement in the [001] direction, (b) the uniform chain case in the [001] direction, (c) the long-short arrangement in the [001] direction, (d) the short-long arrangement in the [110] direction, (e) the uniform chain case in the [110] direction, (f) the long-short arrangement in the [110] direction, (g) the short-long arrangement in the [111] direction, (h) the uniform chain case in the [111] direction, (i) the long-short arrangement in the [111] direction. For (c), (f) and (i), the dotted lines are for the states localized at the acceptors at the end of the chain, which are always the highest states among the states involved to form the total energy ground state here. All the lines in (a), (c), (d), (e), (f), (g), the dotted line in (b) and the solid lines in (i) are doubly-degenerate.}\label{f-13}
\end{figure*}

The hole-hole repulsion term now strongly influences the distribution of the holes: although the parabolic potential due to the negative acceptor cores found in our earlier work\cite{Zhu2020LCoAOMfDPAAiS} is still present,  the holes are no longer concentrated in the middle of the chain but are kept apart by their mutual Coulomb repulsion and have a nearly uniform distribution along the chain.  This suggests that our small-separation case is already on the insulating side of the Mott transition, so the HL approximation can naturally be applied and may be expected to give good results.

For chains of 4 acceptors, the ground-state total energy was obtained from all the methods mentioned in \S\ref{model} along three high-symmetry directions, and is shown in Figure \ref{f-4}. Both the HL and UHF methods are reasonable approximations to the full CI result in all directions, with the HL approach offering a better agreement with the full CI calculation. The difference between the full CI and the HL results reduces as the arrangement changes from short-long to long-short; the HL approximation should be more accurate for chains with more acceptors, especially when the average separation between each pair of acceptors is larger. The average separation grows as $d_1$ increases; the result can then be understood by noting that the accuracy of the HL method for a pair remains roughly constant from $d=1a_0$ to $d=1.5a_0$ (see the right column of Figure \ref{f-1}) but then improves from $d=1.5a_0$ to $d=2a_0$. The UHF approximation also becomes more accurate as more acceptors are included, but the significant discrepancies in the energy of a pair with separations around $1.5a_0$  (Figure \ref{f-1}) are reflected in significant errors in the middle of Figure \ref{f-4}, where $d_1\approx d_2\approx 1.5a_0$. We also computed results for chains of 6 acceptors, using the HL and UHF methods only; the behaviour of the total energies was similar.

We analyse the full CI ground-state eigenvector by looking at the dominant components (those with largest absolute values) in the basis of single-acceptor states described in \S\ref{theory}.  We can separate the 4 degenerate states of an isolated acceptor into two groups,  those derived from $m_F=\pm\frac{3}{2}$ and  those from $m_F=\pm\frac{1}{2}$. We refer to the ground state as `un-hybridized' if the dominant components contain either $m_F=\pm\frac{3}{2}$ or $m_F=\pm\frac{1}{2}$ single-acceptor states (but not both), while we refer to it as `hybridized' if they contain both types of single-accpetor states.

In Figure \ref{f-5}, we show the behaviour of the 50 highest-energy (hence, most favourable) states of the full CI calculation under different arrangements of the bonds along three high-symmetry directions. For the [001] direction, the ground state is non-degenerate on the left-hand (short-long) side of the picture, while it joins three other states and forms a 4-fold-degenerate state on the right-hand side (long-short arrangement side) which is followed in energy by a 8-fold-degenerate state and another 4-fold-degenerate state as shown in Figure \ref{f-5} (b). We observe that among the dominant components, only the states on the acceptors at the end of the chain change between these states; the dimensionality 16 of these highest manifolds comes from the 4 levels on one end multiplied by 4 levels on the other end, implying the existence of a manifold of edge states.  The situation is similar for the other directions; we analyse the structure of this manifold in more detail in \S\ref{edge-states}. It also can be seen that the ground state crosses with the highest exited states between $d_1=1.4a_0$ and $d_1=1.5a_0$ in the [001] direction; the dominant components of the ground state are unhybridized to the left of the dotted line but become hybridized to the right of it. We will refer to the separation where the crossing (or anti-crossing) between the states happens as the `crossing point', and the separation where dominant component of the ground state changes as the `changing point'.  We see that within the resolution of the step size used ($0.1a_0$), the crossing point and the changing point are the same in the [001] direction.

For the UHF calculations we can understand the overall state most clearly in terms of the behaviour of the Fock matrix eigenvalues, shown for different directions in Figure \ref{f-6} (4-acceptor chain) and Figure \ref{f-11} (6-acceptor chain). Here the states are usually doubly degenerate (corresponding to Kramers degeneracy under time-reversal symmetry) but show splittings for certain acceptor arrangements where the symmetry is lower (see \S\ref{symmetry-breaking}).  The four highest states in Figure~\ref{f-6}, and the six highest in Figure~\ref{f-11}, will be occupied by holes. In all cases there is a large gap between filled and empty states due to the effect of the strong hole-hole repulsion. Compared with Figure \ref{f-2} for a dimer, the two significant differences are (i) the splitting of degenerate states, and (ii) the crossing between filled states in the [001] direction in Figure \ref{f-6} (b).  In general we find that the self-consistency cycle in the UHF method breaks the symmetry of the system, with different sets of eigenvectors of the Fock matrix corresponding to the same total energy; we analyse this symmetry breaking further in \S\ref{symmetry-breaking}. The crossing occurs close to the changing point identified in the CI calculation, so the change in the single-acceptor energy levels in the dominant component of the CI ground state is related to a change in the ordering of single-electron states in UHF. For the 6-acceptor chain, it can be seen from Figure \ref{f-11} that another crossing appears around $d_1=1.7a_0$, implying another similar crossing between the total energy ground state and higher excited states which is not able to be shown due to the limit number of acceptors. 

The HL approach for the 4-acceptor chain (not shown) gives similar results to the CI method, including a 4-fold-degenerate ground state when $d_1>d_2$ and the presence of a changing point where the composition of the ground state changes; however, the changing point now appears between $d_1=1.3a_0$ and $d_1=1.4a_0$, while the crossing point is still around $d_1=1.4a_0$.  This suggests that the HL method is a good approximation for both the ground state and low-lying excited states, and preserves some of the main features of the energy spectrum. For the 6-acceptor chain there is only one obvious crossing between the ground state and the first excited states, as the degenerate states appear for significantly smaller values of $d_1$ than before. But we now see two changing points for the eigenvectors: one is between $d_1=1.3a_0$ and $d_1=1.4a_0$, the other is between $d_1=1.6a_0$ and $d_1=1.7a_0$. 

To understand in more detail the behavior of the energy gap, we show in Figure \ref{f-12} (a) the difference between the total-energy ground state and first excited state in the [001] direction as a function of $d_1$. There are two regions of particular interest; the first is the neighbourhood of the crossing/changing point where the gap reduces and then increases again ($d_1=1.3a_0$ to $d_1=1.4a_0$). The minimum gap for 4 acceptors is around $1.4a_0$ for both the CI case (solid line) and the HL case (dashed line), but shifts to shorter separations for 6 acceptors (dotted line). To show the details of the crossings among the first few states, a good choice is to show the energy difference between the ground-state and excited states as the energies shift dramatically from the short-long arrangement to the long-short arrangement according to Figure \ref{f-5} (a). In this way, the crossings between excited states are shown as usual, while the crossing between the ground state and excited states will be reflected by the value of the difference. Here for the convenience of the further discussion, the ground state before the changing point is called as $\left|\phi_0\right\rangle$, while the ground state after the changing point is called as $\left|\phi_0^{'}\right\rangle$. In Figure \ref{f-12} (b), we show the energy difference between the ground-state and first 15 excited states for the full CI calculation, where we find a small gap between excited states around $1.4a_0$, which appears to make the `crossings' here anti-crossings as $\left|\phi_0^{'}\right\rangle$ is found above this gap before the changing point. It is also reasonable to believe the others in the full CI calculation and HL approach are anti-crossings. As there is a band of excited states with similar energies in Figure \ref{f-12} (b), it is helpful to follow the energy difference between the ground state and the excited state that crosses with it, rather than the minimum gap; in Figure \ref{f-12} (c), we show the energy difference between the previous and new ground states during the anti-crossing. This suggests that the true anti-crossing is between $d_1=1.40a_0$ and $d_1=1.41a_0$, a slightly larger value than in the HL approach. The second region of interest is the right-hand side (large $d_1$), where the 4-fold-degenerate manifold of ground states in the 6-acceptor system forms for smaller values of $d_1$ than in the 4-acceptor system; alternatively, for a given $d_1>d_2$,  the degeneracy of the ground state becomes better as more acceptors are involved (the same is true for the following 8-fold-degenerate and 4-fold-degenerate manifolds). This is what would be expected if the degeneracy arises from almost independent sets of localised edge states at either end of the chain (see \S\ref{edge-states}).


Figure \ref{f-6.1} shows that the magnitude of the expectation value of the angular momentum vector on each acceptor in the symmetry-broken UHF solution.  At the smallest values of $d_1$ (the short-long case) the angular momentum is zero everywhere, whereas for large $d_1$ (the long-short case) it is dominantly located at the ends of the chain.  To see if this is related to possible non-trivial edge states, we show the hole distributions from each eigenvector of the Fock matrix for different arrangements in the three high-symmetry directions in Figure \ref{f-13}. Here `short-long' refers to $d_1=1a_0$, $d_2=2a_0$, and `long-short' to $d_1=2a_0$, $d_2=1a_0$. The one-hole states do not localize at any particular acceptor under the short-long or uniform arrangements; however, for the long-short case, two states localize at the ends of the chain (the dotted lines in Figure \ref{f-13} (c), (f) and (i)), while the others have a nearly uniform distribution across the middle. The states localized at the ends (the dotted lines in Figure \ref{f-13} (c), (f) and (i)) are always the lowest (i.e. least favourable) states occupied by holes, which may imply the existence of the non-trivial edge states occurring in the long-short case (since the charge rearrangements we previously identified in the non-interacting case \cite{Zhu2020LCoAOMfDPAAiS} in response to the parabolic potential no longer force the states localized at the end of the chain to be the highest ones and intervene to shift the edge states to the short-long limit).

\subsubsection{Large-separation case ($d_1+d_2=6a_0$)}\label{4alargeresult}

\begin{figure}
\centering
\includegraphics[scale=0.25]{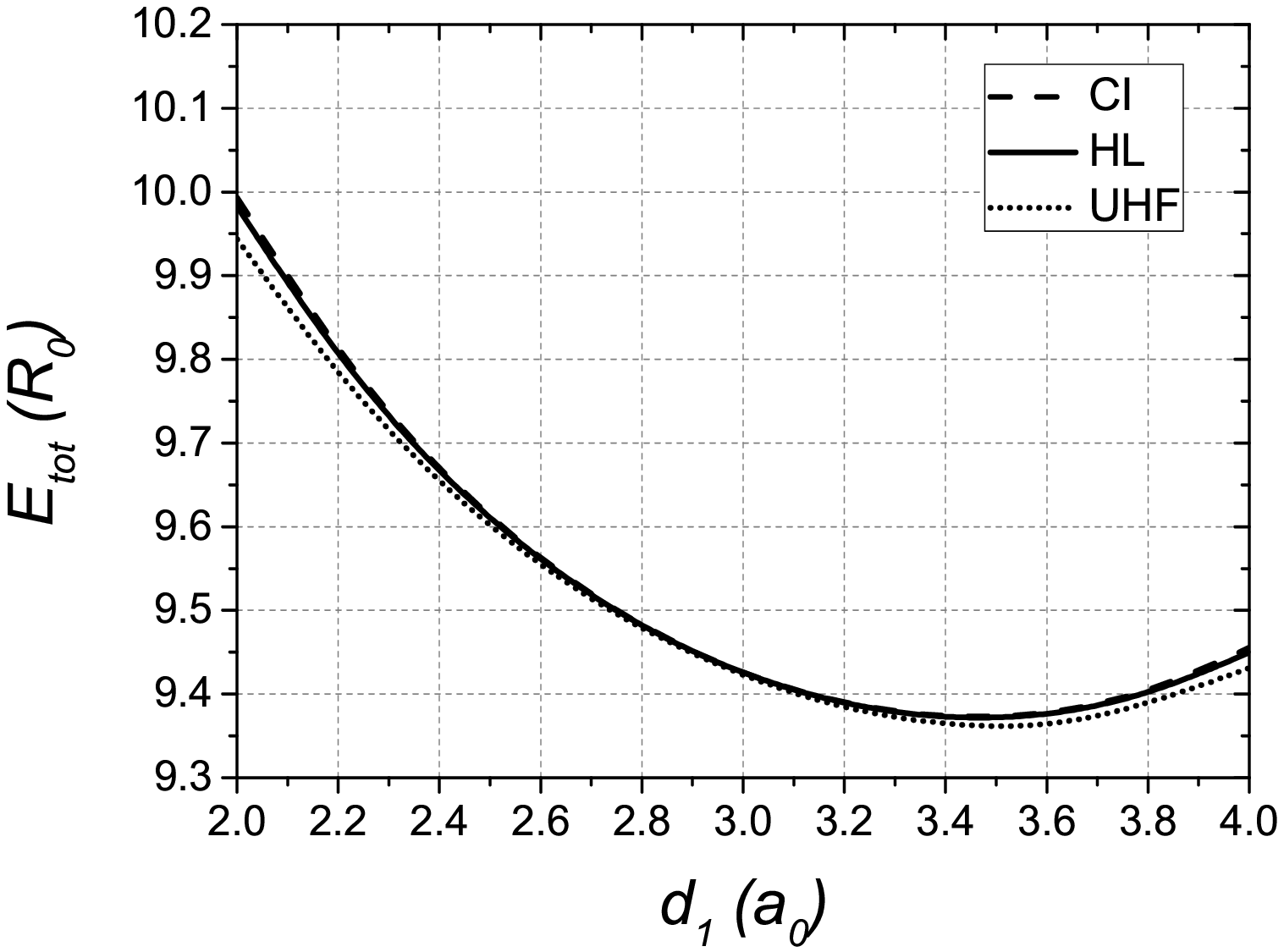}\\(a) [001]\\
\includegraphics[scale=0.25]{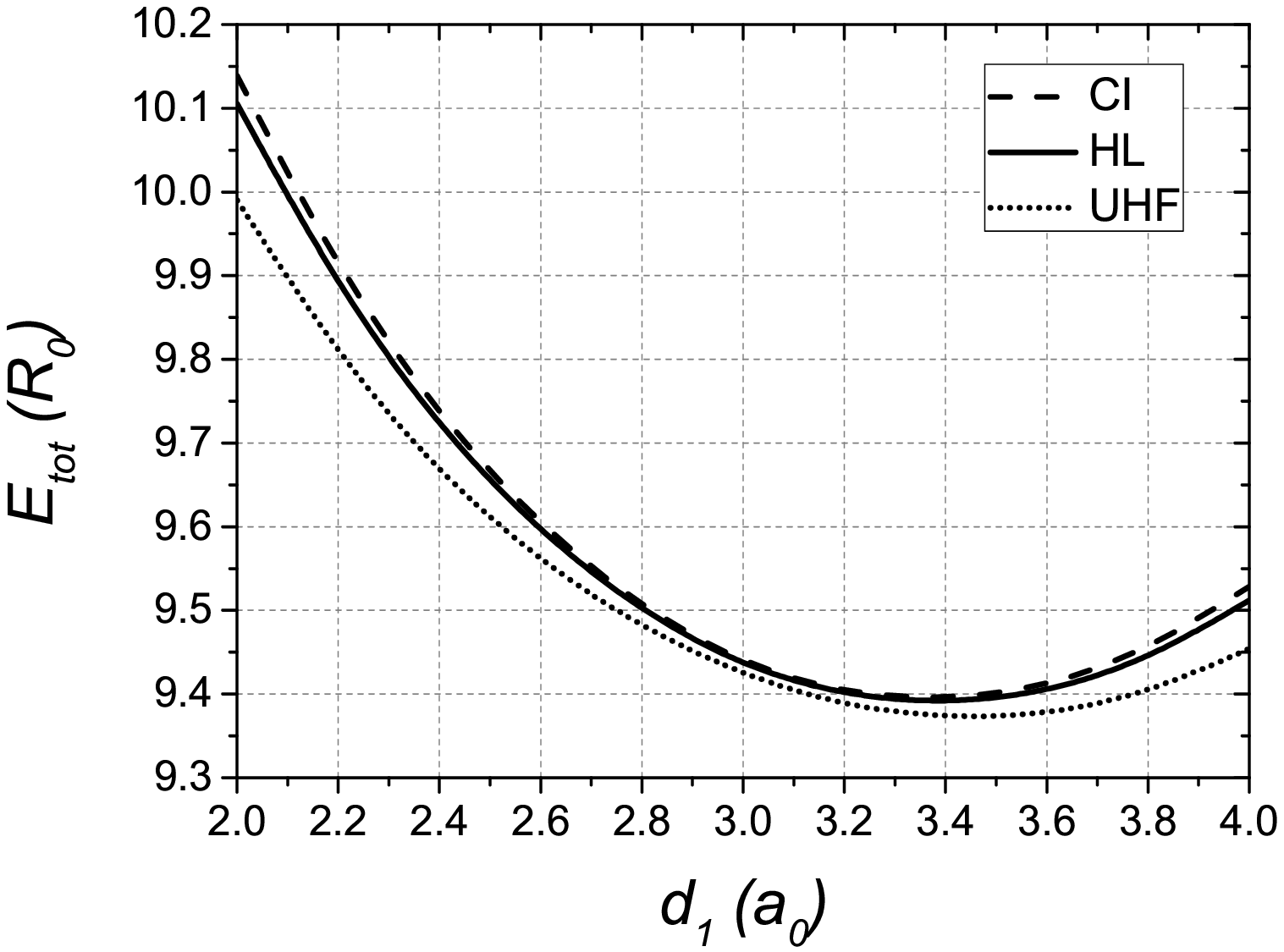}\\(b) [110]\\
\includegraphics[scale=0.25]{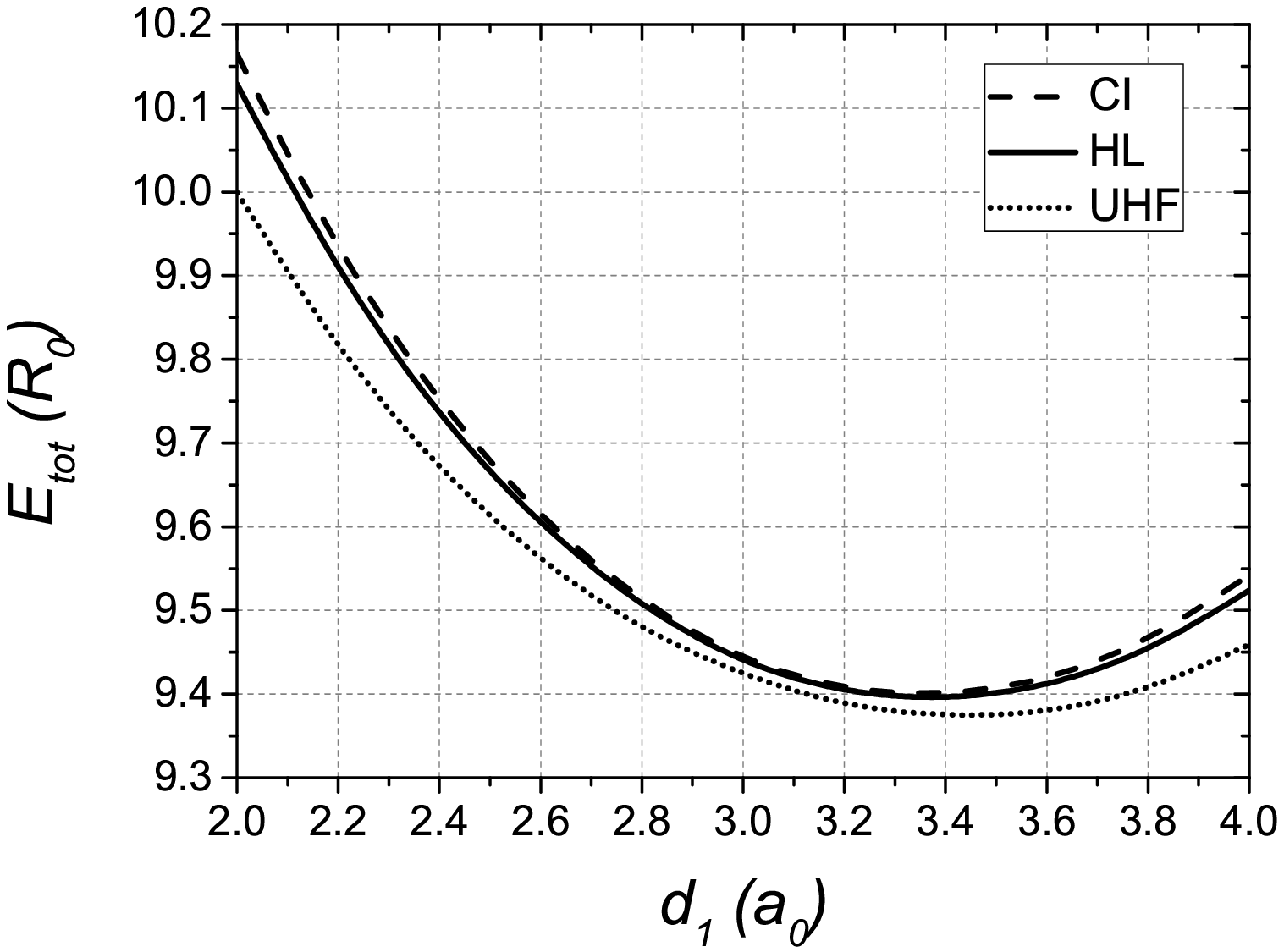}\\(c) [111]
\caption{The total energy of the ground state in three high-symmetry directions for the large-separation case ($d_1+d_2=6a_0$) of the 4-acceptor linear chain: (a) the [001] direction, (b) the [110] direction, (c) the [111] direction. The dashed line is for the full CI calculation, the solid line is for the HL approximation, the dotted line is for the UHF method.}\label{f-7}
\end{figure}
\begin{figure}
\centering
\includegraphics[scale=0.25]{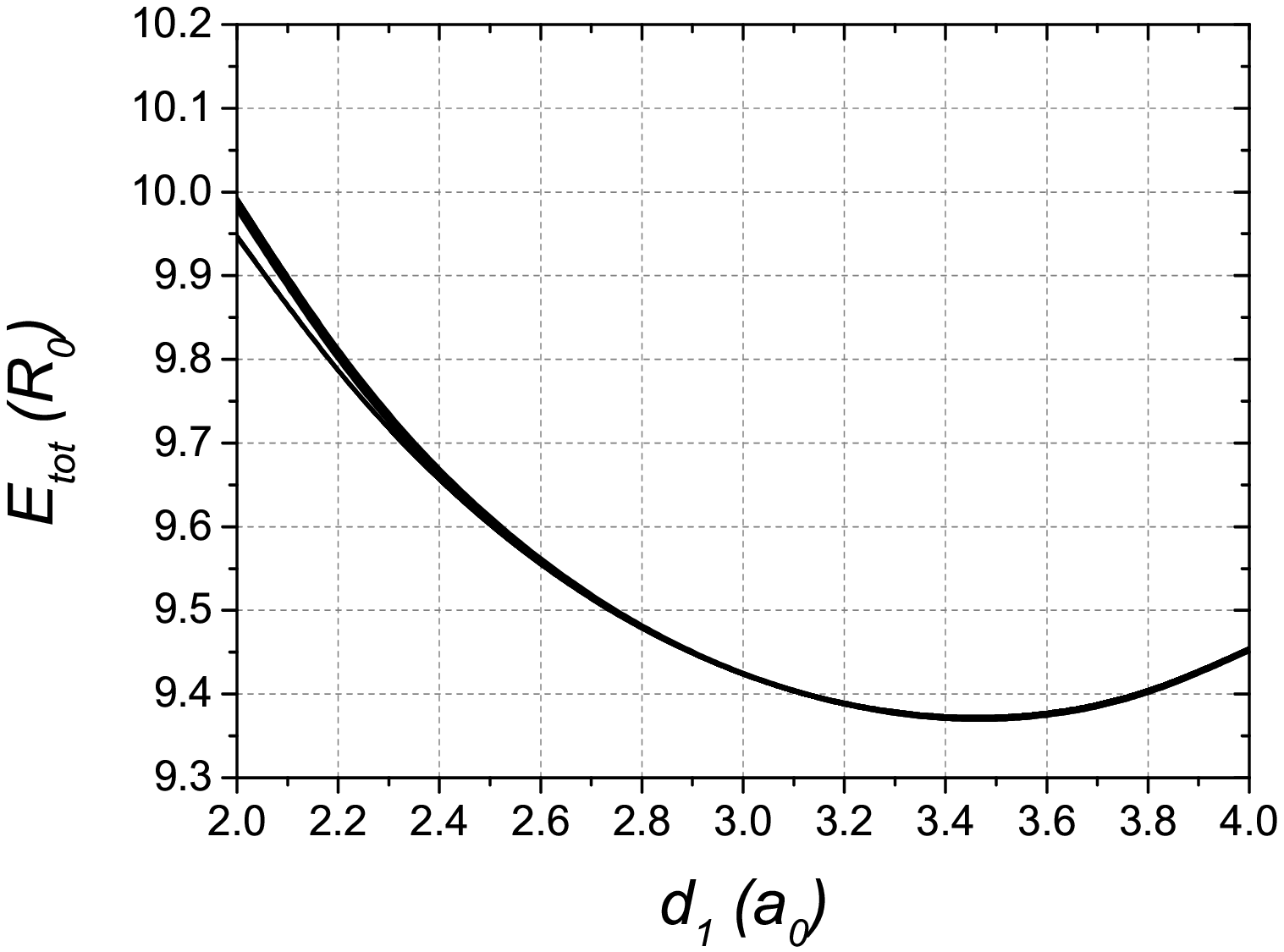}\\(a) [001]\\
\includegraphics[scale=0.25]{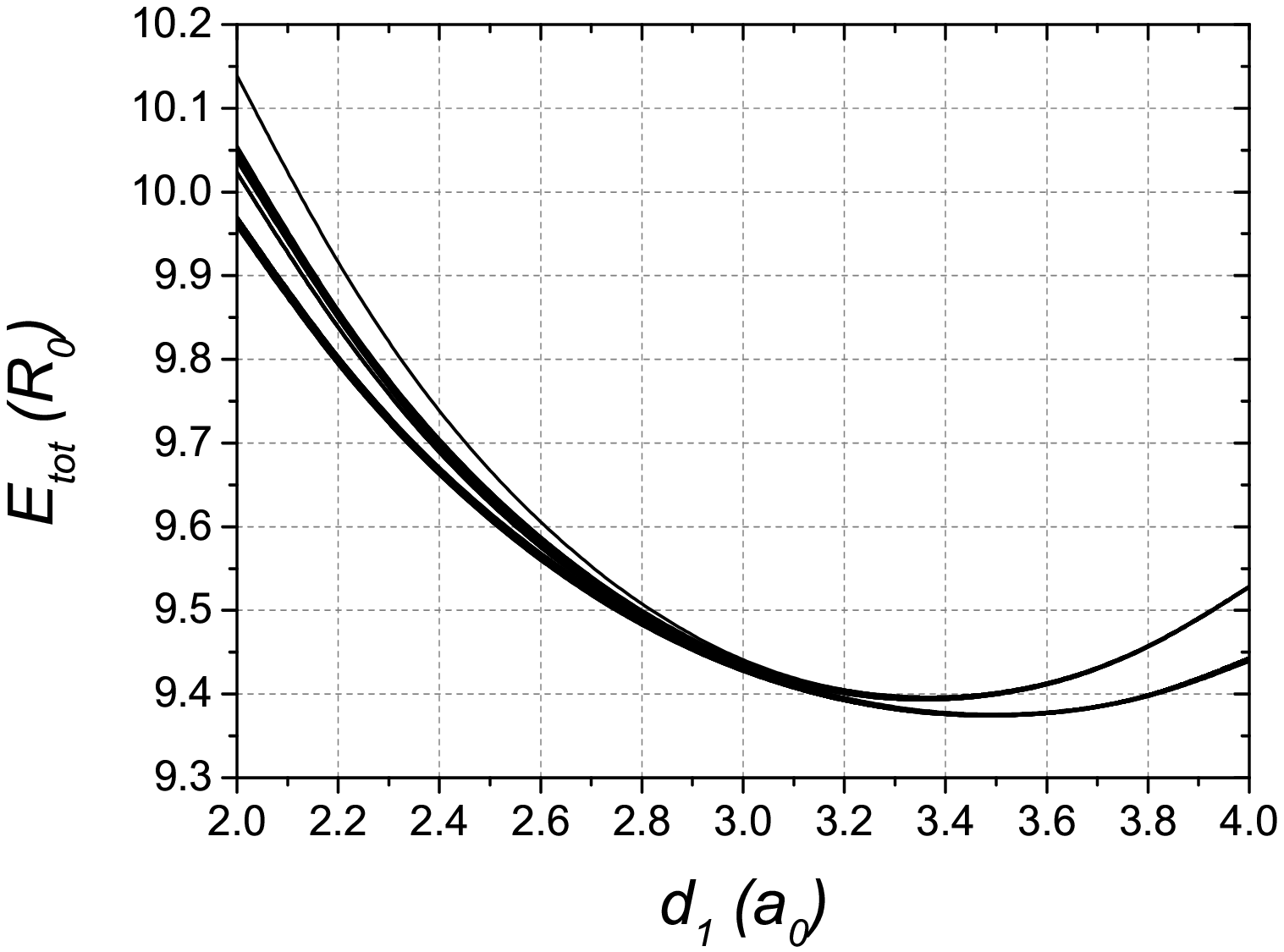}\\(b) [110]\\
\includegraphics[scale=0.25]{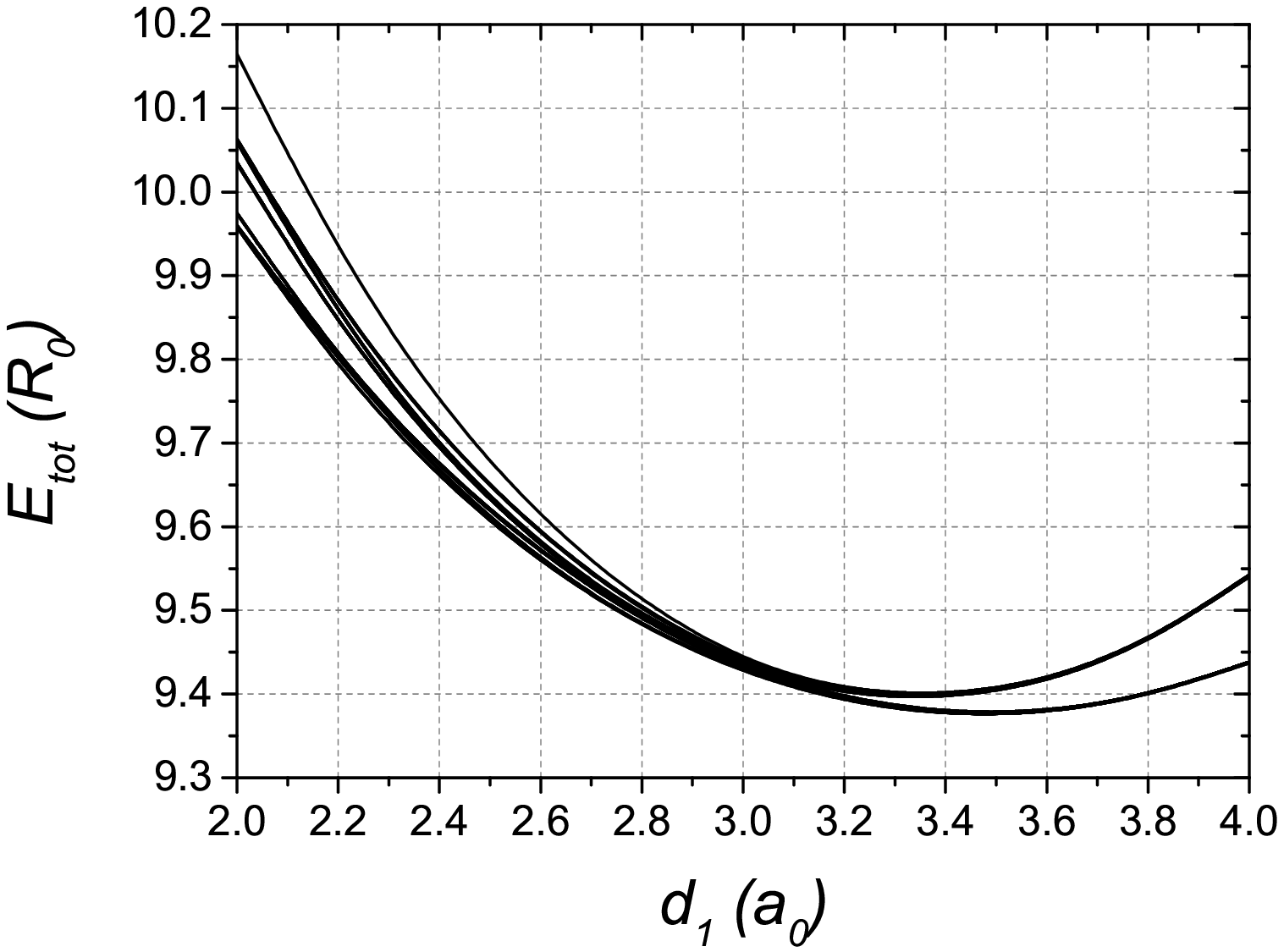}\\(c) [111]
\caption{The total energies of the highest 50 states of the full CI result in three high-symmetry directions for the large-separation case ($d_1+d_2=6a_0$) of the 4-acceptor linear chain: (a) the [001] direction, (b) the [110] direction, (c) the [111] direction.}\label{f-8}
\end{figure}
\begin{figure}
\centering
\includegraphics[scale=0.25]{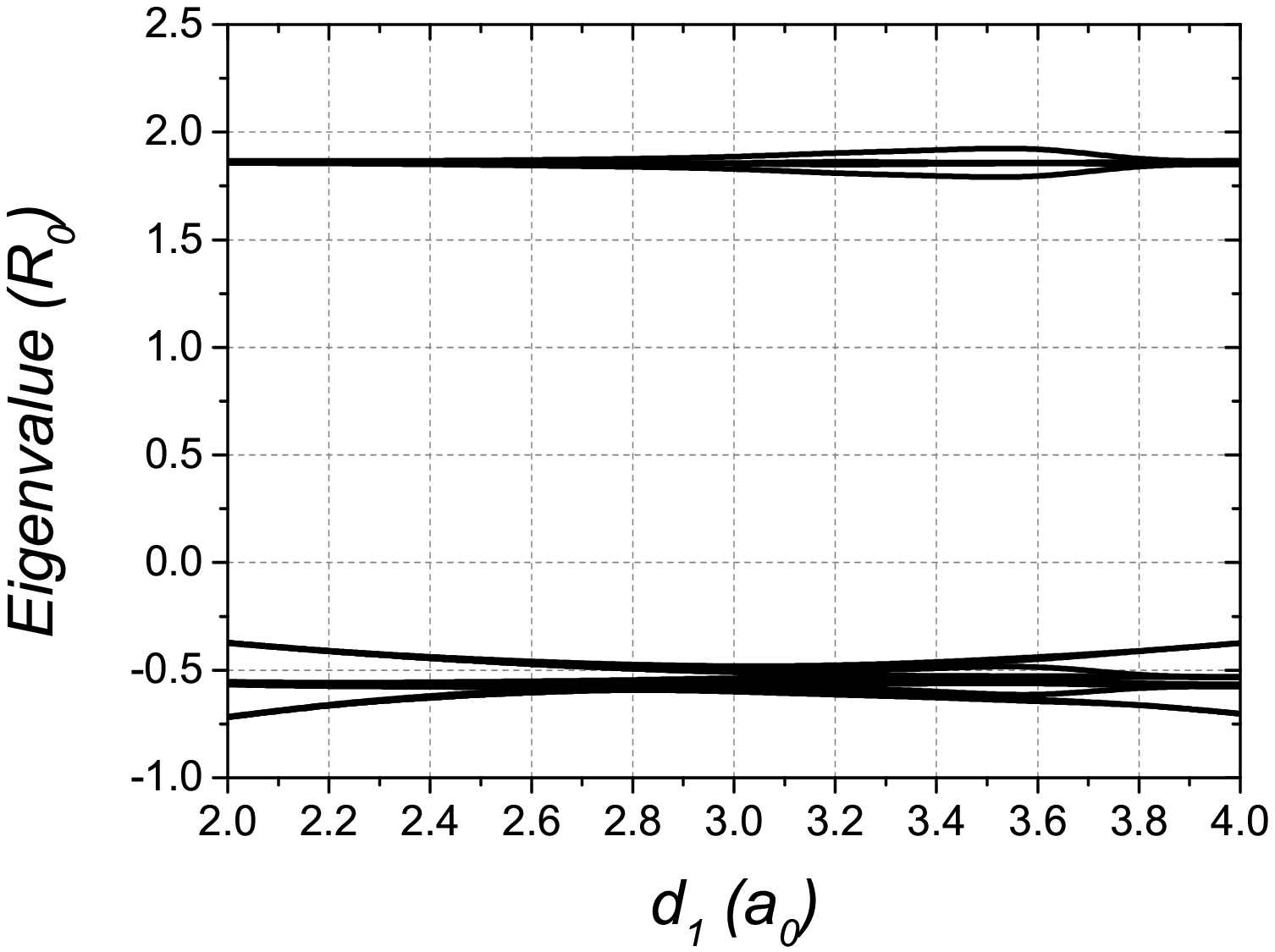}\\(a) [001]\\
\includegraphics[scale=0.25]{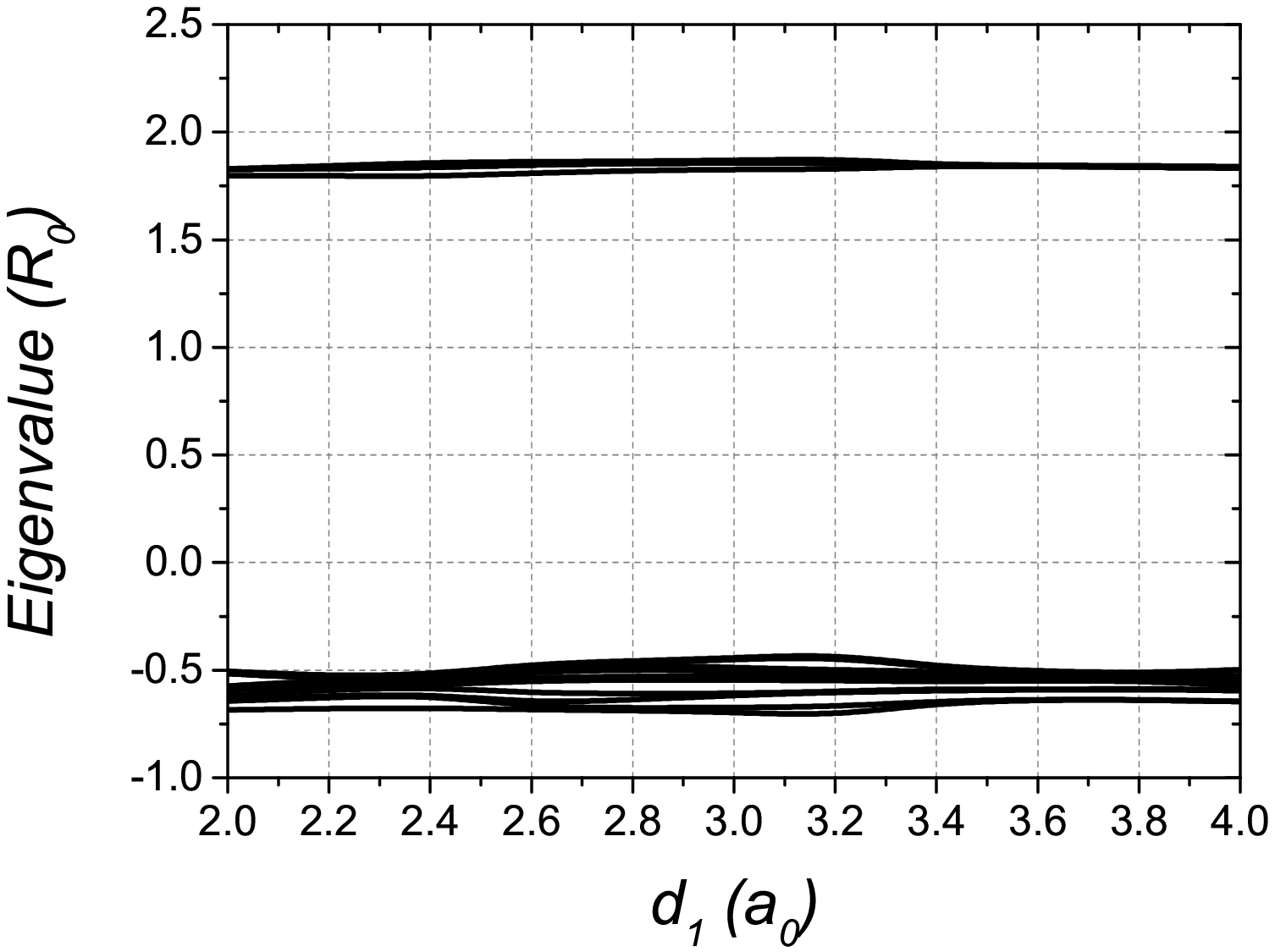}\\(b) [110]\\
\includegraphics[scale=0.25]{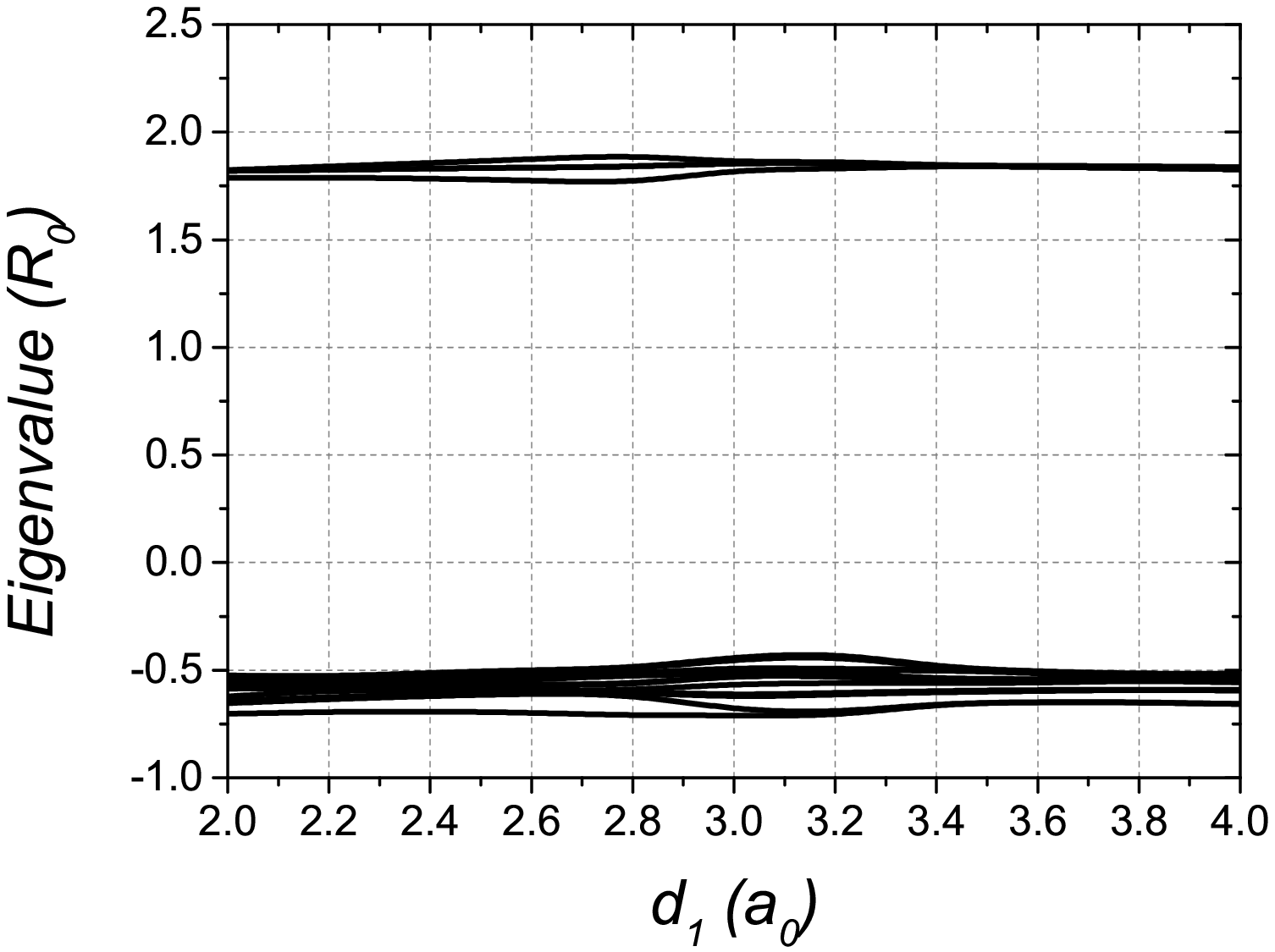}\\(c) [111]
\caption{The Fock matrix eigenvalues obtained from the UHF method in three high-symmetry directions for the large-separation case ($d_1+d_2=6a_0$) of the 4-acceptor linear chain: (a) the [001] direction, (b) the [110] direction, (c) the [111] direction.}\label{f-9}
\end{figure}
\begin{figure}
\centering
\includegraphics[scale=0.25]{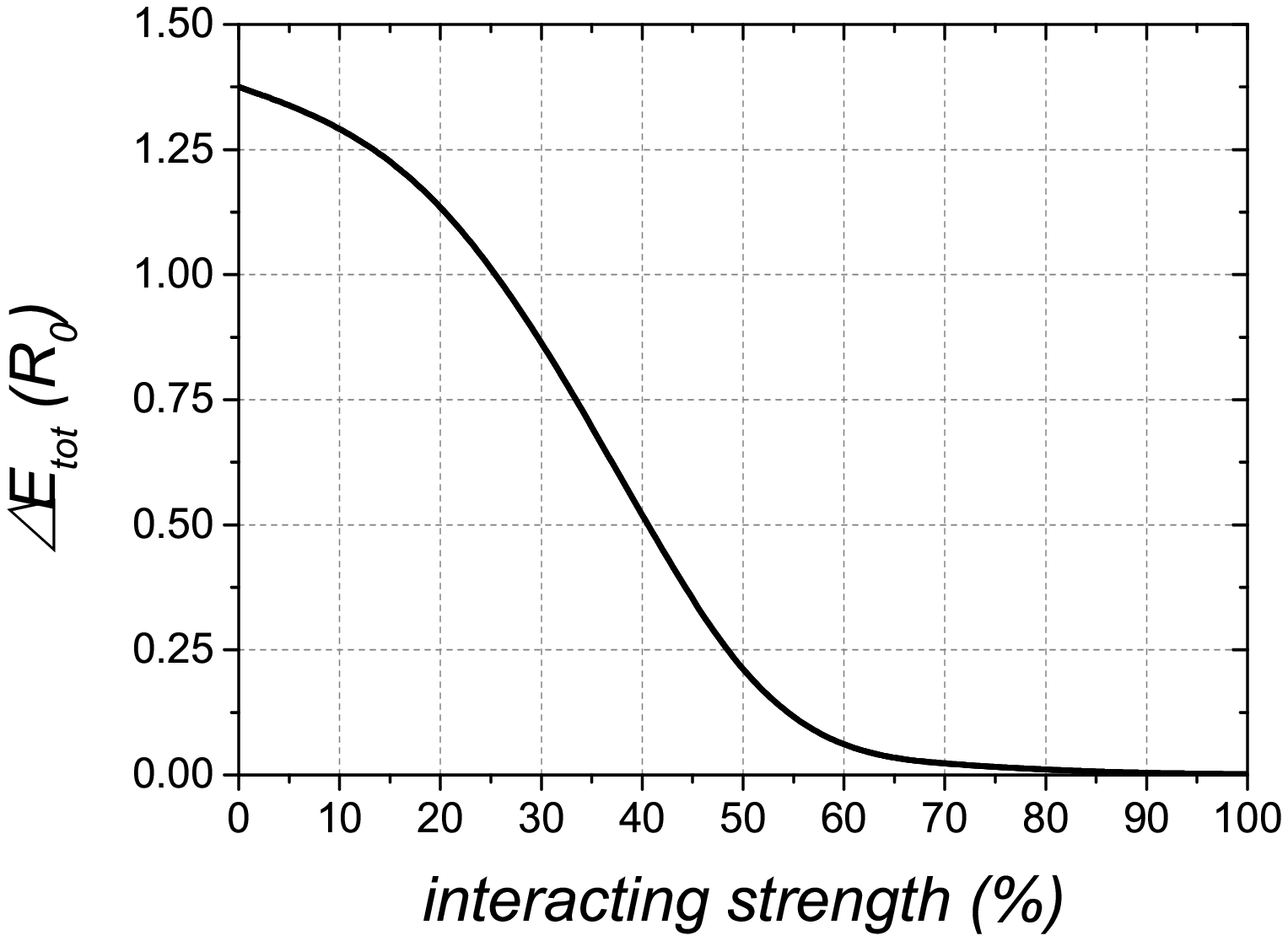}\\(a) $d_1=2a_0, d_2=1a_0$\\
\includegraphics[scale=0.25]{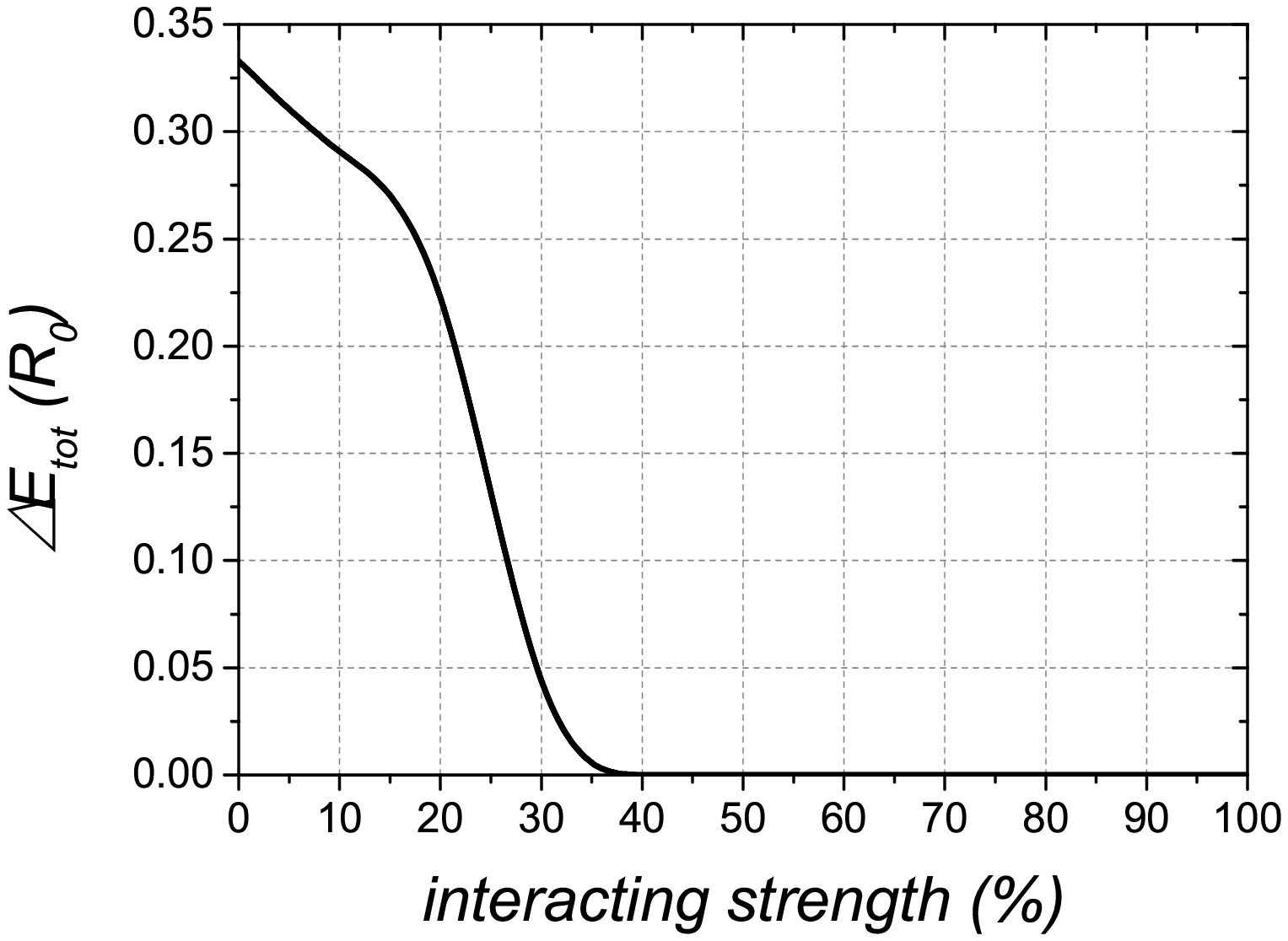}\\(b) $d_1=4a_0, d_2=2a_0$
\caption{The energy difference between the ground state and 3 closest excited states as a function of Coulomb interaction strength (expressed as a percentage) in the long-short limit of the 4-acceptor linear chain in the [001] direction: (a) the small-separation case ($d_1=2a_0, d_2=1a_0$), (b) the large-separation case ($d_1=4a_0, d_2=2a_0$).}\label{f-10}
\end{figure}

For the large-separation case, we show the behaviour of the ground-state total energy obtained from all the methods mentioned in \S\ref{model} for a 4-acceptor chain along high-symmetry directions in Figure \ref{f-7}. The HL and UHF results are closer to the full CI results than in the small separation case, consistent with the better agreement found between the methods for larger separations in the case of pairs (Figure \ref{f-1}).  The best agreement is around the uniform chain ($d_1=d_2=3a_0$); once again, the HL approach offers a better approximation than UHF.

The highest 50 energy states from the full CI result are shown in Figure \ref{f-8} and and the Fock matrix eigenvalues in Figure \ref{f-9}. In all three directions the ground-state is non-degenerate on short-long side (small $d_1$), although this is not clearly visible from Figure \ref{f-8}(a) for the [001] direction;  as found for smaller spacings in \S\ref{4asmallresult}, the ground state joins three other states in each case and forms a 4-fold-degenerate manifold on the right-hand side (large $d_1$). This time there is no change in the character of the ground state and no (anti-)crossing visible among the states in Figure \ref{f-8} or Figure \ref{f-9}; instead, the Fock eigenvalues show a group of four occupied states strongly separated from the unoccupied ones by the self-consistent potential.  There are some small splittings visible among the eigenvalues in Figure \ref{f-9} at particular geometries; these are due to the loss of symmetry in the UHF solution, as discussed in \S\ref{4asmallresult}. 

To compare the 4-fold-degenerate many-hole ground states obtained in the long-short limit for the small- and large-separation cases, and to understand how they relate to our previous results for non-interacting holes\cite{Zhu2020LCoAOMfDPAAiS}, we show in Figure \ref{f-10} the energy difference between the ground state and 3 closest excited states as a function of Coulomb interaction strength for a 4-acceptor linear chain in the [001] direction (interpolating between the non-interacting and fully-interacting cases). We choose the 4-acceptor system because it provides a more straightforward comparison to the one-hole edge states of the noninteracting system, as there will be fewer other states complicating the picture \cite{Zhu2020LCoAOMfDPAAiS}. In both cases, there is a gap in the non-interacting limit, because one-hole edge states move apart in the long-short limit to join two different bulk bands as shown in our previous paper\cite{Zhu2020LCoAOMfDPAAiS}; the 4-fold-degenerate ground state forms once the interaction strength exceeds a critical value, which is smaller in the large-separation case than in the small separation-case. This can be understood because the energy scale set by the non-interacting part of the Hamiltonian is weaker in the large-separation case, so a smaller hole-hole interaction is sufficient to overcome the parabolic confining potential.

\subsubsection{Symmetry breaking in the UHF calculation}\label{symmetry-breaking}
\begin{table}
\centering
\caption{The magnetic symmetry groups of the UHF ground states in different arrangements for the three high-symmetry directions in Hermann-Mauguin notation. Here the prime denotes operations that are only symmetries when accompanied by time reversal; the symbols $m$ and $m'$ are abbreviations for $\frac{1}{m}$ and $\frac{1}{m'}$ respectively.}\label{t-3}
\begin{tabular}{|c|c|c|c|}
\hline
$d_1$&[001]&[110]&[111]\\
\hline
Hamiltonian&$\frac{4}{m}\frac{2}{m}\frac{2}{m}1'$&$\frac{2}{m}\frac{2}{m}\frac{2}{m}1'$&$\bar{3}\frac{2}{m}1'$\\
\hline
\multicolumn{4}{|c|}{Small-separation case ($d_1+d_2=3a_0$):}\\
\hline
1.0$a_0$&$\frac{4}{m}\frac{2}{m}\frac{2}{m}1'$&$\frac{2}{m}\frac{2}{m}\frac{2}{m}1'$&$\bar{3}\frac{2}{m}1'$\\
\hline
1.1$a_0$&$\frac{4}{m'}\frac{2}{m'}\frac{2}{m'}$&$\frac{2}{m}\frac{2}{m}\frac{2}{m}1'$&$\bar{3}\frac{2}{m}1'$\\
\hline
1.2$a_0$&$\frac{4}{m'}\frac{2}{m'}\frac{2}{m'}$&$\frac{2'}{m}\frac{2'}{m}\frac{2}{m'}$&$\bar{3}\frac{2}{m}1'$\\
\hline
1.3$a_0$&$\frac{4}{m'}\frac{2}{m'}\frac{2}{m'}$&$\frac{2'}{m}\frac{2'}{m}\frac{2}{m'}$&$\frac{2}{m'}$\\
\hline
1.4$a_0$-1.5$a_0$&$4m'm'$&$\frac{2'}{m}\frac{2'}{m}\frac{2}{m'}$&$\frac{2}{m'}$\\
\hline
1.6$a_0$-1.7$a_0$&$\frac{4}{m'}\frac{2}{m'}\frac{2}{m'}$&$\frac{2'}{m}\frac{2'}{m}\frac{2}{m'}$&$\frac{2}{m'}$\\
\hline
1.8$a_0$&$4m'm'$&$\frac{2}{m}\frac{2'}{m'}\frac{2'}{m'}$&$\frac{2}{m'}$\\
\hline
1.9$a_0$-2.0$a_0$&$\frac{4}{m}\frac{2'}{m'}\frac{2'}{m'}$&$\frac{2}{m}\frac{2'}{m'}\frac{2'}{m'}$&$\frac{2'}{m'}$\\
\hline
\multicolumn{4}{|c|}{Large-separation case ($d_1+d_2=6a_0$):}\\
\hline
2.0$a_0$-3.2$a_0$&$\frac{4}{m}\frac{2'}{m'}\frac{2'}{m'}$&$\frac{2}{m}\frac{2'}{m'}\frac{2'}{m'}$&$\frac{2'}{m'}$\\
\hline
3.4$a_0$-3.6$a_0$&$\frac{4}{m}\frac{2'}{m'}\frac{2'}{m'}$&$2'mm'$&$m'$\\
\hline
3.8$a_0$-4.0$a_0$&$4m'm'$&$2'mm'$&$m'$\\
\hline\end{tabular}
\end{table}

To investigate the symmetry breaking, we determined the symmetry of the one-hole reduced density matrices, both in the full CI case and after the convergence of the UHF calculation; the results are shown using the Hermann-Mauguin notation for magnetic point groups in the upper part of Table \ref{t-3} for the small-separation case, and in the lower part for the large-separation case.  We observe that for small separations, the UHF solution always begins (for small $d_1$) with the same symmetry as the CI calculation (and the core Hamiltonian).  This is a `grey' magnetic group that contains the time-reversal operation $1'$, meaning that no magnetic moment has developed.  The group then loses some symmetry elements as $d_1$ increases, as magnetic moments develop; it would be more accurate to describe these missing symmetry operations as `hidden' rather than `lost', because they map different members of a manifold of degenerate self-consistent solutions to the UHF equations, each individually having lower symmetry, onto one another. At the points in the [001] direction where the symmetry is lowest ($d_1=1.4a_0, 1.5a_0$ and 1.8$a_0$), the convergence of the SCF procedure is poorer than for other separations; Figure \ref{f-6} (a) and (b) shows that these correspond to the location of the crossing points between different many-body ground-state compositions.  Bearing in mind that the crossing points for 6 acceptors (Figure \ref{f-11} (a) and (b)) are in slightly different positions than for 4 acceptors, and the slight change of the crossing point location shown in Figure \ref{f-12}, it is reasonable to believe that $d_1=1.8a_0$ is the true location of a second crossing point which is not shown in Figure \ref{f-6} (a) and (b) due to the limited number of acceptors. This suggests that the further reduction in symmetry near these points may also be related to the crossings (or potential crossings) between the occupied eigenvalues of the Fock matrix. We also show the behaviour of the total magnetic angular momentum for each acceptor in the different chain orientations in Figure \ref{f-6.1}; the breaking of symmetry is reflected by splitting into two or (at the lowest-symmetry arrangements the [001] direction) four different inequivalent sets. The magnetization pattern shows that non-zero magnetization becomes increasingly concentrated at the ends of the chain as $d_1$ increases, which is also true in the large-separation case. The 6-acceptor system behaves similarly to the 4-acceptor system, so we do not show the results here.

In the large-separation case, the symmetry is broken with respect to the underlying Hamiltonian at \textit{all} separations.  As previously, the broken symmetries are not really lost, but now map different solutions within the manifold of degenerate states  (all having non-zero magnetic moments) into one another.   For values of $d_1$ greater than some critical value, (which depends on the direction), the symmetry is further reduced; comparing with Figure \ref{f-8}, we see this further reduction occurs when the 4-fold-degenerate ground states in the full CI calculation show very small energy differences between each other so they are hard to distinguish in the UHF calculation.  By checking each data point, the switching from ferromagnetically aligned case to antiferromagnetically aligned case is found across the central ($d_2$) bond as it shortens.

\subsection{Linear chain with periodic boundary conditions}\label{infiniteresult}

\begin{figure}
\centering
\includegraphics[scale=0.3]{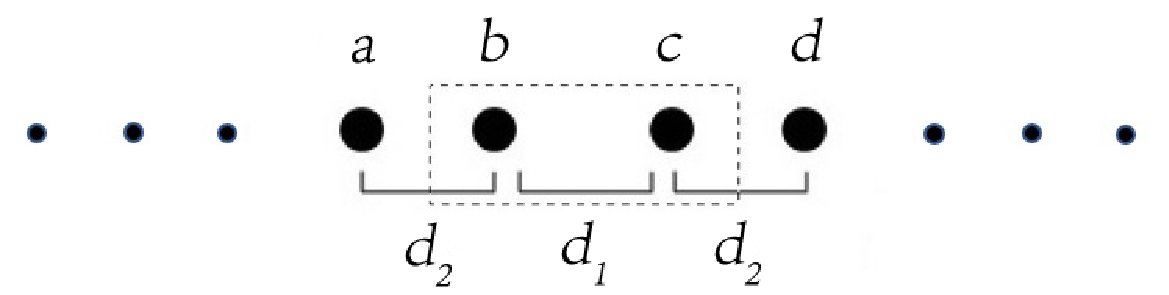}
\caption{The schematic of the linear chain with periodic boundary conditions. a, b, c, d are the labels of acceptors, $d_1<d_2$ is known as the `short-long arrangement', $d_1>d_2$ is known as the `long-short arrangement'.}\label{f-16}
\end{figure}

We now turn to periodic boundary conditions.  A schematic of the system is shown in Figure \ref{f-16}; $(a, b, c, d)$ label four adjacent acceptors, with $b, c$ in the same unit cell, and $d_1$, $d_2$ are the separations. (We have swapped the separation labels relative to the convention used in in our previous paper\cite{Zhu2020LCoAOMfDPAAiS}.)  Approaches based on full diagonalization (full CI calculation and the HL approach) are not extensive and hence not useful with periodic boundary conditions as discussed in \S\ref{ci}, but the UHF method is still suitable.  Since the behaviour of finite chains is found to be quite similar in the small- and large-separation cases, we report results for infinite chains only for smaller separations ($d_1+d_2=3a_0$).


\begin{figure*}
\centering
\includegraphics[scale=0.25]{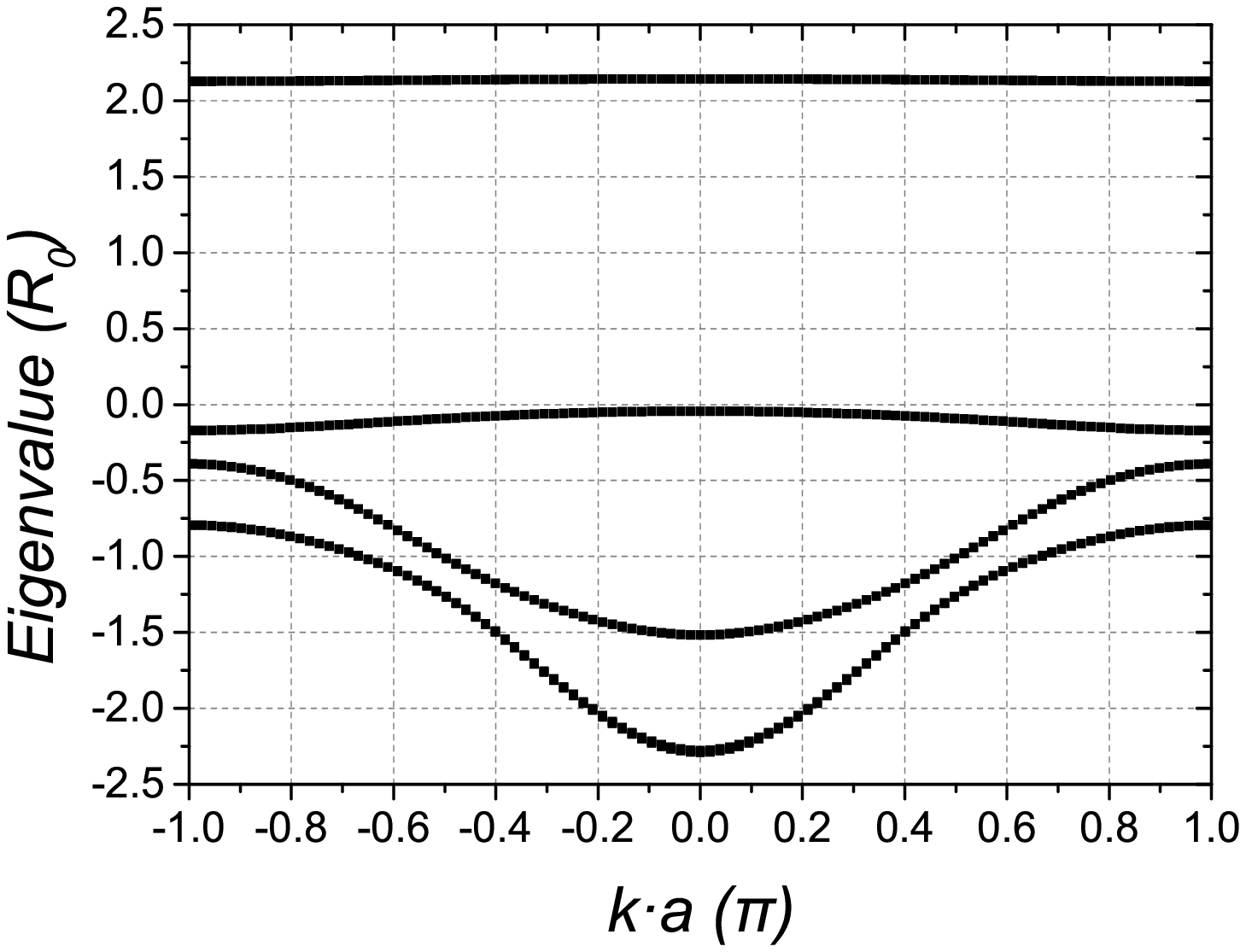}
\includegraphics[scale=0.25]{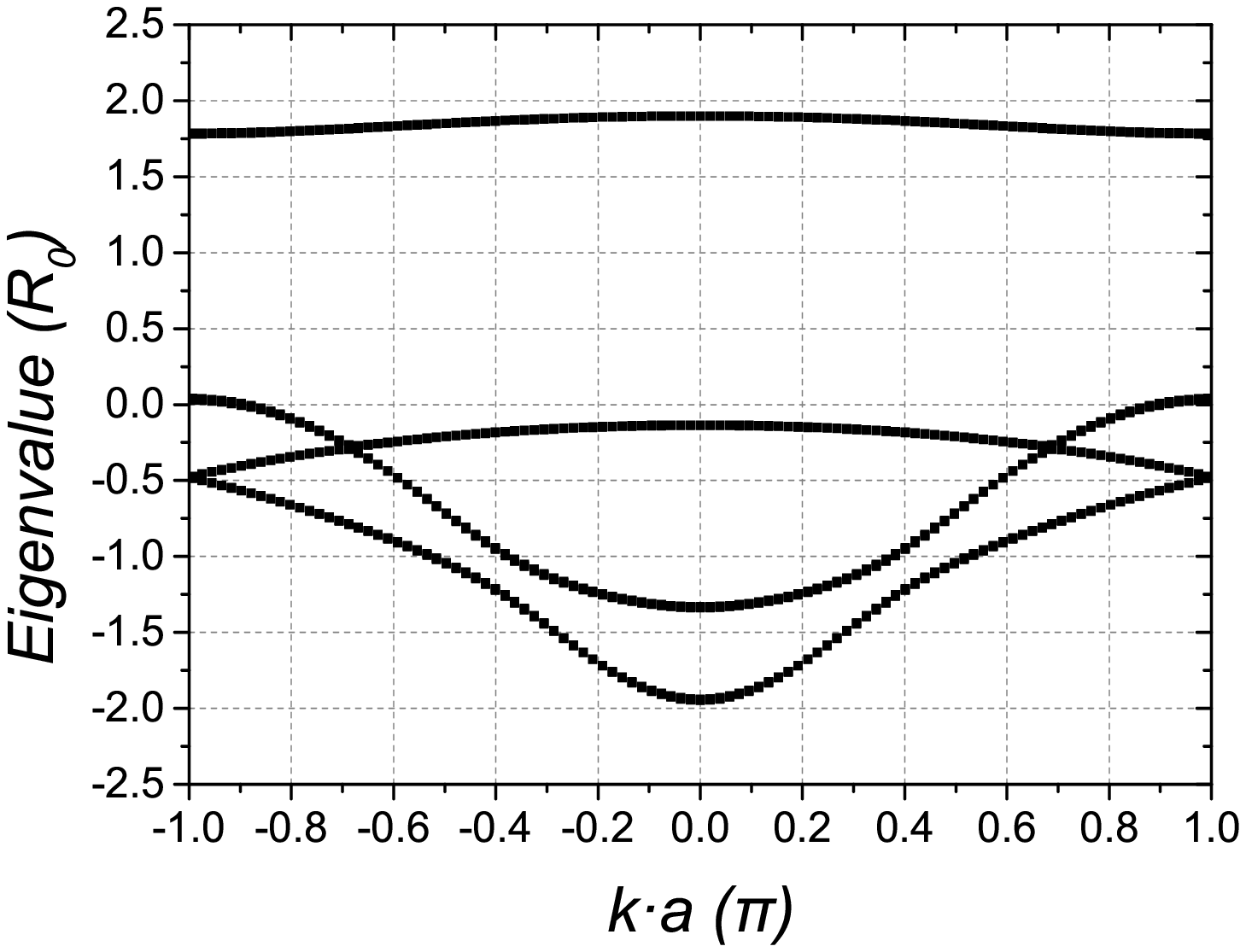}\\(a) [001] \qquad\qquad\qquad\qquad\qquad\qquad\qquad\qquad\qquad\quad (b) [001]\\
\includegraphics[scale=0.25]{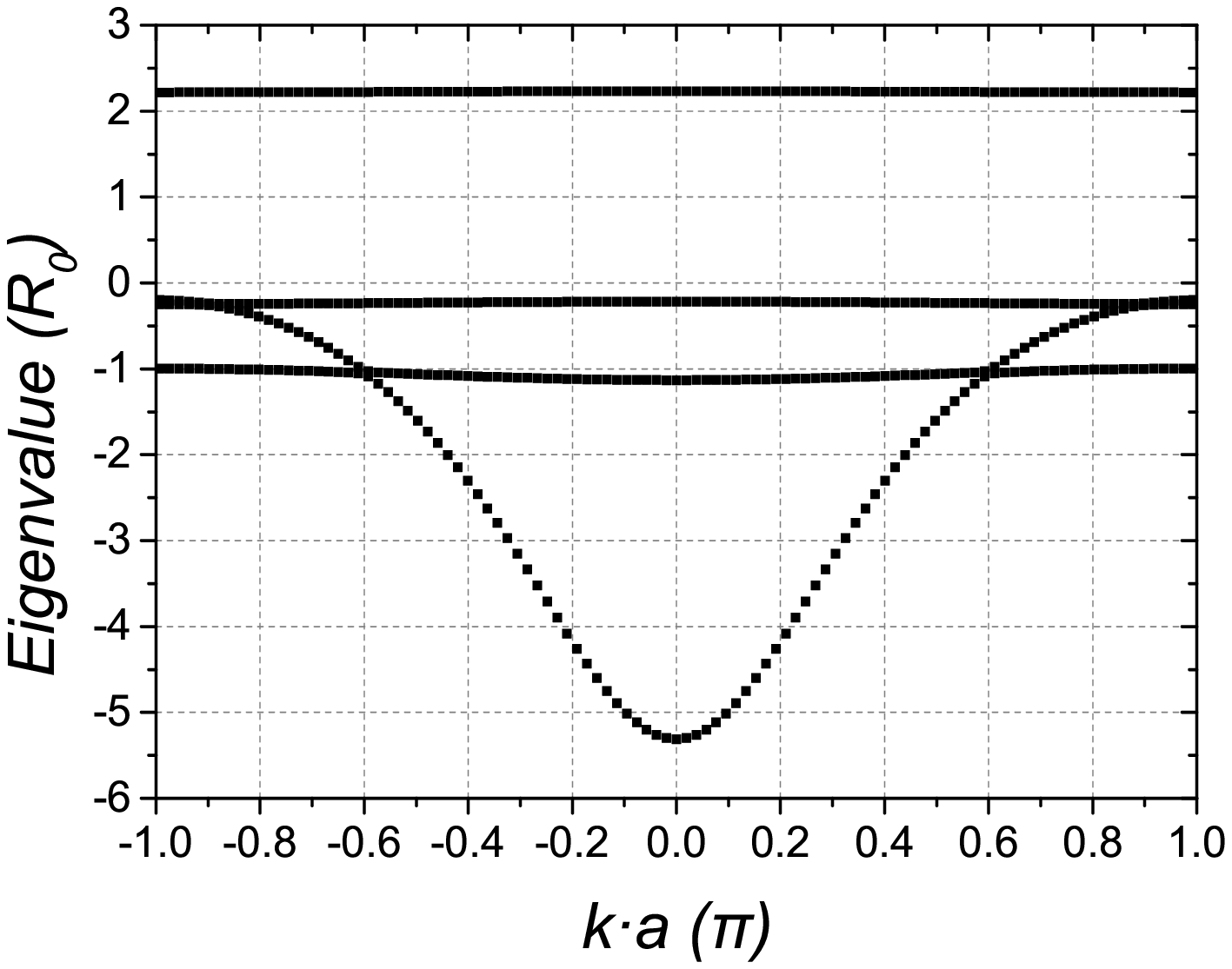}
\includegraphics[scale=0.25]{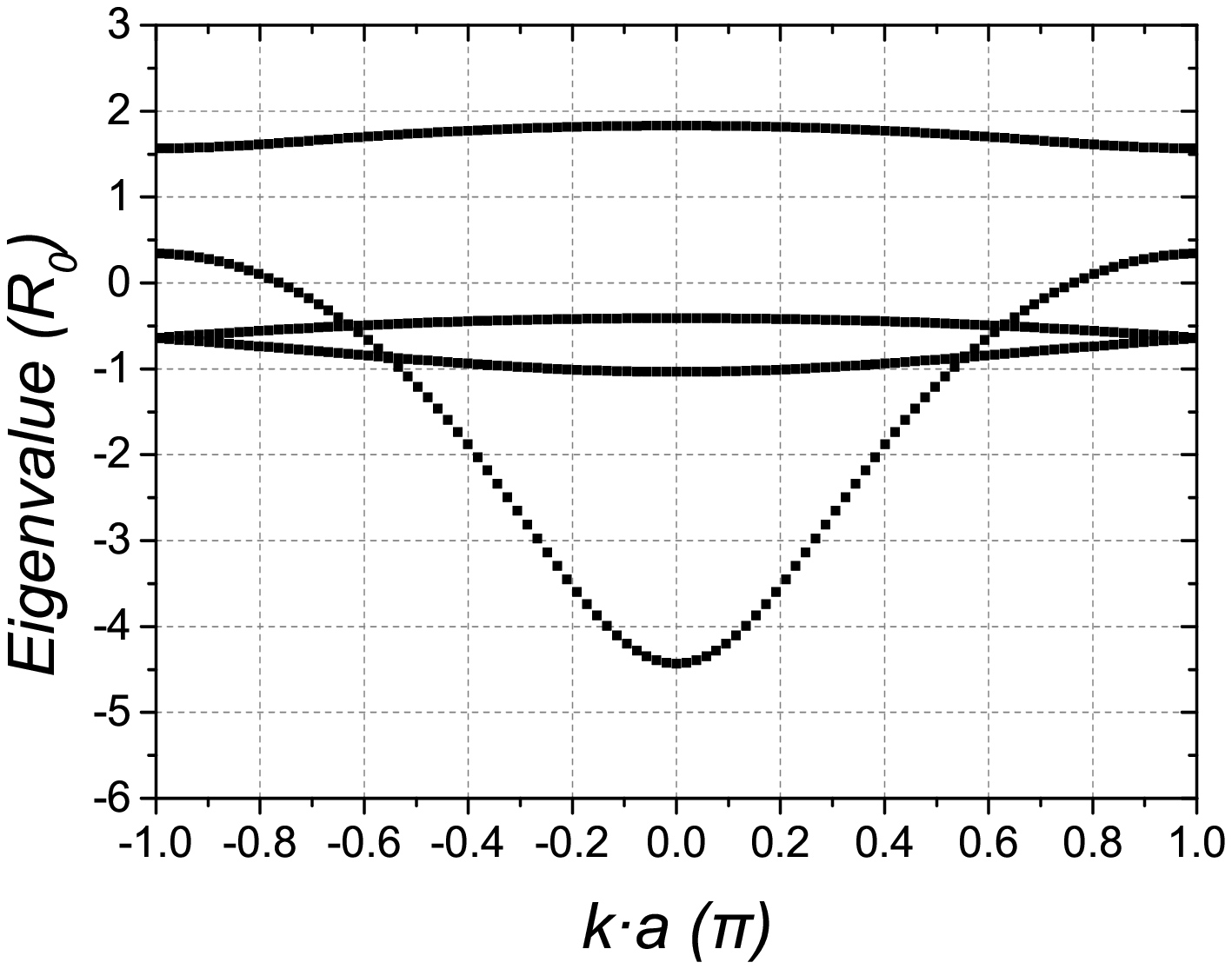}\\(c) [110] \qquad\qquad\qquad\qquad\qquad\qquad\qquad\qquad\qquad\quad (d) [110]\\
\includegraphics[scale=0.25]{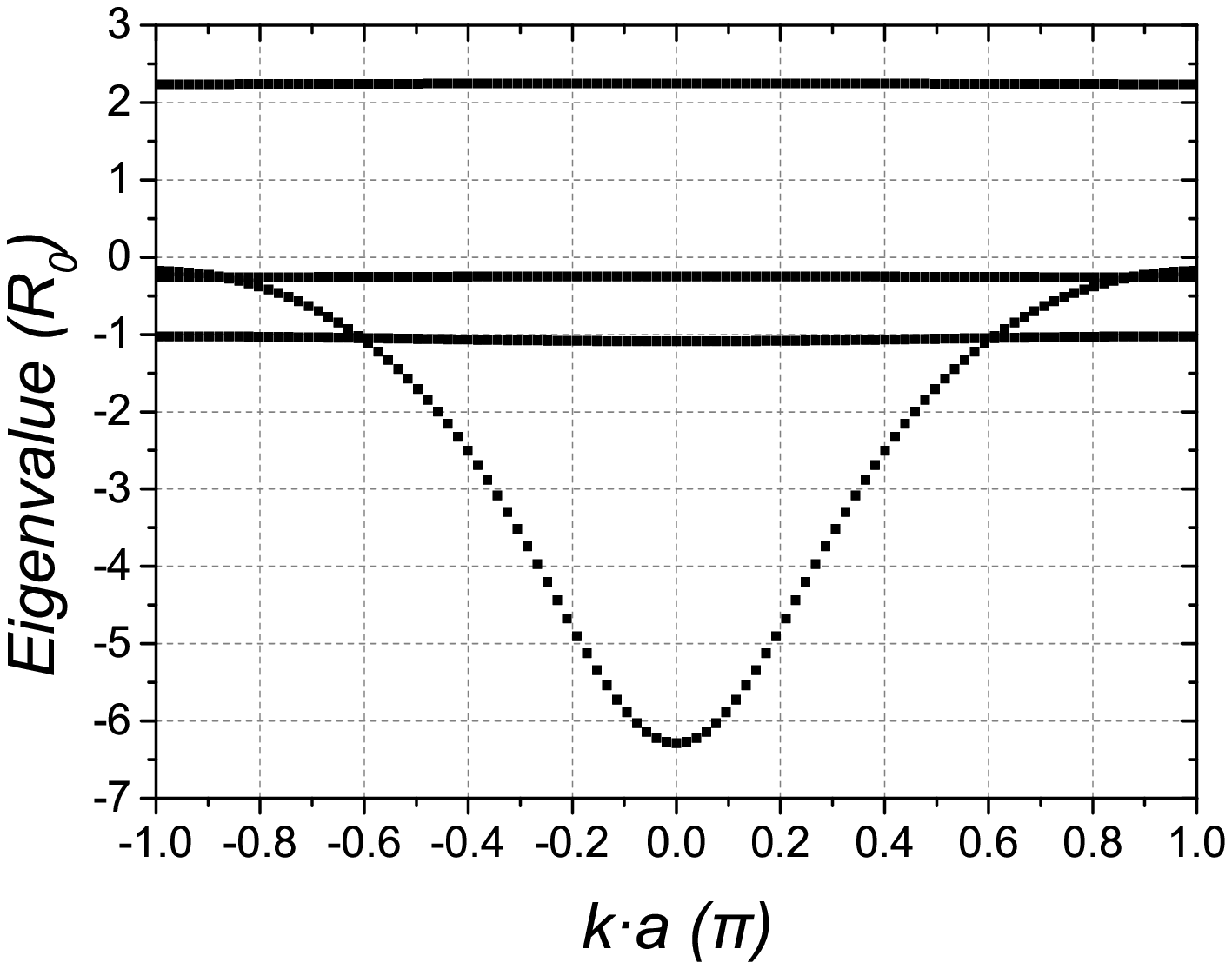}
\includegraphics[scale=0.25]{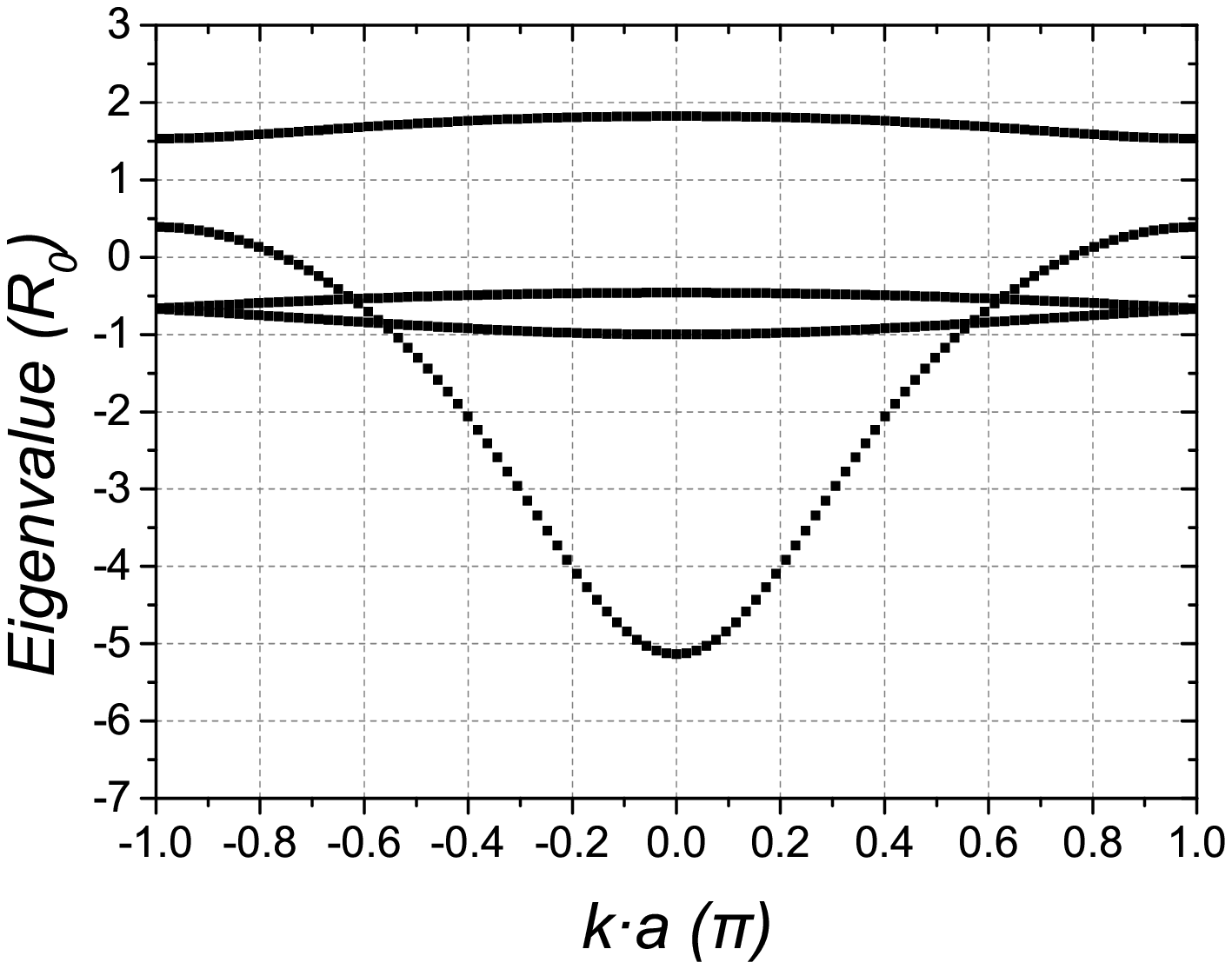}\\(e) [111] \qquad\qquad\qquad\qquad\qquad\qquad\qquad\qquad\qquad\quad (f) [111]
\caption{The band structures of the Fock matrix eigenvalues under different arrangements in three typical directions when $d_1+d_2=3a_0$: (a) the short-long arrangement in the [001] direction, (b) the uniform chain case in the [001] direction, (c) the short-long arrangement in the [110] direction, (d) the uniform chain case in the [110] direction, (e) the short-long arrangement in the [111] direction, (f) the uniform chain case in the [111] direction.}\label{f-18}
\end{figure*}

Figure \ref{f-18} shows the band structures of the Fock matrix eigenvalues. We only show the results for the 'short-long' arrangement ($d_1<d_2$) in each high-symmetry direction, along with the uniform chains ($d_1=d_2$), as the short-long arrangements are equivalent to long-short under periodic boundary conditions.  All the single-hole states are doubly-degenerate, so the two states at the top of each picture will be filled (two holes per cell).  There is a large gap between the filled and empty bands in the short-long dimerised arrangement; for uniform chains, the bands move closer but this gap does not close, showing the existence of a cell-doubling perturbation from the self-consistent field.  This is related to the broken symmetries found in the corresponding finite chain calculations: as shown in Table \ref{t-3}, we found the inversion symmetry is broken (becomes hidden) for some uniform-chain cases. It is reasonable that this also occurs under periodic boundary conditions, leading to an inequivalence of the two atoms in the cell even for a uniform chain and implying that the band structure of the two-atom cell cannot be obtained by simply folding the bands for the one-atom cell.

\subsection{Structure of the edge states}\label{edge-states}
In order to understand the nature of the edge states, we examine the many-hole states from the full CI calculation and compare them to the UHF single-particle states, for both small-separation and large-separation cases in the 4-acceptor finite chain. Both methods show edge states localized at the acceptors at the end of the chain in the long-short arrangement ($d_1>d_2$); however, the signatures are different. The CI method shows a manifold of almost degenerate states spanned by a basis of the form
\begin{equation}
\ket{\psi_{m,n}}=\ket{\psi^A_m}\otimes\ket{\psi^{\mbox{bulk}}}\otimes\ket{\psi^B_n}\label{e-a1},
\end{equation}
where $A$ labels the left end of the chain (acceptor a in Figure \ref{f-3} (a)), $B$ labels the right end (acceptor d in Figure \ref{f-3} (a)), and $\ket{\psi^{\mbox{bulk}}}$ is a common state residing in the interior of the chain (acceptors b and c in Figure \ref{f-3} (a)).  The indices $m$ and $n$ label different states of the ends, and the pair $(m,n)$ together label a member of the almost degenerate manifold.  The transformation from state $\ket{\Psi_{m,n}}$ to $\ket{\Psi_{m',n'}}$ can therefore be carried out by a unitary operator
\begin{align}
\hat{U}=\hat{U}^A&\otimes\hat{1}^{\mbox{bulk}}\otimes\hat{U}^B\quad\notag\\
&\mbox{with}\notag\\
\quad\hat{U}^A=&\ket{\psi^A_{m'}}\bra{\psi^A_m};\notag\\
\quad\hat{U}^B=&\ket{\psi^B_{n'}}\bra{\psi^B_m}.\label{e-a2}
\end{align}
For finite chains, the eigenstates are particular linear combinations of the $\ket{\psi_{m,n}}$ which are almost (but not quite) degenerate; the splittings decay to zero as $d_1$ is increased, or as the chain becomes longer (see Figure \ref{f-12}).  It is therefore important to look at the whole space spanned by the $\ket{\psi_{m,n}}$, especially when the splittings become very small.  The UHF method instead picks out a single symmetry-broken many-hole ground state in which one pair of occupied single-particle states is localized at the chain ends (acceptors a and d in Figure \ref{f-3} (a)) while the other pair is spread over the interior (acceptors b and c in Figure \ref{f-3} (a)).  The single-hole edge states can be written as linear combinations of particular one-hole kets $\ket{\phi^A}$ and $\ket{\phi^B}$ localized at either end.  


We can also examine the symmetries of the edge states $\ket{\psi^A_m}, \ket{\psi^B_n}$ in the light of the classification of the topological phases of one-dimensional interacting fermions proposed by Reference~\onlinecite{Turner2011TPoODFaEPoV}; in the long-short limit we find the characteristic phases are $(\mu=0,\phi=0,\kappa=\pi)$,  hence the state is topologically non-trivial with 4-fold degeneracy, while in the short-long arrangement they are $(\mu=0,\phi=0,\kappa=0)$ (topologically trivial, non-degenerate).   However, we find some differences between the small- and large-separation cases. For the 4-acceptor chain, when $d_1+d_2=3a_0$, $\ket{\psi^A_m}, \ket{\psi^B_n}$ involve only $m_F=\pm1/2$ states in the [001] direction, while $\ket{\psi^{\mbox{bulk}}}$ includes only $m_F=\pm3/2$ states. This is because in the long-short arrangement case, the system can be considered as two single acceptors at the chain ends and a closely-coupled pair of acceptors between them. In that case, $\ket{\psi^{\mbox{bulk}}}$ is dominated by the central pair, while $\ket{\psi^A_m}, \ket{\psi^B_n}$ are dominated by the single-acceptor ends. Since the doubly-degenerate occupied bands at the top of Figure \ref{f-18} (a) and (b) in the [001] direction are always formed predominantly from linear combinations of the $\pm\frac{3}{2}$ states on the two acceptors in the cell, and a single acceptor perturbed by another acceptor always has a ground state of $m_F=\pm1/2$ symmetry, it is reasonable that $\ket{\psi^{\mbox{bulk}}}$ and $\ket{\psi^A_m}, \ket{\psi^B_n}$ only involve $m_F=\pm3/2$ and $\pm1/2$ states respecively. When $d_1+d_2=6a_0$, although $\ket{\psi^{\mbox{bulk}}}$ is similar, the $\{\ket{\psi^A_m}, \ket{\psi^B_n}\}$ involve also the superpositions $\pm\{\ket{3/2}, \ket{1/2}\}$ in two of the four degenerate states. In the large-separation case the degeneracy is more nearly exact, so the properties of individual eigenstates are not clearly defined and we should consider the space spanned by all four degenerate states together. For the 6-acceptor system (which we treat in the HL approximation), we always find the edge states composed purely of $m_F=\pm1/2$ states at the end of the chain (as for 4 acceptors).


We can also calculate the Zak phase for the occupied UHF states in the infinite system by using (\ref{e-2-3-1}) in \S\ref{zak}. We find that Zak phase is 0 for all arrangements in all directions, even although we find the edge states in the finite chains have non-trivial symmetries; this is consistent with the preservation of a gap in the single-particle UHF energy spectrum for all arrangements. The Zak phase is calculated by using the single-hole UHF eigenvectors, and its correspondence with the topological properties of an interacting system is still unclear; it is not surprising that it fails to describe the topological properties of the interacting system in the same way, as was previously noted for the bosonic case \cite{PhysRevLett.110.260405}. In the absence of a rigorously defined topological quantum number for an infinite system with interactions, the direct study of the quantum numbers characterising the edge states of the finite system, introduced by Turner \textit{et al.}\cite{Turner2011TPoODFaEPoV},  provides a better insight into their topological nature.

\section{Conclusion}\label{conclusion}

In this paper, we constructed multi-hole models for neutral, one-dimensional multi-acceptor chains based on three different methods: full configuration interaction, the Heitler-London approximation , and the unrestricted Hartree-Fock method.  The HL approximation solves some of the problems with the CI method, but only the UHF method is able to cope with infinite chains under periodic boundary conditions.

From reference calculations on a pair of acceptors, we found that both the HL approach and the UHF method give good approximations to the ground state of the full CI calculation, with the HL approach offering a better result in the regimes studied (which are on the insulating side of the Mott transition). The UHF method is less useful for the calculation of excited states, so we use the HL approximation to simplify the calculation of low-lying excitations when interactions are strong. The converged UHF state has a large gap between the filled and empty states, due to the self-consistent potential generated by the hole-hole interactions.

For finite chains, the CI ground state is non-degenerate in the short-long arrangement in all directions, but joins three other states to form a 4-fold-degenerate manifold in the long-short arrangement, which is followed in energy by an 8-fold-degenerate state and another 4-fold-degenerate state. By checking the dominant components of these 16 states, we found that only the levels on the acceptors at the end of the chain change between different members of the manifold; the overall 16-fold degeneracy comes from the product of separate sets of 4 levels on each end acceptor. The topological nature of these edge states is confirmed by the presence of non-trivial phases in the classification of one-dimensional fermion edge states by Turner et al. In the small-separation case where $d_1+d_2=3a_0$, an anti-crossing occurs between the ground state and the next excited states in the [001] direction, resulting in a switch from in unhybridized ground state dominated by $m_F=\pm3/2$ states to a hybridized state where $m_F=\pm1/2$ states are also present; this transition is related to the crossing between the filled UHF single-particle states. The UHF solution loses part of the symmetry of the underlying Hamiltonian; for particular arrangements, we found the further broken symmetries related to the crossing (or potential crossing) of Fock matrix eigenstates in the [001] direction. The loss of symmetry corresponds to the emergence of static moments on each acceptor in the UHF approach.


We obtained the UHF band structures of the Fock matrix eigenvalues. We found there is a large gap between the filled and empty states in a dimerised chain, which does not fully close in the uniform chain, showing the existence of a period-doubling perturbation. Since a gap is maintained throughout the transition from short-long to long-short arrangements, the Zak phase is constant (and equal to zero), despite the observation of non-trivial many-body edge states in the long-short case. Hence, this method does not capture the formation of edge states, while the previous method introduced by Turner et al can well characterise their topological properties. The nature of the bulk-edge correspondence in such interacting systems requires further investigation.

\section{Acknowledgement}

We wish to acknowledge the support of the Engineering and Physical Sciences Research Council and UK Research and Innovation under the ADDRFSS programme (grant EP/M009564/1). We thank Nguyen Le, Ben Murdin, Neil Curson, and Gabriel Aeppli for helpful and inspiring discussions.





\bibliography{ref}

\begin{thebibliography}{30}
\expandafter\ifx\csname natexlab\endcsname\relax\def\natexlab#1{#1}\fi
\expandafter\ifx\csname bibnamefont\endcsname\relax
  \def\bibnamefont#1{#1}\fi
\expandafter\ifx\csname bibfnamefont\endcsname\relax
  \def\bibfnamefont#1{#1}\fi
\expandafter\ifx\csname citenamefont\endcsname\relax
  \def\citenamefont#1{#1}\fi
\expandafter\ifx\csname url\endcsname\relax
  \def\url#1{\texttt{#1}}\fi
\expandafter\ifx\csname urlprefix\endcsname\relax\def\urlprefix{URL }\fi
\providecommand{\bibinfo}[2]{#2}
\providecommand{\eprint}[2][]{\url{#2}}

\bibitem[{\citenamefont{Koiller et~al.}(2001)\citenamefont{Koiller, Hu, and
  Das~Sarma}}]{Koiller2002EiSBQCA}
\bibinfo{author}{\bibfnamefont{B.}~\bibnamefont{Koiller}},
  \bibinfo{author}{\bibfnamefont{X.}~\bibnamefont{Hu}}, \bibnamefont{and}
  \bibinfo{author}{\bibfnamefont{S.}~\bibnamefont{Das~Sarma}},
  \bibinfo{journal}{Physical review letters} \textbf{\bibinfo{volume}{88}},
  \bibinfo{pages}{027903} (\bibinfo{year}{2001}).

\bibitem[{\citenamefont{Pantelides and
  Sah}(1974)}]{Pantelides1974ToLSiSINRUaOM}
\bibinfo{author}{\bibfnamefont{S.~T.} \bibnamefont{Pantelides}}
  \bibnamefont{and} \bibinfo{author}{\bibfnamefont{C.~T.} \bibnamefont{Sah}},
  \bibinfo{journal}{Phys. Rev. B} \textbf{\bibinfo{volume}{10}},
  \bibinfo{pages}{621} (\bibinfo{year}{1974}).

\bibitem[{\citenamefont{Salfi et~al.}(2016{\natexlab{a}})\citenamefont{Salfi,
  Tong, Rogge, and Culcer}}]{Salfi_2016}
\bibinfo{author}{\bibfnamefont{J.}~\bibnamefont{Salfi}},
  \bibinfo{author}{\bibfnamefont{M.}~\bibnamefont{Tong}},
  \bibinfo{author}{\bibfnamefont{S.}~\bibnamefont{Rogge}}, \bibnamefont{and}
  \bibinfo{author}{\bibfnamefont{D.}~\bibnamefont{Culcer}},
  \bibinfo{journal}{Nanotechnology} \textbf{\bibinfo{volume}{27}},
  \bibinfo{pages}{244001} (\bibinfo{year}{2016}{\natexlab{a}}).

\bibitem[{\citenamefont{Hendrickx et~al.}(2021)\citenamefont{Hendrickx, Lawrie,
  Russ, van Riggelen, de~Snoo, Schouten, Sammak, Scappucci, and
  Veldhorst}}]{Hendrickx2021AFQGQP}
\bibinfo{author}{\bibfnamefont{N.~W.} \bibnamefont{Hendrickx}},
  \bibinfo{author}{\bibfnamefont{W.~I.~L.} \bibnamefont{Lawrie}},
  \bibinfo{author}{\bibfnamefont{M.}~\bibnamefont{Russ}},
  \bibinfo{author}{\bibfnamefont{F.}~\bibnamefont{van Riggelen}},
  \bibinfo{author}{\bibfnamefont{S.~L.} \bibnamefont{de~Snoo}},
  \bibinfo{author}{\bibfnamefont{R.~N.} \bibnamefont{Schouten}},
  \bibinfo{author}{\bibfnamefont{A.}~\bibnamefont{Sammak}},
  \bibinfo{author}{\bibfnamefont{G.}~\bibnamefont{Scappucci}},
  \bibnamefont{and}
  \bibinfo{author}{\bibfnamefont{M.}~\bibnamefont{Veldhorst}},
  \bibinfo{journal}{Nature} \textbf{\bibinfo{volume}{591}},
  \bibinfo{pages}{580} (\bibinfo{year}{2021}), ISSN \bibinfo{issn}{1476-4687}.

\bibitem[{\citenamefont{Wang et~al.}(2021)\citenamefont{Wang, Marcellina,
  Hamilton, Cullen, Rogge, Salfi, and Culcer}}]{Wang2021OOPfUHCGHSOQ}
\bibinfo{author}{\bibfnamefont{Z.}~\bibnamefont{Wang}},
  \bibinfo{author}{\bibfnamefont{E.}~\bibnamefont{Marcellina}},
  \bibinfo{author}{\bibfnamefont{A.~R.} \bibnamefont{Hamilton}},
  \bibinfo{author}{\bibfnamefont{J.~H.} \bibnamefont{Cullen}},
  \bibinfo{author}{\bibfnamefont{S.}~\bibnamefont{Rogge}},
  \bibinfo{author}{\bibfnamefont{J.}~\bibnamefont{Salfi}}, \bibnamefont{and}
  \bibinfo{author}{\bibfnamefont{D.}~\bibnamefont{Culcer}},
  \bibinfo{journal}{npj Quantum Information} \textbf{\bibinfo{volume}{7}},
  \bibinfo{pages}{54} (\bibinfo{year}{2021}), ISSN \bibinfo{issn}{2056-6387}.

\bibitem[{\citenamefont{Luttinger}(1956)}]{Luttinger1956QToCRiSGT}
\bibinfo{author}{\bibfnamefont{J.~M.} \bibnamefont{Luttinger}},
  \bibinfo{journal}{Phys. Rev.} \textbf{\bibinfo{volume}{102}},
  \bibinfo{pages}{1030} (\bibinfo{year}{1956}).

\bibitem[{\citenamefont{Baldereschi and
  Lipari}(1973)}]{Baldereschi1973SMoSASiS}
\bibinfo{author}{\bibfnamefont{A.}~\bibnamefont{Baldereschi}} \bibnamefont{and}
  \bibinfo{author}{\bibfnamefont{N.~O.} \bibnamefont{Lipari}},
  \bibinfo{journal}{Phys. Rev. B} \textbf{\bibinfo{volume}{8}},
  \bibinfo{pages}{2697} (\bibinfo{year}{1973}).

\bibitem[{\citenamefont{Baldereschi and
  Lipari}(1974)}]{Baldereschi1974CCttSMoSAS}
\bibinfo{author}{\bibfnamefont{A.}~\bibnamefont{Baldereschi}} \bibnamefont{and}
  \bibinfo{author}{\bibfnamefont{N.~O.} \bibnamefont{Lipari}},
  \bibinfo{journal}{Phys. Rev. B} \textbf{\bibinfo{volume}{9}},
  \bibinfo{pages}{1525} (\bibinfo{year}{1974}).

\bibitem[{\citenamefont{Lipari and Baldereschi}(1978)}]{Lipari1978IoASiS}
\bibinfo{author}{\bibfnamefont{N.}~\bibnamefont{Lipari}} \bibnamefont{and}
  \bibinfo{author}{\bibfnamefont{A.}~\bibnamefont{Baldereschi}},
  \bibinfo{journal}{Solid State Communications} \textbf{\bibinfo{volume}{25}},
  \bibinfo{pages}{665–668} (\bibinfo{year}{1978}).

\bibitem[{\citenamefont{C.~Durst et~al.}(2017)\citenamefont{C.~Durst,
  E.~Castoria, and N.~Bhatt}}]{C.Durst2017HLMfAAIiDS}
\bibinfo{author}{\bibfnamefont{A.}~\bibnamefont{C.~Durst}},
  \bibinfo{author}{\bibfnamefont{K.}~\bibnamefont{E.~Castoria}},
  \bibnamefont{and} \bibinfo{author}{\bibfnamefont{R.}~\bibnamefont{N.~Bhatt}},
  \bibinfo{journal}{Physical Review B} \textbf{\bibinfo{volume}{96}},
  \bibinfo{pages}{155208} (\bibinfo{year}{2017}).

\bibitem[{\citenamefont{Durst et~al.}(2020)\citenamefont{Durst, Yang-Mejia, and
  Bhatt}}]{Durst2020QIbAPiDS}
\bibinfo{author}{\bibfnamefont{A.~C.} \bibnamefont{Durst}},
  \bibinfo{author}{\bibfnamefont{G.}~\bibnamefont{Yang-Mejia}},
  \bibnamefont{and} \bibinfo{author}{\bibfnamefont{R.~N.} \bibnamefont{Bhatt}},
  \bibinfo{journal}{Phys. Rev. B} \textbf{\bibinfo{volume}{101}},
  \bibinfo{pages}{035202} (\bibinfo{year}{2020}).

\bibitem[{\citenamefont{Zhu et~al.}(2020)\citenamefont{Zhu, Wu, and
  Fisher}}]{Zhu2020LCoAOMfDPAAiS}
\bibinfo{author}{\bibfnamefont{J.}~\bibnamefont{Zhu}},
  \bibinfo{author}{\bibfnamefont{W.}~\bibnamefont{Wu}}, \bibnamefont{and}
  \bibinfo{author}{\bibfnamefont{A.~J.} \bibnamefont{Fisher}},
  \bibinfo{journal}{Phys. Rev. B} \textbf{\bibinfo{volume}{101}},
  \bibinfo{pages}{085303} (\bibinfo{year}{2020}).

\bibitem[{\citenamefont{Schofield et~al.}(2003)\citenamefont{Schofield, Curson,
  Simmons, Ruess, Hallam, Oberbeck, and Clark}}]{Schofield2003APPoSDiS}
\bibinfo{author}{\bibfnamefont{S.~R.} \bibnamefont{Schofield}},
  \bibinfo{author}{\bibfnamefont{N.~J.} \bibnamefont{Curson}},
  \bibinfo{author}{\bibfnamefont{M.~Y.} \bibnamefont{Simmons}},
  \bibinfo{author}{\bibfnamefont{F.~J.} \bibnamefont{Ruess}},
  \bibinfo{author}{\bibfnamefont{T.}~\bibnamefont{Hallam}},
  \bibinfo{author}{\bibfnamefont{L.}~\bibnamefont{Oberbeck}}, \bibnamefont{and}
  \bibinfo{author}{\bibfnamefont{R.~G.} \bibnamefont{Clark}},
  \bibinfo{journal}{Physical review letters} \textbf{\bibinfo{volume}{91}},
  \bibinfo{pages}{136104} (\bibinfo{year}{2003}).

\bibitem[{\citenamefont{Dwyer et~al.}(2021)\citenamefont{Dwyer, Baek, Farzaneh,
  Dreyer, Williams, and Butera}}]{Dwyer2021}
\bibinfo{author}{\bibfnamefont{K.~J.} \bibnamefont{Dwyer}},
  \bibinfo{author}{\bibfnamefont{S.}~\bibnamefont{Baek}},
  \bibinfo{author}{\bibfnamefont{A.}~\bibnamefont{Farzaneh}},
  \bibinfo{author}{\bibfnamefont{M.}~\bibnamefont{Dreyer}},
  \bibinfo{author}{\bibfnamefont{J.~R.} \bibnamefont{Williams}},
  \bibnamefont{and} \bibinfo{author}{\bibfnamefont{R.~E.}
  \bibnamefont{Butera}}, \bibinfo{journal}{arXiv2103.07529}
  (\bibinfo{year}{2021}).

\bibitem[{\citenamefont{Clauws et~al.}(1988)\citenamefont{Clauws, Broeckx,
  Rotsaert, and Vennik}}]{Clauws1989OSoSISiGaS}
\bibinfo{author}{\bibfnamefont{P.}~\bibnamefont{Clauws}},
  \bibinfo{author}{\bibfnamefont{J.}~\bibnamefont{Broeckx}},
  \bibinfo{author}{\bibfnamefont{E.}~\bibnamefont{Rotsaert}}, \bibnamefont{and}
  \bibinfo{author}{\bibfnamefont{J.}~\bibnamefont{Vennik}},
  \bibinfo{journal}{Physical review. B, Condensed matter}
  \textbf{\bibinfo{volume}{38}}, \bibinfo{pages}{12377} (\bibinfo{year}{1988}).

\bibitem[{\citenamefont{Vinh et~al.}(2013)\citenamefont{Vinh, Redlich, van~der
  Meer, Pidgeon, Greenland, Lynch, Aeppli, and Murdin}}]{Vinh2013TRDoSATiS}
\bibinfo{author}{\bibfnamefont{N.~Q.} \bibnamefont{Vinh}},
  \bibinfo{author}{\bibfnamefont{B.}~\bibnamefont{Redlich}},
  \bibinfo{author}{\bibfnamefont{A.~F.~G.} \bibnamefont{van~der Meer}},
  \bibinfo{author}{\bibfnamefont{C.~R.} \bibnamefont{Pidgeon}},
  \bibinfo{author}{\bibfnamefont{P.~T.} \bibnamefont{Greenland}},
  \bibinfo{author}{\bibfnamefont{S.~A.} \bibnamefont{Lynch}},
  \bibinfo{author}{\bibfnamefont{G.}~\bibnamefont{Aeppli}}, \bibnamefont{and}
  \bibinfo{author}{\bibfnamefont{B.~N.} \bibnamefont{Murdin}},
  \bibinfo{journal}{Physical Review X} \textbf{\bibinfo{volume}{3}},
  \bibinfo{pages}{011019} (\bibinfo{year}{2013}).

\bibitem[{\citenamefont{Dai et~al.}(1992)\citenamefont{Dai, Zhang, and
  P.~Sarachik}}]{Dai1992ECoMSntMIT}
\bibinfo{author}{\bibfnamefont{P.}~\bibnamefont{Dai}},
  \bibinfo{author}{\bibfnamefont{Y.}~\bibnamefont{Zhang}}, \bibnamefont{and}
  \bibinfo{author}{\bibfnamefont{M.}~\bibnamefont{P.~Sarachik}},
  \bibinfo{journal}{Physical review. B, Condensed matter}
  \textbf{\bibinfo{volume}{45}}, \bibinfo{pages}{3984} (\bibinfo{year}{1992}).

\bibitem[{\citenamefont{Litvinenko et~al.}(2015)\citenamefont{Litvinenko,
  Bowyer, Greenland, Stavrias, Li, Gwilliam, Villis, Matmon, Pang, Redlich
  et~al.}}]{Litvinenko2015CCaDoOWiSwEaORO}
\bibinfo{author}{\bibfnamefont{K.}~\bibnamefont{Litvinenko}},
  \bibinfo{author}{\bibfnamefont{E.}~\bibnamefont{Bowyer}},
  \bibinfo{author}{\bibfnamefont{P.}~\bibnamefont{Greenland}},
  \bibinfo{author}{\bibfnamefont{N.}~\bibnamefont{Stavrias}},
  \bibinfo{author}{\bibfnamefont{J.}~\bibnamefont{Li}},
  \bibinfo{author}{\bibfnamefont{R.}~\bibnamefont{Gwilliam}},
  \bibinfo{author}{\bibfnamefont{B.}~\bibnamefont{Villis}},
  \bibinfo{author}{\bibfnamefont{G.}~\bibnamefont{Matmon}},
  \bibinfo{author}{\bibfnamefont{M.}~\bibnamefont{Pang}},
  \bibinfo{author}{\bibfnamefont{B.}~\bibnamefont{Redlich}},
  \bibnamefont{et~al.}, \bibinfo{journal}{Nature communications}
  \textbf{\bibinfo{volume}{6}}, \bibinfo{pages}{6549} (\bibinfo{year}{2015}).

\bibitem[{\citenamefont{van~der Heijden et~al.}(2018)\citenamefont{van~der
  Heijden, Kobayashi, House, Salfi, Barraud, Lavi{\'e}ville, Simmons, and
  Rogge}}]{Heijden2018RaCotSOSoTCAAiaST}
\bibinfo{author}{\bibfnamefont{J.}~\bibnamefont{van~der Heijden}},
  \bibinfo{author}{\bibfnamefont{T.}~\bibnamefont{Kobayashi}},
  \bibinfo{author}{\bibfnamefont{M.~G.} \bibnamefont{House}},
  \bibinfo{author}{\bibfnamefont{J.}~\bibnamefont{Salfi}},
  \bibinfo{author}{\bibfnamefont{S.}~\bibnamefont{Barraud}},
  \bibinfo{author}{\bibfnamefont{R.}~\bibnamefont{Lavi{\'e}ville}},
  \bibinfo{author}{\bibfnamefont{M.~Y.} \bibnamefont{Simmons}},
  \bibnamefont{and} \bibinfo{author}{\bibfnamefont{S.}~\bibnamefont{Rogge}},
  \bibinfo{journal}{Science Advances} \textbf{\bibinfo{volume}{4}},
  \bibinfo{pages}{eaat9199} (\bibinfo{year}{2018}).

\bibitem[{\citenamefont{Corna et~al.}(2018)\citenamefont{Corna, Bourdet,
  Maurand, Crippa, Kotekar-Patil, Bohuslavskyi, Lavieville, Hutin, Barraud,
  Jehl et~al.}}]{Corna2018EDESRMbSVOCiaSQD}
\bibinfo{author}{\bibfnamefont{A.}~\bibnamefont{Corna}},
  \bibinfo{author}{\bibfnamefont{L.}~\bibnamefont{Bourdet}},
  \bibinfo{author}{\bibfnamefont{R.}~\bibnamefont{Maurand}},
  \bibinfo{author}{\bibfnamefont{A.}~\bibnamefont{Crippa}},
  \bibinfo{author}{\bibfnamefont{D.}~\bibnamefont{Kotekar-Patil}},
  \bibinfo{author}{\bibfnamefont{H.}~\bibnamefont{Bohuslavskyi}},
  \bibinfo{author}{\bibfnamefont{R.}~\bibnamefont{Lavieville}},
  \bibinfo{author}{\bibfnamefont{L.}~\bibnamefont{Hutin}},
  \bibinfo{author}{\bibfnamefont{S.}~\bibnamefont{Barraud}},
  \bibinfo{author}{\bibfnamefont{X.}~\bibnamefont{Jehl}}, \bibnamefont{et~al.},
  \bibinfo{journal}{npj quantum information} \textbf{\bibinfo{volume}{4}},
  \bibinfo{pages}{6} (\bibinfo{year}{2018}).

\bibitem[{\citenamefont{Crippa et~al.}(2018)\citenamefont{Crippa, Maurand,
  Bourdet, Kotekar-Patil, Amisse, Jehl, Sanquer, Lavi\'eville, Bohuslavskyi,
  Hutin et~al.}}]{Crippa2018ESDbMMiSOQ}
\bibinfo{author}{\bibfnamefont{A.}~\bibnamefont{Crippa}},
  \bibinfo{author}{\bibfnamefont{R.}~\bibnamefont{Maurand}},
  \bibinfo{author}{\bibfnamefont{L.}~\bibnamefont{Bourdet}},
  \bibinfo{author}{\bibfnamefont{D.}~\bibnamefont{Kotekar-Patil}},
  \bibinfo{author}{\bibfnamefont{A.}~\bibnamefont{Amisse}},
  \bibinfo{author}{\bibfnamefont{X.}~\bibnamefont{Jehl}},
  \bibinfo{author}{\bibfnamefont{M.}~\bibnamefont{Sanquer}},
  \bibinfo{author}{\bibfnamefont{R.}~\bibnamefont{Lavi\'eville}},
  \bibinfo{author}{\bibfnamefont{H.}~\bibnamefont{Bohuslavskyi}},
  \bibinfo{author}{\bibfnamefont{L.}~\bibnamefont{Hutin}},
  \bibnamefont{et~al.}, \bibinfo{journal}{Phys. Rev. Lett.}
  \textbf{\bibinfo{volume}{120}}, \bibinfo{pages}{137702}
  (\bibinfo{year}{2018}).

\bibitem[{\citenamefont{Crippa et~al.}(2019)\citenamefont{Crippa, Ezzouch,
  Aprá, Amisse, Lavieville, Hutin, Bertrand, Vinet, Urdampilleta, Meunier
  et~al.}}]{Crippa2019GRDRaCCoaSQiS}
\bibinfo{author}{\bibfnamefont{A.}~\bibnamefont{Crippa}},
  \bibinfo{author}{\bibfnamefont{R.}~\bibnamefont{Ezzouch}},
  \bibinfo{author}{\bibfnamefont{A.}~\bibnamefont{Aprá}},
  \bibinfo{author}{\bibfnamefont{A.}~\bibnamefont{Amisse}},
  \bibinfo{author}{\bibfnamefont{R.}~\bibnamefont{Lavieville}},
  \bibinfo{author}{\bibfnamefont{L.}~\bibnamefont{Hutin}},
  \bibinfo{author}{\bibfnamefont{B.}~\bibnamefont{Bertrand}},
  \bibinfo{author}{\bibfnamefont{M.}~\bibnamefont{Vinet}},
  \bibinfo{author}{\bibfnamefont{M.}~\bibnamefont{Urdampilleta}},
  \bibinfo{author}{\bibfnamefont{T.}~\bibnamefont{Meunier}},
  \bibnamefont{et~al.}, \bibinfo{journal}{Nature Communications}
  \textbf{\bibinfo{volume}{10}}, \bibinfo{pages}{2776} (\bibinfo{year}{2019}).

\bibitem[{\citenamefont{Salfi et~al.}(2016{\natexlab{b}})\citenamefont{Salfi,
  Mol, Rahman, Klimeck, Simmons, C.~L.~Hollenberg, and
  Rogge}}]{Salfi2016QSotHMwDAiS}
\bibinfo{author}{\bibfnamefont{J.}~\bibnamefont{Salfi}},
  \bibinfo{author}{\bibfnamefont{J.}~\bibnamefont{Mol}},
  \bibinfo{author}{\bibfnamefont{R.}~\bibnamefont{Rahman}},
  \bibinfo{author}{\bibfnamefont{G.}~\bibnamefont{Klimeck}},
  \bibinfo{author}{\bibfnamefont{M.}~\bibnamefont{Simmons}},
  \bibinfo{author}{\bibfnamefont{L.}~\bibnamefont{C.~L.~Hollenberg}},
  \bibnamefont{and} \bibinfo{author}{\bibfnamefont{S.}~\bibnamefont{Rogge}},
  \bibinfo{journal}{Nature Communications} \textbf{\bibinfo{volume}{7}},
  \bibinfo{pages}{11342} (\bibinfo{year}{2016}{\natexlab{b}}).

\bibitem[{\citenamefont{Zak}(1989)}]{Zak1989BPfEBiS}
\bibinfo{author}{\bibfnamefont{J.}~\bibnamefont{Zak}}, \bibinfo{journal}{Phys.
  Rev. Lett.} \textbf{\bibinfo{volume}{62}}, \bibinfo{pages}{2747}
  (\bibinfo{year}{1989}).

\bibitem[{\citenamefont{Clementi and Davis}(1966)}]{Clementi1966ESoLMS}
\bibinfo{author}{\bibfnamefont{E.}~\bibnamefont{Clementi}} \bibnamefont{and}
  \bibinfo{author}{\bibfnamefont{D.}~\bibnamefont{Davis}},
  \bibinfo{journal}{Journal of Computational Physics}
  \textbf{\bibinfo{volume}{1}}, \bibinfo{pages}{223} (\bibinfo{year}{1966}).

\bibitem[{\citenamefont{Szab{\'o} and Ostlund}(1982)}]{Szabo1982MQCItAEST}
\bibinfo{author}{\bibfnamefont{A.}~\bibnamefont{Szab{\'o}}} \bibnamefont{and}
  \bibinfo{author}{\bibfnamefont{N.}~\bibnamefont{Ostlund}},
  \emph{\bibinfo{title}{Modern Quantum Chemistry: Introduction to Advanced
  Electronic Structure Theory}} (\bibinfo{publisher}{Macmillan},
  \bibinfo{year}{1982}).

\bibitem[{\citenamefont{Le et~al.}(2020)\citenamefont{Le, Fisher, Curson, and
  Ginossar}}]{Le2020TPoaDFHMfSNL}
\bibinfo{author}{\bibfnamefont{N.~H.} \bibnamefont{Le}},
  \bibinfo{author}{\bibfnamefont{A.~J.} \bibnamefont{Fisher}},
  \bibinfo{author}{\bibfnamefont{N.~J.} \bibnamefont{Curson}},
  \bibnamefont{and} \bibinfo{author}{\bibfnamefont{E.}~\bibnamefont{Ginossar}},
  \bibinfo{journal}{npj Quantum Information} \textbf{\bibinfo{volume}{6}},
  \bibinfo{pages}{24} (\bibinfo{year}{2020}).

\bibitem[{\citenamefont{Fidkowski and Kitaev}(2010)}]{Fidkowski2010EoIotTCoFFS}
\bibinfo{author}{\bibfnamefont{L.}~\bibnamefont{Fidkowski}} \bibnamefont{and}
  \bibinfo{author}{\bibfnamefont{A.}~\bibnamefont{Kitaev}},
  \bibinfo{journal}{Phys. Rev. B} \textbf{\bibinfo{volume}{81}},
  \bibinfo{pages}{134509} (\bibinfo{year}{2010}).

\bibitem[{\citenamefont{Turner et~al.}(2011)\citenamefont{Turner, Pollmann, and
  Berg}}]{Turner2011TPoODFaEPoV}
\bibinfo{author}{\bibfnamefont{A.~M.} \bibnamefont{Turner}},
  \bibinfo{author}{\bibfnamefont{F.}~\bibnamefont{Pollmann}}, \bibnamefont{and}
  \bibinfo{author}{\bibfnamefont{E.}~\bibnamefont{Berg}},
  \bibinfo{journal}{Phys. Rev. B} \textbf{\bibinfo{volume}{83}},
  \bibinfo{pages}{075102} (\bibinfo{year}{2011}).

\bibitem[{\citenamefont{Grusdt et~al.}(2013)\citenamefont{Grusdt, H\"oning, and
  Fleischhauer}}]{PhysRevLett.110.260405}
\bibinfo{author}{\bibfnamefont{F.}~\bibnamefont{Grusdt}},
  \bibinfo{author}{\bibfnamefont{M.}~\bibnamefont{H\"oning}}, \bibnamefont{and}
  \bibinfo{author}{\bibfnamefont{M.}~\bibnamefont{Fleischhauer}},
  \bibinfo{journal}{Phys. Rev. Lett.} \textbf{\bibinfo{volume}{110}},
  \bibinfo{pages}{260405} (\bibinfo{year}{2013}).

\end{thebibliography}

\end{document}